\documentclass[12pt]{article}
\usepackage{epsf}

\usepackage{epsfig,graphics}
\usepackage {graphicx}
\usepackage {epsfig}
\usepackage {subfigure}
\usepackage {tabularx} 
\usepackage{rotate}	
\usepackage{slashed}
\usepackage{amsmath}
\usepackage{amsfonts}
\usepackage{amssymb}
\usepackage{graphicx}
\usepackage{cite}

\newcommand{\bmat}{\left(\begin{array}}
\newcommand{\emat}{\end{array}\right)}

\def\yzero{\smash{\hbox{$y\kern-4pt\raise1pt\hbox{${}^\circ$}$}}}

\def\beq{\begin{equation}}
\def\eeq{\end{equation}}
\def\beqa{\begin{eqnarray}}
\def\eeqa{\end{eqnarray}}

\def\-{\hphantom{-}}

\def\s2{\frac{1}{\sqrt2}}

\def\beq{\begin{equation}}
\def\eeq{\end{equation}}
\def\beqa{\begin{eqnarray}}
\def\eeqa{\end{eqnarray}}

\def\IF{\relax{\rm I\kern-.18em F}}
\def\II{\relax{\rm I\kern-.18em I}}
\def\IP{\relax{\rm I\kern-.18em P}}
\def\IC{\relax\hbox{\kern.25em$\inbar\kern-.3em{\rm C}$}}
\def\IR{\relax{\rm I\kern-.18em R}}

\def\Dsl{\,\raise.15ex\hbox{/}\mkern-13.5mu D} %this one can be subscripted
\def\IZ{Z\kern-.4em  Z}

\newcommand{\Abb}[2]{A_{\left[\bar{#1}\right.}A_{\bar{#2}]}}

\newcommand{\Abn}[2]{A_{\left[\bar{#1}\right.}A_{#2]}}

\newcommand{\Gb}[0]{G_{\bar 1 \bar 2 \bar 3}}
\newcommand{\Gn}[0]{G_{123}}
\newcommand{\Sb}[1]{S_{\bar #1 \bar #1}}
\newcommand{\Sn}[1]{S_{#1 #1}}
\catcode`\@=11   
\newdimen\@rotdimen
\newbox\@rotbox  

\def\@vspec#1{\special{ps:#1}}%  passes #1 verbatim to the output
\def\@rotstart#1{\@vspec{gsave currentpoint currentpoint translate
   #1 neg exch neg exch translate}}% #1 can be any origin-fixing transformation
\def\@rotfinish{\@vspec{currentpoint grestore moveto}}% gets back in synch 
\def\@rotr#1{\@rotdimen=\ht#1\advance\@rotdimen by\dp#1%
   \hbox to\@rotdimen{\hskip\ht#1\vbox to\wd#1{\@rotstart{90 rotate}%
   \box#1\vss}\hss}\@rotfinish}
\def\@rotl#1{\@rotdimen=\ht#1\advance\@rotdimen by\dp#1%
   \hbox to\@rotdimen{\vbox to\wd#1{\vskip\wd#1\@rotstart{270 rotate}%
   \box#1\vss}\hss}\@rotfinish}%
\def\@rotu#1{\@rotdimen=\ht#1\advance\@rotdimen by\dp#1%
   \hbox to\wd#1{\hskip\wd#1\vbox to\@rotdimen{\vskip\@rotdimen
   \@rotstart{-1 dup scale}\box#1\vss}\hss}\@rotfinish}%
\def\@rotf#1{\hbox to\wd#1{\hskip\wd#1\@rotstart{-1 1 scale}%
   \box#1\hss}\@rotfinish}%
\def\rotate{\@ifnextchar[{\@rotate}{\@rotate[l]}}
\def\@rotate[#1]#2{\setbox\@rotbox=\hbox{#2}\@nameuse{@rot#1}\@rotbox}

\catcode`\@=12
\topmargin
-1.5cm
\textwidth
15.5cm
\textheight
23.5cm
\oddsidemargin
0.7cm
\evensidemargin
0.7cm

\begin{document}

%----------------------------------------------------------------------%
%  numbering equations with section number
%----------------------------------------------------------------------%
\makeatletter
\@addtoreset{equation}{section}
\makeatother
\renewcommand{\theequation}{\thesection.\arabic{equation}}
%----------------------------------------------------------------------%
%  title page
%----------------------------------------------------------------------%
\pagestyle{empty}
%\vspace*{1.0in}
\vspace{-0.2cm}
\rightline{ IFT-UAM/CSIC-14-29}
%\rightline{FTUAM-14-XX}
%\rightline{\tt hep-th/xxxxxxx}
\vspace{1.2cm}
\begin{center}

%\vspace{0.5cm}

\LARGE{Flux-induced Soft Terms on Type IIB/F-theory Matter Curves and \\
 Hypercharge Dependent Scalar Masses} 
\\[13mm]
  \large{Pablo G. C\'amara$^{a}$,   Luis E. Ib\'a\~nez$^{b} $ and Irene  Valenzuela$^{b}$  \\[6mm]}
\small{
$^a$ Departament de F\'{\i}sica Fonamental and Institut de Ci\`encies\\[-0.3em] 
del Cosmos, Universitat de Barcelona, Mart\'{\i} i Franqu\'es 1, 08028 Barcelona, Spain\\
 $^b$  Departamento de F\'{\i}sica Te\'orica
and Instituto de F\'{\i}sica Te\'orica UAM/CSIC,\\[-0.3em]
Universidad Aut\'onoma de Madrid,
Cantoblanco, 28049 Madrid, Spain 
\\[8mm]}
\small{\bf Abstract} \\[7mm]
\end{center}
\begin{center}
\begin{minipage}[h]{15.22cm}
Closed string fluxes induce generically 
SUSY-breaking soft terms on supersymmetric 
type IIB orientifold compactifications with  D3/D7 branes.
This was studied in the past by inserting those fluxes on the 
DBI+CS  actions  for adjoint D3/D7 fields, 
where D7-branes had no magnetic fluxes.  In the present work we generalise those
computations to the phenomenologically more relevant  case  of chiral
bi-fundamental fields laying at 7-brane intersections and F-theory local
matter curves. We also include  the effect of 7-brane  magnetic flux as well as 
 more general closed string  backgrounds, including the effect of distant 
 D3(${\overline {D3}}$)-branes.
  We discuss several applications 
 of our results.  We find that squark/slepton masses become in general 
 flux-dependent in F-theory GUT's.  Hypercharge-dependent non-universal scalar masses 
  with a characteristic sfermion hierarchy  $m_E^2<m_L^2 <m_Q^2  < m_D^2 < m_U^2$ are  obtained.
 There are also   flavor-violating  soft terms both for
 matter fields living at intersecting 7-branes or on D3-branes at singularities. 
 They point at a very heavy sfermion  spectrum to avoid FCNC constraints.
 We also discuss the possible microscopic description  of the 
 fine-tuning of the EW Higgs boson in compactifications with a MSSM spectrum.

\end{minipage}
\end{center}
\newpage
%----------------------------------------------------------------------%
%  Resetting of counters
%----------------------------------------------------------------------%
\setcounter{page}{1}
\pagestyle{plain}
\renewcommand{\thefootnote}{\arabic{footnote}}
\setcounter{footnote}{0}
%----------------------------------------------------------------------%
%  Paper begins
%----------------------------------------------------------------------%

%\end{document}

%\end{document}

\tableofcontents

\newpage 
%&&&&&&&&&&&&&&&&&&&&&&&&&&&&&&&&&
\section{Introduction}
%&&&&&&&&&&&&&&&&&&&&&&&&&&&&&&&&&

Supersymmetry is probably the most elegant and attractive symmetry beyond the Standard Model (SM) which has been proposed so far.
One of its more relevant properties is the stability of scalars against radiative corrections, which makes a 
SUSY version of the SM (like the MSSM) a possible solution to the hierarchy/naturality problem. The recent discovery at LHC of a
scalar particle with the properties of the Higgs particle and a mass around $m_H\simeq 126$  GeV \cite{:2012gu,:2012gk} is consistent with the expectations of the 
MSSM, which predict $m_H\leq 130$ GeV.  On the other hand such a value for the Higgs mass requires  a quite massive spectrum of SUSY particles,
with squarks typically  heavier than $3-5$~TeV at least and possibly much higher.  This is is also in accord with the fact that no trace
of SUSY particles has been observed at LHC at 8~TeV. Still SUSY remains singled out as one of the most attractive ideas
to stabilise the hierarchy of scales. This is particularly so within the context of String Theory, in which
supersymmetry is a built-in ingredient.

There is a variety of ways in order to obtain chiral $\mathcal{N}=1$ SUSY string compactifications, many of them related by different
string dualities.  In recent years much effort has been dedicated to the study of type IIB orientifold compactifications
with unbroken $\mathcal{N}=1$, $D=4$ SUSY \cite{BOOK}. Its non-perturbative extension, F-theory  \cite{Vafa:1996xn}
compactified on Calabi-Yau (CY) 4-folds has been in particular extensively analysed. 
One can construct local F-theory SU(5) unified models with a number of phenomenologically interesting properties \cite{Donagi:2008ca,Beasley:2008dc,Beasley:2008kw,Donagi:2008kj,Heckman:2010bq,Weigand:2010wm,Leontaris:2012mh,Maharana:2012tu}, 
including gauge coupling unification and a large top quark  Yukawa coupling. Furthermore,  closed string  RR and NS fluxes in type IIB
compactifications may lead, when combined with other dynamical effects,  to a fixing of all the moduli 
\cite{Giddings:2001yu,Kachru:2003aw,fluxes}.

A crucial ingredient to make contact with low-energy physics is the structure of the SUSY breaking soft terms.  In trying to study
those,  two complementary paths have been followed:

\begin{itemize}

\item {\it Bottom-up local approach}. In this case one studies the physics of a local set of D7-branes (or D3-branes), without a full  knowledge 
of the complete compact space.  SUSY-breaking is felt by the D-branes as induced by the closed string backgrounds in the
vicinity of the branes. These backgrounds include  RR and NS 3-form fluxes as well as a 5-form flux, dilaton and 
metric backgrounds. They parametrize our ignorance of the full compactification details.
The SUSY-breaking soft terms may be obtained by expanding the DBI+CS 7-brane action 
around its location including closed string background insertions.

\item {\it   $\mathcal{N}=1$ supergravity effective action}. Here one starts from the effective supergavity action in terms of the
K\"ahler potential of the moduli fields and the K\"ahler metric of the matter fields. The superpotential includes a
Gukov-Vafa-Witten moduli-dependent piece fixing the complex dilaton and the complex structure moduli as well as 
non-perturbative superpotentials included to fix the K\"ahler moduli. 

\end{itemize}

Both approaches have advantages and shortcomings. The first gives us a microscopic description of the
origin of the soft terms but no information on the global structure of the compactification, including how
closed string moduli are fixed. On the other hand the effective supergravity approach requires a detailed 
knowledge of the K\"ahler potential of the moduli as well as the matter metrics and the allowed
non-perturbative effects. Having a full control of these latter aspects in specific compactifications
is a challenge.  In this paper we will follow the first bottom-up strategy to study  
SUSY-breaking soft terms induced by  closed and open string backgrounds on localised sets
of  bulk and/or intersecting 7-branes. 

In Ref.~\cite{Camara:2004jj} (see also \cite{Lust:2004fi}) soft terms induced by closed string 3-form $G_3$ fluxes on bulk D7-branes were obtained
by starting with the DBI+CS action and inserting closed string backgrounds. The matter fields transform in the adjoint
representation,  so  that the results are not of direct phenomenological interest.  Fully realistic  models require two additional
features to induce chirality:  i) intersecting 7-branes and ii) open string magnetic fluxes on the
world-volume of the branes. As we said, a particularly interesting class of models are
local F-theory SU(5) GUT's,  in which SM fermions live in  1-complex dimensional {\it matter curves} which may be
interpreted as 7-brane intersections.  In these models chirality is induced by open string fluxes. Furthermore, the
breaking from SU(5) to the SM gauge group appears through non-vanishing hypercharge fluxes $ \langle F_Y\rangle\not= 0$.
This novel way to break a GUT symmetry requires certain conditions on the structure of U(1) couplings to
axions and it is not directly available in other classes of string compactifications like the  heterotic string.
In this paper we address the computation of soft terms for chiral fields at matter curves. We also study the effect from
magnetic fluxes, including hypercharge, on the obtained soft terms.

First, we revisit in section \ref{sec2} the structure of soft terms for the world-volume adjoint fields of 7-branes. 
We start with the DBI+CS D7-brane action and switch on both ISD and IASD closed string fluxes.
Here we generalise the results of \cite{Camara:2004jj} by allowing for the simultaneous presence of both classes of flux.
We then switch on in addition open string magnetic fluxes and compute their effect on the soft terms, which appear 
in this case at quadratic order on the open string fluxes.  Although the addition of magnetic fluxes may induce chirality starting
with bulk 7-branes, 
in the context of F-theory GUT model building, the divisor $S$  wrapped by the 
GUT 7-branes is rigid. That means that the adjoint fields  $\Phi$ parametrizing the 7-brane location are absent and
no Yukawa coupling $\Phi\times A\times A$ is present. This is an additional reason to
consider the SM fields to be localised at intersecting 7-branes.

Chiral fields at  F-theory  matter  curves generically appear when the 7-brane geometric mode $\Phi$ gets 
a position-dependent vev on the divisor $S$.  At these  complex dimension-one curves the symmetry of the F-theory singularity of the
base ($A_4$ for an SU(5) GUT symmetry) is enhanced to a larger one (SU(6) or SO(10) for SU(5) GUT's).  One can write down a topologically twisted
action for the 6-dimensional theory on the matter curves \cite{Beasley:2008dc}, with  matter fields localised on the curves with an
exponential damping in the traverse directions. If there are appropriate magnetic fluxes on the matter curves,
upon reduction to 4D a chiral spectrum of GUT families appears  localised on the curves.

In section \ref{sec3} we study the soft terms induced by the closed string background on the chiral fields localized on these matter 
curves. To do this we combine the results obtained  for adjoint fields in section \ref{sec2} together with
knowledge of the structure of the local wave-functions on the matter curves. 
In addition, we also compute the leading open string flux corrections to those soft terms.
These leading corrections in this case turn out to be linear on the open string flux, rather than quadratic as in the case of adjoints. We also compare with a simple effective
single K\"ahler-modulus supergravity action and see how, in the simplest situations, ISD fluxes 
with no magnetic fluxes correspond 
to modulus-dominated SUSY breaking soft terms with matter fields of  {\it modular weight 1/2}. 

In section \ref{sec4} we discuss the possible effect of other distant sources that may be present in a 
complete compactification, on the local set of 7-branes. Those effects can be understood in terms of 
the  backreaction of the distant sources on the local closed string background.  Unlike D3-brane models, D7-branes have no D3-brane charge and 
to leading order they do not feel the presence of distant ${\overline {\rm D3}}$-branes. However, once magnetic fluxes are switched on the world-volume of 7-branes,
there is an  induced  D3-brane charge and distant ${\overline {\rm D3}}$-branes contribute to the soft scalar masses to quadratic order in the magnetic fluxes.  We also discuss additional compactification effects
in a simple example.

In section \ref{applic} we apply the obtained results in several directions. First we consider the local SO(12) configuration  of Ref.~\cite{yukawasSO(12)}, in
which a set of matter curves associated to the SU(5) matter fields is given, with an appropriate set of
magnetic fluxes consistent with local SU(5) chirality, hypercharge fluxes breaking SU(5) down to the SM gauge group and
doublet-triplet splitting. This local setting is appropriate for the study of Yukawa couplings 
of type ${\bf 10}\times {\bf \bar 5}\times {\bf \bar 5}_H$ which give rise to charged lepton and D-quark masses.
Such scheme  gives us an explicit arena in which the flux dependence of the soft terms
can be computed in some detail. We compute the magnetic flux corrections for  matter fields and find that
they depend both on the U(1) fluxes giving rise to chirality as well as on  hypercharge fluxes.  Magnetic fluxes 
may give substantial non-universal corrections, that may be as large as $50\%$ of the uncorrected squared masses.
We also find a hierarchy of sfermion  masses with 
$m_E^2<m_L^2 <m_Q^2  < m_D^2 < m_U^2$. This ordering is quite particular  and is different from what
RGE or MSSM D-term contributions originates.  We also compute the flux-dependent corrections to
trilinear scalar terms. In subsection \ref{sec52} we address the scalar soft masses  for  a local  F-theory $E_6$ setting
\cite{Font:2013ida},  
which is the appropriate enlarged symmetry for the generation of U-quark masses in F-theory.

As a second  application we consider the generation of flavor-violating soft terms induced by fluxes. As noted in \cite{Camara:2013fta},
if the closed string 3-form flux background strongly varies over the 4-cycle $S$, flavor-dependent soft scalar masses
appear both for squarks and sleptons.  In subsection \ref{nonuniv} we describe this effect within the context of the
SO(12) local model of Ref.~\cite{yukawasSO(12)}. We also consider the generation of $(m_{LR}^2)_{ij}$ non-diagonal transitions 
coming from trilinear couplings.  In general, the scale of soft terms is strongly constrained by experimental  limits on FCNC  
transitions, which point towards very massive squark and slepton spectra, in the multi-TeV regime or above.
In subsection \ref{sec53} we also briefly compare the flavor non-universalities here discussed with those arising in
MSSM-like  models of D3-branes at singularities. We point out that the backreation of localised sources in 
generic compactifications also tends to induce substantial flavor non-universalities in this case.

As  a final application, in subsection \ref{sec54} we discuss the different terms contributing to the Higgs mass in
a setting with the Higgs living at intersecting 7-branes. We do   this in view of the possibility that
the Higgs mass in a high-scale SUSY-breaking context could remain light due to a fine-tuning, trying to identify the microscopic origin of this
fine-tuning. One sees that there is a wealth of corrections contributing to the Higgs mass in such a setting.
One important point is  that the mass is directly sensitive to the {\it local} value of the flux densities rather than
to the integrated (integer) fluxes. In addition  it depends on the full geometry of the compactification, including the 
precise location of other localised objects.   

Finally, we leave section \ref{sec6}  for the  discussion and outlook.

\section{Soft terms on type IIB orientifolds with bulk matter fields}
\label{sec2}

In this section we review and extend the local computation of flux-induced SUSY-breaking soft-terms that was performed in Ref.~\cite{Camara:2004jj} for
7-brane scalars transforming in the adjoint represention of the gauge group. However, we 
consider slightly more general configurations than in \cite{Camara:2004jj}, allowing for the simultaneous presence of imaginary self-dual (ISD) and imaginary anti self-dual (IASD) 3-form fluxes as well as for magnetization on the 7-branes.
Even though only ISD fluxes provide for  solutions  to the 10D classical equations of motion, complete compactifications addressing moduli fixing typically  include
additional non-perturbative ingredients that generically induce IASD fluxes and other closed string backgrounds. That is why it is interesting to keep trace also of those.

More precisely, we consider closed string backgrounds of the general form
\begin{align}
ds^2 &= Z(x^m)^{-1/2} \eta_{\mu\nu} d\hat{x}^\mu d\hat{x}^\nu + Z(x^m)^{1/2} ds^2_{\rm CY} \label{back1}\\
\tau &= \tau(x^m) \nonumber\\
G_3 &= \frac{1}{3!}G_{lmn}(x^m)dx^l\wedge dx^m\wedge dx^n \nonumber\\
\chi_4 &= \chi(x^m)d\hat{x}^0\wedge d\hat{x}^1\wedge d\hat{x}^2 \wedge d\hat{x}^3 \nonumber \\
F_5 &= d\chi_4 + *_{10}d\chi_4 \nonumber
\end{align}
with $\tau = C_0 +ie^{-\phi}$ the complex axio-dilaton, $G_3=F_3-\tau H_3$ (with $F_3$ and $H_3$ the RR and NSNS flux respectively) and $ds_{\rm CY}^2$ the Ricci-flat metric of the underlying Calabi-Yau. Hatted coordinates are along the non-compact directions. 

At any point in the internal space the background can be decomposed according to the SU(3)-structure 
preserved by the compactification. In general the relation between local and global parameters of the 
background is however highly non-trivial, except for simple cases like toroidal compactifications 
where the local SU(3)-structure can be  straightforwardly extended into a global one.  

From the viewpoint of the local SU(3)-structure the antisymmetric flux $G_3$ transforms as a $\mathbf{20}=\overline{\mathbf{10}}+\mathbf{10}$, 
with the $\overline{\mathbf{10}}$ and $\mathbf{10}$ representations corresponding respectively to the ISD $G_3^+$ and 
IASD $G_3^-$ components of the 3-form flux, defined as
\begin{equation}
G_3^{\pm} = \frac12 (G_3 \mp i*_6 G_3)\ , \qquad *_6 G_3^\pm = \pm i G_3^\pm
\end{equation}
These components can be further decomposed into irreducible 
representations of SU(3). Thus, IASD fluxes in the $\mathbf{10}$ are decomposed according to $\mathbf{10} = \mathbf{6} + \mathbf{3} + \mathbf{1}$, 
where the $\mathbf{6}$ and $\mathbf{3}$ representations read \cite{hep-th/0009211}
\begin{align}
S_{ij} &= \frac12 (\epsilon_{ikl}G_{j\bar k\bar l}+\epsilon_{jkl}G_{i\bar k\bar l})\\
A_{ij} &= \frac12 (\epsilon_{\bar i\bar k\bar l}G_{kl\bar j}-\epsilon_{\bar j\bar k \bar l}G_{kl\bar i})\nonumber
\end{align}
respectively, whereas the singlet is given by the $G_{123}$ component of the flux, proportional to the holomorphic 3-form $\Omega$ of the internal space. Local coordinates 
are complexified according to the local complex structure as $z^m=\frac{1}{\sqrt{2}}(x^{2m+2}+ix^{2m+3})$, 
$m=1,2,3$. Similar definitions apply
in the decomposition of ISD fluxes into $G_{\bar 1\bar 2\bar 3}$, $S_{\bar i\bar j}$ and $A_{ij}$. For simplicity and to avoid cumbersome expressions, in this paper we take $S_{12}=A_{12}=S_{\bar 1\bar 2}=A_{\bar 1\bar 2}=0$. The dependence on these components can be obtained by requiring $SO(4)\times SO(2)$ convariance in our expressions \cite{Camara:2004jj}. Furthermore, the tensors $A_{ij}$ and $A_{\bar i\bar j}$ correspond respectively to (1,2) and (2,1) non-primitive components of the flux, that are incompatible with the cohomology of a Calabi-Yau (although a local component in principle could be allowed). In addition we set $S_{3i}=S_{\bar 3\bar i}=0$, where $z^3$ is the complex direction transverse to the D7-branes, since those flux components generically lead to Freed-Witten (FW) anomalies in the worldvolume of D7-branes, as discussed in \cite{Camara:2004jj}. 

Being defined in the $M_{\rm Pl.}\to \infty$ limit, soft-terms in the effective 8d theory of a stack of 7-branes can be understood 
from the background in a local transverse patch around the stack of 7-branes. Such local background receives in general contributions 
from globally non-trivial fluxes as well as from the backreaction of distant sources, as we discuss in section \ref{sec4}. Thus, we expand the 
background (\ref{back1}) around the stack of 7-branes as
\begin{align}
Z^{-1/2} &= 1+\frac12 K_{mn}y^m y^n + \ldots \label{expan}\\
\tau & = \tau_0 + \frac12 \tau_{mn}y^my^n + \ldots\nonumber \\
\chi &= \textrm{const.} + \frac12 \chi_{mn} y^m y^n+\ldots\nonumber\\
G_{lmn}(x^r)&= G_{lmn}+\ldots \nonumber
\end{align}
where we have denoted by $y^m$ the two coordinates that are transverse to the stack of 7-branes and which for the sake of concreteness in what follows we take to be $x^8$ and $x^9$. Dots in the rhs of eqs.~(\ref{expan}) represent higher order terms in the expansion, and will only contribute to non-renormalizable couplings in the 4d effective action. In the next subsections we make use of this local expansion to compute the flux-induced soft-breaking terms for the adjoint fields of a stack of 
7-branes.

\subsection{Unmagnetized bulk D7-brane fields}
\label{sec21}

We first address the case of unmagnetized 7-branes, leaving the case of magnetized branes for the next subsection. We closely follow the 
procedure developed in \cite{Camara:2004jj}. Thus, we expand the DBI+CS action of D7-branes in transverse coordinates in presence of 
the local background (\ref{back1}) and (\ref{expan}), and make use of the identification
\begin{equation}
z^3 = 2\pi \alpha' \Phi \label{iden}
\end{equation}
to derive an 8d effective action that contains flux-induced SUSY-breaking 
soft terms. Dimensional reduction then leads to a soft-breaking Lagrangian in the 4d effective theory.
	¼
The relevant piece of the D7-brane DBI+CS action for the computation of flux-induced soft terms is given by
\begin{equation}
S=-\mu_7\, \textrm{STr}\left[\int d^8\xi \, e^{-\phi} \sqrt{-\textrm{det}\left(P[E_{ab}]+\sigma F_{ab}\right)}- g_s \int P\left[ -C_6\wedge \mathcal{F}_2+C_8\right]\right]\label{action1}
\end{equation}
where
\begin{equation}
E_{ab}= e^{\phi/2}G_{ab}-B_{ab}\ , \qquad
\sigma=2\pi\alpha' ,\qquad
\mu_7=(2\pi)^{-3}\sigma^{-4}g_s^{-1}\ , \qquad \mathcal{F}_2\equiv B_2-\sigma F_2
\end{equation}
`STr' denotes the symmetrized trace over gauge indices and $P[\, \cdot\, ]$ is the pull-back to the 7-brane worldvolume. Our conventions are such that the metric has signature diag$(- + + + \ldots)$ whereas $dz^1\wedge d\bar z^1\wedge dz^2\wedge d\bar z^2$ has negative signature.

The terms contributing to the determinant in the DBI piece of the action are given by
\begin{equation}
\textrm{det}(P[E_{ab}])=e^{4\phi}\,\textrm{det}\left(g_{ab}+2\sigma^2 D_{(a}\Phi D_{b)}\bar\Phi-e^{-\phi/2}\mathcal{F}_{ab}\right)
\end{equation}
Expanding the determinant as well as the square root in the DBI piece then leads to the following 8d Lagrangian\footnote{The non-conventional sign for the kinetic term of $\Phi$ is due to the particular signature that we have taken of the 8d metric.}
\begin{equation}
\mathcal{L}_{\rm 8d}= \mu_7 e^\phi\,  \textrm{STr}\left(-1-\sigma^2D_a\Phi D_a\bar\Phi-\frac{g_s^{-1}}{4} \mathcal{F}_{ab}\mathcal{F}_{ab}+C_8-C_6\wedge \mathcal{F}_2\right) \ .\label{dbi1}
\end{equation}

In order to proceed further we should relate the dilaton $\phi$, the $B$-field and the RR-fields that appear in this expression to fluctuations of the 8d field $\Phi$ in the limit $M_{\rm Pl}\to \infty$. Let us first address the case of the axio-dilaton. Complexifying the second equation in (\ref{expan}) and making use of eq.~(\ref{iden}) we write
\begin{equation}
\tau = ig_s^{-1} \left(1+ \frac{\sigma^2\tau_{33}}{2} \Phi^2+\frac{\sigma^2\tau_{\bar 3\bar 3}}{2}\bar{\Phi}^2+\sigma^2\tau_{3\bar 3}|\Phi|^2 + \ldots \right)\label{tauex}
\end{equation}
where for simplicity we have fixed $\langle \tau\rangle = ig_s^{-1}$. The 10d supergravity equations of motion then put restrictions 
on the parameters of this expansion. More precisely, from the equation
\begin{equation}
\nabla^2 \tau = \frac{1}{i\, \textrm{Im }\tau}\nabla^M\tau\,\nabla_M\tau + \frac{1}{12i}G_{mnp}G^{mnp} -\frac{4\kappa_{10}^2(\textrm{Im }\tau)^2}{\sqrt{-g}}\frac{\delta S_7}{\delta\bar\tau}\label{nablatau}
\end{equation}
we get the constraint
\begin{equation}
\tau_{3\bar 3} = \frac{1}{2i}\left(G_{123}G_{\bar 1\bar 2\bar 3}+\frac{1}{4}\sum_{k=1}^3S_{kk}S_{\bar k\bar k}\right)\label{tau33}
\end{equation}
and therefore in the presence of both ISD and IASD 3-form fluxes the dilaton is generically non-constant. In this expression we have assumed that localised distant 7-brane sources do not contribute to the soft terms, and thus have ignored last term in eq.~(\ref{nablatau}). This is the case if there are no anti-D7-brane charges present in the compactification, as we assume in what follows.

Similarly, from the equation
\begin{equation}
dB_2 = -\frac{\textrm{Im }G_3}{\textrm{Im }\tau}
\end{equation}
we obtain for the $B$-field components
\begin{align}
B_{12}&=\frac{g_s\sigma}{2i}\left[(G_{\bar 1\bar 2\bar 3})^*\Phi - \frac12 S_{\bar 3\bar 3}\bar \Phi-G_{123}\Phi+\frac12(S_{33})^*\bar\Phi\right]\label{b12}\\
B_{1\bar 2}&=\frac{g_s\sigma}{4i}\left[-S_{\bar 2\bar 2}\Phi + (S_{\bar 1 \bar 1})^*\bar \Phi - S_{11}\bar\Phi + (S_{22})^*\Phi\right]\nonumber
\end{align}
And from the equations
\begin{align}
dC_6&=H_3\wedge C_4 - *_{10}\, \textrm{Re }G_3\label{bi}\\
dC_8&=H_3\wedge C_6-*_{10}\, \textrm{Re }\tau\nonumber
\end{align}
we get respectively for the RR 6-form and 8-form potentials
\begin{align}
C_{\hat 0\hat 1\hat 2\hat 3 12}&=\frac{\sigma}{2i}\left[-(G_{\bar 1\bar 2\bar 3})^*\Phi-G_{123}\Phi+\frac12 S_{\bar 3 \bar 3}\bar \Phi+\frac12(S_{33})^*\bar\Phi\right]\label{ic6}\\
C_{\hat 0\hat 1\hat 2\hat 3 1\bar 2}&=\frac{\sigma}{4i}\left[S_{\bar 2\bar 2}\Phi+(S_{22})^*\Phi-(S_{\bar 1\bar 1})^*\bar\Phi-S_{11}\bar \Phi\right]\nonumber
\end{align}
and
\begin{align}
C_{\hat 0\hat 1\hat 2\hat 3 1\bar 1 2\bar 2}&=-\frac{g_s\sigma^2}{16}\left[\left(-2G_{123}S_{33}+(S_{11}S_{22})^*-S_{\bar 1\bar 1}S_{\bar 2\bar 2}+2(G_{\bar 1\bar 2\bar 3}S_{\bar 3\bar 3})^*\right)\Phi^2 \ + \ \textrm{c.c.}\right. \nonumber\\ 
&\left.+\left(|S_{33}|^2-|S_{\bar 3\bar 3}|^2-4|G_{\bar 1\bar 2\bar 3}|^2+|S_{\bar 1\bar 1}|^2+|S_{\bar 2\bar 2}|^2+4|G_{123}|^2-|S_{22}|^2-|S_{11}|^2\right)|\Phi|^2\right]\nonumber\\
&+\frac{ig_s\sigma^2}{4}\left[\tau_{3 3}+(\tau_{\bar 3\bar 3})^*\right]\Phi^2\ + \ \textrm{c.c.}\label{ic8}
\end{align}
In particular the non-constant contribution (\ref{tau33}) of the axio-dilaton is crucial for $dC_8$ to be locally integrable when ISD and IASD 3-form fluxes are simultaneously present.

Plugging eqs.~(\ref{tau33}), (\ref{b12}), (\ref{ic6}) and (\ref{ic8}) into the 8d Lagrangian (\ref{dbi1}) and rescaling the fields in order to
have canonically normalized 8d kinetic terms, we get the following Lagrangian for the worldvolume bosons of 7-branes
\begin{multline}
\mathcal{L}_{\rm B}= \textrm{Tr}\left(D_a\Phi D_a\bar\Phi-\frac{1}{4g_{\rm 8}^2} F_{ab}F_{ab}-\frac{g_s}{4}\left[2|G_{\bar 1\bar 2\bar 3}|^2+\frac12|S_{\bar 3\bar 3}|^2+\frac12|S_{11}|^2+\frac12|S_{22}|^2\right]|\Phi|^2\right.\\
\left.+\frac{g_s}{4}\left[(S_{\bar 3\bar 3})^*((G_{\bar 1\bar 2\bar 3})^*-G_{123})+\frac12(S_{11})^*((S_{22})^*-S_{\bar 2\bar 2})-(G_{\bar 1\bar 2\bar 3})^*S_{33}
-\frac12(S_{22})^*S_{\bar 1\bar 1}-2i\tau_{33}\right]\Phi^2\right.\\
\left. +\textrm{h.c.}-\frac{ig_s^{1/2}}{2}\Phi\left[(S_{22})^* \left(\partial_{\bar 1}A^{\bar 2}-\partial_2 A^1+g_{\rm 8}[A^1, A^{\bar 2}]\right)+(S_{11})^*\left(\partial_1 A^2-\partial_{\bar 2}A^{\bar 1}+g_{\rm 8}[A^{\bar 1},A^2] \right)\right.\right.\\
\left.\left.+(S_{\bar 3\bar 3})^*\left(\partial_{1}A^{\bar 2}-\partial_2 A^{\bar 1}+g_{\rm 8}[A^{\bar 1}, A^{\bar 2}]\right)+2(G_{\bar 1\bar 2\bar 3})^*\left(\partial_{\bar 1}A^{ 2}-\partial_{\bar 2} A^{1}+g_{\rm 8}[A^{1}, A^{2}]\right)\right]+\textrm{h.c.}\right)\label{8df}
\end{multline}
where the 8d gauge coupling constant $g_{\rm 8}$ is given by
\begin{equation}
g_{\rm 8}=g_s^{1/2}(2\pi)^{5/2}\alpha' \ .
\end{equation}
The closed string background therefore may induce scalar masses as well as trilinear couplings for the fields in the 
worldvolume of the 7-branes. 
Apart from the terms that were derived in \cite{Camara:2004jj}, sourced by purely ISD or IASD 3-form fluxes, there 
are extra contributions to the B-term coming from the simultaneous presence of ISD and IASD 3-form fluxes as well as from the 
non-constant complex axion-dilaton. These contributions can arise in non Calabi-Yau compactifications, but also may result from the backreaction of 
non-perturbative effects in more conventional compactifications \cite{Koerber:2007xk, Baumann:2010sx, Dymarsky:2010mf}. Observe also the presence of quadratic derivative couplings induced by the 3-form fluxes. These couplings 
were already noticed in \cite{Camara:2009xy}, where it was shown that represent a mixing between massive modes due to Majorana mass terms induced by 3-form fluxes. Such mixing however does not affect the lightest mode of each KK tower and those derivative couplings therefore can be safely neglected for the purposes of this paper.\footnote{Note that even if the mixing were involving the lightest modes, it could be still safely neglected based on the large separation between the SUSY-breaking and Planck scales.}

Flux-induced masses for the 8d fermions localized in the worldvolume of 7-branes can be computed similarly, starting in this case with the DBI+CS fermionic
action. Following closely the procedure described in \cite{Camara:2004jj}, we obtain
\beq {\mathcal{L}}_{\rm F}=\frac{g_s^{1/2}}{2\sqrt{2}}
\textrm{Tr}[(G_{\bar{1}\bar{2}\bar{3}})^*\lambda\lambda+
 \frac{1}{2}(S_{\bar{3}\bar{3}})^*\Psi^3\Psi^3+
\frac{1}{2}S_{11}\Psi^1\Psi^1+\frac{1}{2}S_{22}\Psi^2\Psi^2]+\textrm{h.c.}
\label{fermionix}
\eeq
where $\lambda$ is the 8d gaugino and $\Psi^i$, $i=1,2,3$, the three additional complex fermions that live in the worldvolume of the 7-branes, and that in flat space form the fermionic content of an $\mathcal{N}=4$ vector supermultiplet.

Having the bosonic and fermionic 8d Lagrangians for the lightest fields of 7-branes, we can obtain the 
4d soft SUSY-breaking Lagrangian by dimensional reduction. For the case of fields transforming in the adjoint representation of 
the gauge group dimensional reduction is straightforward, since their internal wavefunctions are constant over the 4-cycle. In the
same notation for the 4d soft-term Lagrangian of Ref.~\cite{Camara:2003ku} 
\begin{multline}
\mathcal{L}_{\rm soft}\ =\ -(m^2)_{ij}\phi^i\bar\phi^j\ -\ \left(\frac{1}{3!}A_{ijk}\phi^i\phi^j\phi^k\ +\ \frac12 B_{ij}\phi^i\phi^j \ -\ \frac12 M\lambda\lambda\ +  \frac12 \mu_{ij}\psi^i\psi^j \right. \\
\left. -  \ \frac12 C_{ijk}\bar \phi^i\bar \phi^j \phi^k\ +\ \textrm{h.c.}\right)\label{notationsoft}
\end{multline}
and for the case of 7-branes wrapping a $T^4$, we obtain the  4d soft-terms
\beqa
m_{1\bar{1}}^2\ & = & m_{2\bar{2}}^2 \ = 0 \ \ ;\ \
B_{ij}\ =\ 0\ \ ,\ i,j\not=3
\nonumber \\
m_{3\bar{3}}^2\ & =&\
\frac{g_s}{2}\left(\vert G_{\bar{1}\bar{2}\bar{3}}
\vert^2+\frac{1}{4}\vert S_{\bar{3}\bar{3}}\vert^2 +\frac{1}{4}
\vert S_{11}\vert^2+\frac{1}{4}
\vert S_{22}\vert^2\right)
\nonumber
\\
B_{33}\ &=& \ \frac{g_s}{2}\left(-
(G_{{\bar 1}{\bar 2}{\bar 3}}S_{{\bar 3}{\bar {3}} })^*-
\frac{1}{2}(S_{{2}{2}}S_{{1}{1}})^*+ (S_{\bar 3\bar 3})^*G_{123}\right.\nonumber\\
&&\quad  + \left.\frac12(S_{11})^*S_{\bar 2\bar 2}+(G_{\bar 1\bar 2\bar 3})^*S_{33}
+\frac12(S_{22})^*S_{\bar 1\bar 1}+2i\tau_{33}\right)
\nonumber
\\
A^{ijk}\ &=& \ -h^{ijk} {{g_s^{1/2}}\over {\sqrt{2}}}\
(G_{\bar{1}\bar{2}\bar{3}})^*
 \nonumber
\\
C^{ijk}\ &=& -\frac{g_s^{1/2}}{2\sqrt{2}}
\left[h^{jk1}S_{11}+h^{jk2}S_{22}-h^{jk3}(S_{\bar{3}\bar{3}})^* \right]
\nonumber\\
M^a\ &=& \  {{g_s^{1/2}}\over {\sqrt{2}}}\
(G_{\bar{1}\bar{2}\bar{3}})^* \nonumber\\
\mu_{33} \ & =& \  -{{g_s^{1/2}}\over {2\sqrt{2}}} (S_{{\bar 3}{\bar 3}})^*
 \nonumber\\
\mu _{ii} \ & =& \  -{{g_s^{1/2}}\over {2\sqrt{2}}} S_{{ i}{i}}\  , \qquad i=1,2  \ \ ,
\nonumber
\label{softgen7}
\eeqa
with
\begin{equation}
h_{ijk}=2\epsilon_{ijk}\sqrt{2}g_{\rm YM}
\label{yukbulk}
\end{equation}
the Yukawa coupling and $g_{\rm YM}=g_{\rm 8}/\sqrt{\textrm{Vol}(T^4)}$ the 4d gauge coupling constant.
Note that only the geometric field $\Phi$ gets a mass at this level, whereas the $A_i,A_{\bar i}$ remain massless.
This is expected from 8d gauge invariance.  This will be relevant in our generalisation to the 
matter curve case below. 

Although for concreteness here we have reduced the 8d theory in a 4-torus, we could have equally performed dimensional 
reduction in a different type of 4-cycle, obtaining analogous expressions for the soft terms of a stack of 7-branes that wraps 
such 4-cycle. For that aim, note that no knowledge of the metric of the 4-cycle is required, but only its topological features. Whereas the 7-brane field content 
will change according to the homology of the 4-cycle, we expect expressions for the soft terms not far from those obtained here in the toroidal case. This will be even more certain in the case of soft terms for bifundamental matter fields discussed in the next section, since the wave-functions of those fields are localized also along some of the directions of the 4-cycle.

\subsection{Magnetized bulk D7-brane fields}
\label{magnet}

We now consider the addition of magnetic fluxes on the world-volume of D7-branes. Namely, we consider 
the presence of a local magnetic background
\begin{equation}
\langle F_2\rangle = i(F_+ + F_-)\, dz^1\wedge d\bar z^1 + i(F_+-F_-)\, dz^2\wedge d\bar z^2
\end{equation}
where the D-term equations will in general require the vanishing of the self-dual component $F_+$. The magnetic flux 
$F_\mp$ induces a charge of  $D3({\overline {D3}})$-brane in the worldvolume of D7-branes, and 
therefore it is expected the presence of flux-induced D3-brane soft-terms proportional to the 
magnetic background, apart from the soft-terms described in the previous subsection for unmagnetized D7-branes. 

In more precise terms, the effect of magnetic fluxes on the D7-branes can be understood in terms of two different mechanisms. On one side the magnetic flux sources new renormalizable couplings in the 4d Lagrangian that originate from 
higher order couplings on which two or more of the gauge field-strengths are taken to be background. On the other side, the magnetic flux deforms the internal wavefunctions of charged fields and induces the mixing of massive modes in order to minimize the additional source of potential energy introduced by the flux. In this subsection we address the first of these effects. This is the only relevant one for the soft masses and B-terms of geometric moduli $\Phi$ in magnetized non-intersecting D7-branes. In this subsection we compute those contributions to soft masses. 
Then, in the next section we address the more interesting case of chiral-matter bifundamental fields, where the effect of the magnetization in the internal wavefunctions turns out to be the leading effect. 

The microscopic computation of soft masses for magnetized bulk D7-brane fields follows the same steps than in the previous subsection. We work to quadratic order in the magnetization.
The relevant piece of the DBI+CS action is again given by eq.~(\ref{action1}) with the addition of the CS coupling to the
RR 4-form, that becomes also relevant in presence of magnetization,
\begin{multline}
S=-\mu_7\int d^8\xi \, \textrm{STr}\left[ e^{-\phi} \sqrt{-\textrm{det}\left(P[E_{\mu\nu}]+\sigma F_{\mu\nu}\right)}\right]\\
+\mu_7 g_s \int \textrm{STr}\left(P\left[\frac12 C_4\wedge \mathcal{F}_2\wedge \mathcal{F}_2- C_6\wedge \mathcal{F}_2+C_8\right]\right) \ .
\label{action}
\end{multline}
It is convenient to factorize the determinant that appears in the DBI piece of the action in Minkowski and 4-cycle pieces as
\begin{equation}
\textrm{det}(P[E_{\mu\nu}])=g_s^{4}\,\textrm{det}\left(\eta_{\mu\nu}+2Z\sigma^2\partial_\mu\Phi\partial_\nu\bar\Phi+Z^{1/2}g_s^{-1/2}\sigma F_{\mu\nu}\right)\textrm{det}\left(g_{ab}-Z^{-1/2}g_s^{-1/2}\mathcal{F}_{ab}\right) \  ,
\end{equation}
from which we get
\begin{multline}
\textrm{det}(P[E_{\mu\nu}])=-g_s^4-g_s^3Z\sigma^2\left(1+\frac{Z^{-1}g_s^{-1}}{2}\mathcal{F}_{ab}\mathcal{F}_{ab}\right)\left(2g_s\partial_\mu\Phi\partial_\mu\bar\Phi-\frac{1}{2} F_{\mu\nu}F_{\mu\nu}\right)\\
-\frac12 g_s^3Z^{-1}\mathcal{F}_{ab}\mathcal{F}_{ab}
+\frac{g_s^2}{4}Z^{-2}\mathcal{F}_{ab}\mathcal{F}_{bc}\mathcal{F}_{cd}\mathcal{F}_{da}
-\frac{g_s^2}{8}Z^{-2}\left(\mathcal{F}_{ab}\mathcal{F}_{ab}\right)^2  \ .
\end{multline}  
Plugging this expression into eq.~(\ref{action}) and expanding the square root that appears in DBI part of the action, we find
\begin{multline}
\mathcal{L}_{\rm 8d}=\mu_7 \int_{\Sigma_4}\, e^\phi\, \textrm{STr}\left[  \left(-1-Z\sigma^2\hat\theta\partial_\mu\Phi\partial_\mu\bar\Phi- \frac{g_s^{-1}}{4}\sigma^2 Z\hat\theta F_{\mu\nu}F_{\mu\nu}-\frac{g_s^{-1}}{4} Z^{-1}\mathcal{F}_{ab}\mathcal{F}_{ab}\right.\right.\\
\left.\left.+\frac{g_s^{-2}}{8}Z^{-2}\mathcal{F}_{ab}\mathcal{F}_{bc}\mathcal{F}_{cd}\mathcal{F}_{da}-\frac{1}{32}g_s^{-2}Z^{-2}\left[\mathcal{F}_{ab}\mathcal{F}_{ab}\right]^2\right)dx^4+C_8-C_6\wedge \mathcal{F}_2+\frac12 \chi\, \mathcal{F}_2\wedge\mathcal{F}_2\right]\label{action2} \ , 
\end{multline}
where we have defined
\begin{equation}
\hat\theta\equiv 1+\frac{Z^{-1}g_s^{-1}}{4}\mathcal{F}_{ab}\mathcal{F}_{ab}=1+Z^{-1}g_s^{-1}\sigma^ 2(F_+^2+F_-^2)\label{hattheta} \ .
\end{equation}
The contribution of magnetic fluxes to the soft masses of the 4d fields that descend from $\Phi$ can be read from this expression. The relevant terms in this equation are
\begin{multline}
\delta\mathcal{L}_{\Phi^2}=\mu_7 \sigma^2\int_{\Sigma_4} dx^4\, e^\phi\, \textrm{STr}\left[-Z\hat\theta\partial_\mu\Phi\partial_\mu\bar\Phi-g_s^{-1} Z^{-1}(F_+^ 2+F_-^2)- \chi\, (F_+^2-F_-^2)  \right.\\
\left.+\frac{g_s^{-2}}{2}Z^{-2}\left(F_{ab}F_{bc}B_{cd}B_{da}+\frac12 F_{ab}B_{bc}F_{cd}B_{da}-\frac18 B_{ab}B_{ab}F_{cd}F_{cd}-\frac14 F_{ab}B_{ab}F_{cd}B_{cd}\right)\right]
\ .
\label{lphi}
\end{multline}
Expanding $e^\phi = (\textrm{Im}\,\tau)^{-1}$ as in eq.~(\ref{tauex}), $Z$ and $\chi$ as
\begin{align}
Z^{-1/2}&= Z_0^{-1/2}+\frac{\sigma^2}{2}\left(K_{33}\Phi^2+(K_{33})^*\bar\Phi^2+2K_{3\bar 3}|\Phi|^ 2\right)+\ldots\label{zz}\\
\chi&=\chi_0+\frac{\sigma^2}{2}\left(\chi_{33}\Phi + (\chi_{33})^*\bar\Phi+2\chi_{3\bar 3}|\Phi|^2\right)+\ldots\nonumber
\end{align}
making use of the identities (\ref{b12})-(\ref{ic8}),\footnote{The identities (\ref{bi})-(\ref{ic8}) in general receive additional contributions from the magnetization, however one may check that those contributions turn into sub-leading corrections to the soft-masses of $\Phi$.} 
dimensionally reducing over a $T^4$ and rescaling the fields to have canonically normalized 4d kinetic terms, we obtain the following additional 
contributions to the soft-masses and B-term (\ref{softgen7}) induced by the magnetization in the worldvolume of D7-branes
\begin{align}
\delta m_{3\bar 3}^2 &= -\sigma^2\left( \frac{ 1}{8}[4|G_{\bar 1\bar 2\bar 3}|^2+|S_{\bar 3\bar 3}|^2+|S_{\bar 1\bar 1}|^2+|S_{\bar 2\bar 2}|^2+2|S_{11}|^2+2|S_{22}|^2]\right.\nonumber\\
&\qquad\left.-\frac14\textrm{Re}\left(S_{11}S_{\bar 1\bar 1}+S_{22}S_{\bar 2\bar 2}\right) - 2g_s^{-1}K_{3\bar 3}-\chi_{3\bar 3} -(\textrm{Im}\,\tau)_{3\bar 3}\right)F_+^2 \nonumber\\
&\qquad - \sigma^2\left( \frac{1}{8}[|S_{11}|^2+|S_{22}|^2+4|G_{123}|^2+|S_{33}|^2+2|S_{\bar 3\bar 3}|^2+8|G_{\bar 1\bar 2\bar 3}|^2]\right.\nonumber\\
&\qquad \left. -\textrm{Re}\left(G_{123}G_{\bar 1\bar 2\bar 3}+\frac14 S_{33}S_{\bar 3\bar 3}\right) - 2g_s^{-1}K_{3\bar 3}+\chi_{3\bar 3} -(\textrm{Im}\,\tau)_{3\bar 3}\right)F_-^2\nonumber\\
\delta B_{33} &= \sigma^2\left( \frac{1}{4}[2(S_{\bar 3\bar 3}G_{\bar 1\bar 2\bar 3})^*+S_{\bar 1\bar 1}S_{\bar 2\bar 2}+2(S_{11}S_{22})^*-2S_{\bar 2\bar 2}(S_{11})^*-2S_{\bar 1\bar 1}(S_{22})^*\right.\nonumber\\
&\left.\qquad -2G_{123}(S_{\bar 3\bar 3})^*-2S_{33}(G_{\bar 1\bar 2\bar 3})^*] + 2g_s^{-1}K_{33}+\chi_{33} +(\textrm{Im}\,\tau)_{33}\right)F_+^2\nonumber\\
&\qquad+\sigma^2\left( \frac{1}{4}[(S_{11}S_{22})^*+2G_{123}S_{33}+4(S_{\bar 3\bar 3}G_{\bar 1\bar 2\bar 3})^*-4S_{33}(G_{\bar 1\bar 2\bar 3})^*-4G_{123}(S_{\bar 3\bar 3})^*\right.\nonumber\\
&\qquad-S_{\bar 2\bar 2}(S_{11})^*-S_{\bar 1\bar 1}(S_{22})^*\left.] + 2g_s^{-1}K_{33}-\chi_{33} +(\textrm{Im}\,\tau)_{33}\right)F_-^2 \ .
\label{corr1}
\end{align}
where we have defined
\begin{equation}
(\textrm{Im}\,\tau)_{33}\ \equiv\ \frac{\tau_{33}-(\tau_{\bar 3\bar 3})^*}{2i}\ , \qquad (\textrm{Im}\,\tau)_{3\bar 3}\ \equiv\ \frac{\tau_{3\bar 3}-(\tau_{\bar 3 3})^*}{2i}
\end{equation}
In particular, note that among the contributions of antiself-dual magnetic fluxes to soft-masses there are terms proportional to
$(2g_s^{-1}K_{3\bar 3}-\chi_{3\bar 3}+(\textrm{Im}\,\tau)_{3\bar 3})$, in agreement with the expressions for soft-masses
in the worldvolume of D3-branes that were obtained in Ref.~\cite{Camara:2003ku}. Similarly among the contributions of self-dual magnetic fluxes 
we identify terms that are proportional to $(2g_s^{-1}K_{3\bar 3}+\chi_{3\bar 3}+(\textrm{Im}\,\tau)_{3\bar 3})$, identified with the expressions for soft-masses in the worldvolume of anti D3-branes.

We can also compute the leading corrections of magnetic fluxes to trilinear couplings of the form $\Phi\times A\times A$. The starting point is again eq.~(\ref{action2}). The relevant terms in that equation are now
\begin{multline}
\mathcal{L}_{\Phi AA,F}=\mu_7\sigma^2\int_{\Sigma_4}dx^4 \textrm{STr}\left(-Z\hat\theta g_s \partial_\mu\Phi\partial_\mu\bar\Phi-\frac{\hat\theta}{4} F_{\mu a}F_{\mu a}-\right.\\
\left.-\frac{1}{2}Z^{-2}g_s^{-1}\sigma B_{ab}F_{bc}F_{cd}F_{da}+\frac{1}{8}Z^{-2}g_s^{-1}\sigma B_{ab}F_{ab}F_{cd}F_{cd}\right)
\label{L}
\end{multline}
Some little algebra shows that
\begin{multline}
-\frac{1}{2}B_{ab}F_{bc}F_{cd}F_{da}+\frac{1}{8}B_{ab}F_{ab}F_{cd}F_{cd}=\\
=\frac{\sigma}{2i}\left( F_-^2[\Phi\Abb{1}{2}\left(-2(\Gb)^*+2\Gn\right)+\bar \Phi \Abb{1}{2}\left(\Sb{3}-(\Sn{3})^*\right)]+\right.\\
 \left.+F_+^2[\Phi \Abn{1}{2}\left(\Sb{2}-(\Sn{2})^*\right)+\bar\Phi\Abn{1}{2}\left(-(\Sb{1})^*+\Sn{1}\right)]\right)+\textrm{h.c.}
\label{depF}
\end{multline}
where we are keeping only terms that contribute to soft trilinear couplings. Plugging this expression into eq.~(\ref{L}), dimensionally
reducing over a $T^4$ and rescaling the fields to have canonically normalized 4d kinetic terms, we get the following corrections to 
trilinear soft couplings from magnetization in the worldvolume of D7-branes
\begin{align}
\delta A^{ijk} &= -h^{ijk}\frac{\sigma^2}{\sqrt{2g_s}}\left[F_-^2\left(G_{123}-2(G_{\bar 1\bar 2\bar 3})^*\right)-F_+^2(G_{\bar 1\bar 2\bar 3})^*\right]\label{corr2}\\
\delta C^{ijk} &= -\frac{\sigma^2}{2\sqrt{2g_s}}\left\{h^{jk1}\left[F_+^2\left((S_{\bar 1\bar 1})^*-2S_{11}\right)-F_-^2S_{11}\right]+h^{jk2}\left[F_+^2\left((S_{\bar 2\bar 2})^*-2S_{22}\right)-F_-^2S_{22}\right]\right.\nonumber\\
&\qquad \left.+h^{jk3}\left[F_-^2\left(2(S_{\bar 3\bar 3})^*-S_{33}\right)+F_+^2(S_{\bar 3\bar 3})^*\right]\right\}\nonumber
\end{align}
These corrections are quadratic in the magnetic fluxes, as expected. 

Dimensional reduction of eq.~(\ref{action}) also includes corrections to the gauge coupling constants upon
replacing $\langle F^2\rangle$ by its vev. In the context of F-theory SU(5) unification, corrections from the hypercharge flux $F_Y$ are particularly
relevant, since they generically induce non-universal thresholds for the three SM gauge coupling constants, which
may have interesting phenomenological implications, see e.g.~\cite{Blumenhagen:2008aw,imrv,iv}. 
These were computed in \cite{Blumenhagen:2008aw,DW} and we will not reproduce them here.  
 Let us also remark that the SM gauginos also become
slightly non-universal once the corresponding gaugino fields are normalized to one. We will not consider these 
gaugino mass 
corrections in what follows, since they are expected to be generically small if gauge coupling unification is to be maintained.

\section{Soft terms for type IIB chiral matter  bifundamental fields}
\label{sec3}

In the previous section we have considered soft-breaking terms for 4d fields that descend from geometric moduli $\Phi$ in 
non-intersecting magnetized branes. We have done this in two steps. First, a 8d field theory with the relevant operators induced by the 
closed string background has been derived in the limit $M_{\rm Pl}\to \infty$. Next, we have dimensionally reduced that 8d theory to obtain
the soft-breaking Lagrangian in 4d. For the case of bulk fields, e.g.~adjoints in non-intersecting magnetized D7-branes,
this last step is straightforward. However, this general procedure can in principle be equally applied in more involved settings,
such as intersecting magnetized D7-branes with 3-form fluxes.

In this section we compute 4d soft-breaking terms
for chiral matter bifundamental fields localised at D7-brane intersections (or \emph{matter curves}). Although the procedure described above is in principle feasible (see e.g. \cite{Camara:2009xy}), in practice it quickly becomes technically too 
involved as the background gets more general. Thus, we instead exploit a short-cut by making use of the general ideas 
behind Higgssing in 4d supersymmetric theories and the 4d soft-breaking Lagrangians for bulk fields obtained in the previous section.

\subsection{Fields at matter curves}
\label{matterfields}

When computing the 4d effective theory of a stack of magnetized/intersecting D7-branes one dimensionally reduces an 8d supersymmetric gauge theory, as we have described in the previous section. In the case of locally vanishing closed string fluxes this 8d theory is simply given by topologically twisted 8d $\mathcal{N}=1$ SYM \cite{Donagi:2008ca, Beasley:2008dc}.\footnote{For simplicity we take the  normal bundle of the magnetized/intersecting D7-branes to be trivial, so that we can ignore the effect of the twist. In the more general case, this can be however easily implemented by shifting the magnetization along the canonical bundle \cite{Conlon:2009qq}. } The bosonic part reads
\begin{equation}
\mathcal{L}_{\rm SYM} = \textrm{Tr}\left(D_a\Phi D_a \bar \Phi - \frac12 ([\Phi,\bar \Phi])^2 - \frac14 F_{ab}F_{ab}\right)\label{lag1}
\end{equation}
where $D_a\Phi = \partial_a \Phi + i[A_a,\Phi]$ and $F_2 = dA+A\wedge A$. To linear order in the fluctuations, the corresponding equations of motion are
\begin{equation}
D_aD^a\Phi = 0 \ , \qquad D_a F^{ab} = 0\label{eqs8d}
\end{equation}
We take here for simplicity  an underlying $U(N)$ gauge symmetry group,
although the results may be extended easily to $SO(N)$ and $E_n$ groups, as we will see later.
$U(N)$  is broken to some product of smaller groups by the magnetization/intersections. The latter are parametrized in terms of backgrounds for $F_2$ and $\Phi$ 
\begin{align}
\langle F_2\rangle &= \left[i(F_+^\alpha + F_-^\alpha)\, dz_1\wedge d\bar z_1 + i(F_+^\alpha-F_-^\alpha)\, dz_2\wedge d\bar z_2\right]\, Q_\alpha \label{magneti}\\
\langle \Phi \rangle &= m_i^\alpha z_i Q_\alpha \nonumber \ ,
\end{align}
where $Q_\alpha$ are the generators of the Cartan subalgebra of $U(N)$. Dimensionally reducing eq.~(\ref{eqs8d}) to 4d amounts to solving the following system of second-order differential equations for the internal wavefunction $\Psi$ of a 4d scalar with mass $m$ and $U(1)_\alpha \subset U(N)$ charges $q_\alpha$ (see e.g.~\cite{Marchesano:2010bs})
\begin{align}
(\mathbb{D}^+\mathbb{D}^- + q_\alpha F_+^\alpha\mathbb{I})\Psi = m^2 \Psi\label{ddeq}\\
(\mathbb{D}^-\mathbb{D}^+ - q_\alpha F_+^\alpha\mathbb{I})\bar\Psi = m^2 \bar\Psi\nonumber %\label{ddeqc}
\end{align}
with
\begin{equation}
\mathbb{D}^\pm \equiv \begin{pmatrix}0 & D_1^\pm & D_2^\pm & D_3^\pm\\
-D_1^\pm & 0 & -D_3^\mp & D_2^\mp \\
-D_2^\pm & D_3^\mp & 0 & -D_1^\mp\\
-D_3^\pm & -D_2^\mp & D_1^\mp & 0
\end{pmatrix} \ , \qquad \Psi = \begin{pmatrix}0\\ a^1\\ a^2\\ \phi \end{pmatrix}
 \ , \qquad \bar \Psi = \begin{pmatrix}0\\  a^{\bar 1}\\  a^{\bar 2}\\ \bar\phi \end{pmatrix}
\end{equation}
and
\begin{align}
D_1^-&=\partial_1-\frac{q_\alpha}{2}(F_+^\alpha+F_-^\alpha)\bar z_1 & D_1^+&=\bar\partial_{\bar 1}+\frac{q_\alpha}{2}(F_+^\alpha+F_-^\alpha) z_1\label{d1}\\
D_2^-&=\partial_2-\frac{q_\alpha}{2}(F_+^\alpha-F_-^\alpha)\bar z_2 & D_2^+&=\bar\partial_{\bar 2}+\frac{q_\alpha}{2}(F_+^\alpha-F_-^\alpha) z_2\nonumber\\
D_3^-&=-q_\alpha [(m_1^\alpha)^*\bar z_1+(m_2^\alpha)^*\bar z_2] & 
D_3^+&=q_\alpha (m_1^\alpha z_1+m_2^\alpha z_2)\nonumber %\label{d3}
\end{align}
In these expressions $a^{1,2}$ and $\phi$ are respectively the components of the internal wavefunction $A$ along the Wilson lines and the geometric scalar
$\Phi$, and $[Q^\alpha,\Psi]=-q^\alpha \Psi$. Besides this local diffeo-algebraic equation, wavefunctions must also satisfy the global periodicity conditions of the 4-cycle $S$.

In order to solve eq.~(\ref{ddeq}) note that $\mathbb{D}^+\mathbb{D}^-$ can be expressed as
\begin{equation}
\mathbb{D}^\pm\mathbb{D}^\mp = -\mathbb{I}\sum_{i=1}^3D_i^\pm D_i^\mp+\mathbb{B}^{\pm}\label{ecus1}
\end{equation}
with
\begin{equation}
\mathbb{B}^\pm=\begin{pmatrix}0 & 0 & 0 & 0\\
0 & [D_2^\pm,D_2^\mp] & [D_2^\mp,D_1^\pm] & [D_3^\mp,D_1^\pm]\\
0 & [D_1^\mp,D_2^\pm] & [D_1^\pm,D_1^\mp] & [D_3^\mp,D_2^\pm]\\
0 & [D_1^\mp,D_3^\pm] & [D_2^\mp,D_3^\pm] & [D_2^\pm,D_2^\mp]+[D_1^\pm,D_1^\mp]
\end{pmatrix}\label{bmatrix}
\end{equation}
Diagonalising this matrix
\begin{equation}
\mathbb{J}^\dagger\cdot \mathbb{B}^+\cdot \mathbb{J}=\textrm{diag}(0,\lambda_1,\lambda_2,\lambda_3)\label{ecus2}
\end{equation}
we get
\begin{equation}
\tilde D_p^- = \sum_j \xi_{p,j} D_j^- \ , \qquad \tilde D_p^+=\sum_j \xi^*_{p,j}D_j^+
\end{equation}
where $\xi_p$ is the $p$-th eigenvector of $\mathbb{B}$. These operators span the algebra of three quantum harmonic oscillators, namely
\begin{equation}
[\tilde{D}_p^+,\tilde{D}_p^-]\ = \ -\lambda_p\ , \qquad p =1,2,3\label{ecus3}
\end{equation}
leading to three KK towers of 4d scalars. The matrix $\mathbb{B}$ has a single negative eigenvalue that, without loss of generality, we take here to be $\lambda_1$. Making use of eqs.~(\ref{ecus1}), (\ref{ecus2}) and (\ref{ecus3}) we can explicitly solve eqs.~(\ref{ddeq}). The wavefunction for the lightest mode of each tower is given by
\begin{equation}
\Psi_p = \xi_p \varphi_p
\end{equation}
with $\varphi_p$ a function on the 4-cycle $S$ satisfying locally 
\begin{equation}
\tilde D_p^-\varphi_p=\tilde D_q^+\varphi_p=0 \ , \qquad p,q=1,2,3\ , \quad q\neq p
\end{equation}
The mass of the lowest mode for each tower of scalars is given by
\begin{equation}
m^2_{\Psi_p}=\lambda_p - \lambda_1 + q_\alpha F^\alpha_+
\end{equation}
And similarly for the complex conjugate degrees of freedom.

To give a concrete example, consider a stack of three D7-branes with gauge group $U(3)$ (see 
\cite{Font:2009gq,Conlon:2009qq,Aparicio:2011jx}),
 wrapping a 4-torus parametrized by the holomorphic condition
\begin{equation}
z_3=0\label{toro1}
\end{equation}
In applications in section \ref{applic} we will consider the  phenomenologically most interesting case of  $SO(12)$  or $E_6$ gauge groups, relevant  for SU(5) F-theory unification. However, this simpler $U(3)$ model suffices to illustrate the main ideas of this section.

Let us tilt one of the D7-branes of the stack an angle so that instead of (\ref{toro1}) it wraps a 4-torus parametrized by the condition
\begin{equation}
z_3-m_az_1 = 0\label{curvaa}
\end{equation}
with $m_a$ a constant of the order of the string scale that determines the number of intersections in the complex 2-torus spanned by $z_1$.
For future reference, we denote this matter curve as $\Sigma_a = \left\{z_1=0\right\}$. 

The original $U(3)$ gauge group is broken as
\begin{align}
U(3)&\ \to\ U(2) \times U(1)\ \to\ SU(2)\times U(1)\label{group}\\
\mathbf{8}&\ \to\ \mathbf{3}^0 + \mathbf{1}^0 + \mathbf{\bar 2}^{+} + \mathbf{2}^{-}\nonumber
\end{align}
where the diagonal $U(1)\subset U(2)$ becomes massive due to the presence of St\"uckelberg couplings. From the point of view of the 8d $U(3)$ SYM theory the breaking (\ref{group}) is encoded in a background for the geometric modulus
\begin{equation}
\langle \Phi\rangle = \frac{1}{\sqrt{6}}m_az_1(Q_1+Q_2-2Q_3)\label{phiback}
\end{equation}
where $Q_\alpha$, $\alpha=1,2,3$, are the Cartan generators of $U(3)$. 
The 8d fields $\Phi$ and $A$ can be decomposed according to (\ref{group}) as
\begin{equation}
\Phi = \left(\begin{tabular}{c|c}$\Phi_{\mathbf{3}^0}$ & $\Phi_{a^+}$ \\
\hline
$\Phi_{a^-}$ & $\Phi_{\mathbf{1}^0}$ \end{tabular}
\right)+\langle \Phi\rangle \, \qquad A=\left(\begin{tabular}{c|c}$A_{\mathbf{3}^0}$ & $A_{a^+}$ \\
\hline
$A_{a^-}$ & $A_{\mathbf{1}^0}$ \end{tabular}
\right)
\end{equation}
with the 4d scalars in the bifundamental representation arising from the $U(3)$ off-diagonal fluctuations.

One may easily check that eq.~(\ref{bmatrix}) gives rise in this case to
\begin{equation}
\mathbb{B}^+=\begin{pmatrix}0 & 0 & 0 & 0\\
0 & 0 & 0 & m_a\\
0 & 0 & 0 & 0\\
0 & m_a & 0 & 0 \end{pmatrix} \ , 
\end{equation}
with eigenvalues 0 and $\pm |m_a|$. Thus, according to our discussion above, the internal wavefunctions for the lightest mode in each of the three KK towers of 4d scalars are given by
\begin{equation}
\Psi_{a_1^+}=\frac{1}{\sqrt{2}}\begin{pmatrix}-1\\ 0\\ 1 \end{pmatrix}\varphi \ , \qquad \Psi_{a_2^+}=\frac{1}{\sqrt{2}}\begin{pmatrix}1\\ 0\\ 1 \end{pmatrix}\varphi\ , \qquad \Psi_{a_3^+}=\begin{pmatrix}0\\ 1\\ 0 \end{pmatrix}\varphi\label{wave1}
\end{equation}
where $\varphi$ is a real function of the coordinates of the 4-cycle $S$, 
locally given by
\begin{equation}
\varphi = f(z_2)\, \textrm{exp}\left[-\frac{|m_a|}{2}|z_1|^2\right]\ , \label{wave2}
\end{equation}
and $f(z_2)$ are holomorphic functions specified by the global properties of the 4-cycle $S$ such that the wavefunctions (\ref{wave1}) are orthonormalized. The exponential factor in (\ref{wave2}) shows in particular the localization of the energy density along the matter curve $\Sigma_a$. The resulting 4d masses for the modes (\ref{wave1}) are respectively
\begin{equation}
m_{a_1^+}^2 = 0 \ , \qquad m_{a_2^+}^2 = 2|m_a|^2 \ , \qquad m_{a_3^+}^2 = |m_a|^2\label{mase}
\end{equation}

Wavefunctions and 4d masses for the charge conjugated sector $a^-$ follow exactly the same expressions (\ref{wave1})-(\ref{mase}), with the role of $\Psi_{a_1^-}$ and $\Psi_{a_2^-}$ exchanged with respect to eq.~(\ref{wave1}). Thus, in total we obtain a massless vector-like pair of 4d charged fields localized in $\Sigma_a$ and transforming in the $\mathbf{\bar 2}^+ + \mathbf{2}^-$ representation of the gauge group, as expected.

This simple setting can be extended in several ways. First, one may consider magnetization in the worldvolume of D7-branes. The effect of magnetization is to modify the wavefunctions (\ref{wave1}) and (\ref{wave2}) and to lift one of the two chiral components of the above vector-like pair of 4d zero modes. Thus, turning on a magnetic flux in the above $U(3)$ D7-brane setting of the form
\begin{equation}
\langle F_2\rangle\ =\ \frac{iF_-^a}{\sqrt{6}}\, (dz_1\wedge d\bar z_1 - dz_2\wedge d\bar z_2) \, (Q_1+Q_2-2Q_3)\label{magfl}
\end{equation}
leads to the modified wavefunctions  
\begin{equation}
\Psi_{a_1^+}=\frac{1}{\sqrt{2\lambda_a(\lambda_a-F_-^a)}}\begin{pmatrix}F_-^a-\lambda_a\\ 0\\ m_a \end{pmatrix}\varphi^- \ , \quad \Psi_{a_2^+}=\frac{1}{\sqrt{2\lambda(\lambda_a+F_-^a)}}\begin{pmatrix}F_-^a+\lambda_a\\ 0\\ m_a \end{pmatrix}\varphi^+\label{wave3}
\end{equation}
where $\lambda_a = \sqrt{(F_-^a)^2 + m_a^2}$ and
\begin{equation}
\varphi^{\pm} = f(z_2)\textrm{exp}\left[-\frac{\lambda_a}{2}|z_1|^2\pm \frac{F_-^a}{2}|z_2|^2\right]
\end{equation}
We have introduced subscript $a$ to refer to quantities associated to curve $\Sigma_a$. Such notation is useful in later sections when several matter curves are present. Similar expressions to (\ref{wave3}) again apply for $a_{1,2}^-$. Note that only one of the two wavefunctions (\ref{wave3}) is normalizable in the presence of the magnetic flux and thus a chiral spectrum is indeed obtained, with local chirality determined by the sign of the magnetization. Besides those, there can be additional chiral fermions localized in other regions of the matter curve, with the total chirality determined by the integral of the magnetic flux along the matter curve (see also \cite{Palti:2012aa} for a discussion of local versus global chirality).

We now want to extend  this simple setting  to consider the effect of closed string 3-form fluxes in the neighbourhood of D7-branes. As we have discussed in the previous section, the effect of 3-form fluxes (and other closed string backgrounds) in the limit $M_{\rm Pl}\to \infty$ is to deform the 8d theory (\ref{lag1}) by adding new renormalizable couplings sourced by the closed string background. Hence, we should consider the more complicate 8d Lagrangian (\ref{8df}), which includes closed string fluxes, 
 instead of (\ref{lag1}). Dimensional reduction of this Lagrangian in presence of non-trivial magnetization and intersections becomes rather complicated. In particular, internal wavefunctions for chiral matter fields such as (\ref{wave1}) and (\ref{wave3}) receive also contributions from the closed string background. 

In what follows, we pursue a simpler route to obtain the 4d soft-breaking Lagrangian for chiral matter fields. 
However before moving to the details, a comment regarding the consistency of the 3-form flux background in presence of intersecting D7-branes is in order. Note that the 3-form flux background has to satisfy some restrictions in order not to induce Freed-Witten anomalies in the worldvolume of the tilted D7-branes. Indeed, the condition for a NSNS 3-form flux not to induce a tadpole for the gauge field in the worldvolume of a stack of D7-branes is given by \cite{Cascales:2003zp}
\begin{equation}
\int_{\Pi_a} P[H_3] = 0\label{fw1}
\end{equation}
for any 3-cycle $\Pi_a\subset S$, as can be easily seen by integrating by parts the D7-brane CS coupling $\int_S B_2\wedge F_2 $. This condition puts constraints on the intersection parameters $m_i^\alpha$ in presence of non-trivial 3-form fluxes. For instance, in the above simple example of a tilted D7-brane wrapping the 4-torus parametrized by eq.~(\ref{curvaa}) it leads to the local constraints
\begin{align}
m_a[(S_{11})^*-S_{\bar 1\bar 1}]-m_a^*[S_{11}-(S_{\bar 1\bar 1})^*]&=0\label{fwcons}\\
m_a[(S_{\bar 3\bar 3})^*-S_{33}]-m_a^*[S_{\bar 3\bar 3}-(S_{33})^*]&=0\nonumber 
\end{align}
and hence the phases of the intersection parameter $m_a$ and those of the complexified 3-form fluxes must be suitably aligned.
Note however that  the  constraint (\ref{fw1}) is a global condition, and for generic 4-folds the toroidal constraints 
(\ref{fwcons}) need not apply locally. Thus, we do not impose them in what follows.

\subsection{Soft terms for fields on matter curves}
\label{sec32}

To compute the expression of soft terms for bifundamental fields localised on matter curves, 
we combine the information  about 4d soft terms for bulk fields obtained  in section \ref{sec2} 
with our discussion on matter field wavefunctions of previous subsection. For simplicity we first consider the case with no magnetic fluxes and only pure ISD closed string fluxes,
namely only the flux components $G_{{\bar i}{\bar j}{\bar k}}$ and $S_{{\bar 3}{\bar 3}}$ are non-vanishing. The effect  of magnetization on the soft terms for bifundamental fields will be discussed in subsection \ref{magnetico}. For simplicity we also assume that closed string fluxes are approximately constant over the
4-cycle $S$, so that they can be factored out when performing dimensional reduction. 
The case of locally varying closed and open string fluxes was considered in 
\cite{Camara:2013fta} and is briefly studied in section \ref{nonuniv}.

The reader may easily check that the soft  scalar terms for D7-brane adjoints that we found in eqs.~(\ref{softgen7}) can be 
rewritten  in terms of a 4d scalar 
potential of the form 
\beq
V_{\rm ISD} \ =\  \vert  M^*\Phi ^*\ +\  F_{\Phi} \vert ^2 \ +\ \vert F_{A_{\bar 1}}  \vert ^2 \ +\vert F_{A_{\bar 2}}\vert ^2 
\label{noscaleadjoint}
\eeq
where $M=g_s^{1/2}(G_{{\bar 1}{\bar 2}{\bar 3}})^*/\sqrt{2}$
 is the gaugino mass, and $F_i$ are the auxiliary fields for the different 4d complex scalar fields,
\beq
{ F}_\Phi= \ \partial_{\Phi}W \ ,\qquad  F_{A_{\bar 1}}= \ \partial_{A_{\bar 1}}W \   ,\qquad F_{A_{\bar 2}}= \ \partial_{A_{\bar 2}}W \ .
\label{auxiliary}
\eeq
In these expressions W is the physical superpotential of the 4d effective theory (with normalised fields) 
and it  includes a $\mu$-term for $\Phi$, with $\mu=-g_s^{1/2}(S_{\bar 3\bar 3})^*/(2\sqrt{2})$,
and a cubic term proportional to the Yukawa coupling, i.e.
\begin{equation}
W\ =\ \frac{\mu}{2}\, \Phi^2 \ +\ h_{123}\, A_{\bar 1}A_{\bar 2}\Phi \ +\ \ldots\label{superp}
\end{equation}

The scalar potential (\ref{noscaleadjoint}) is positive definite. This is consistent with the fact that
ISD fluxes locally preserve a no-scale structure \cite{Giddings:2001yu}. 
In terms of the physical scalar fields we have
\begin{equation}
V_{\rm ISD} \ =\  \left(\vert  M\vert ^2 +  \vert \mu \vert ^2\right) \, |\Phi|^2 +\ M \mu\, \Phi^2\ +\ \mu\, h_{123}^*\, \Phi A^{\bar 1} A^{\bar 2}\ +\ M h_{123}\, \Phi A^1 A^2\ +\ \textrm{h.c.} 
\end{equation}
Indeed, comparing with eq.~(\ref{notationsoft}) we read out the following pattern of soft terms, 
\begin{gather}
m^2_{3\bar 3}=\vert  M\vert ^2 \ +\  \vert \mu \vert ^2\quad ;\quad A_{ijk}=-Mh_{ijk}\\
B_{33}= 2M\mu\quad ;\quad C_{3jk}=-\mu h^*_{3jk}
\end{gather}
This reproduces the result for non-magnetized and non-intersecting 7-branes obtained in eq.~(\ref{softgen7}) when only ISD closed string fluxes are turned on. We will see in subsection \ref{compsugra} that this pattern corresponds to modulus dominance SUSY-breaking in an effective supergravity approach.

Let us now turn to the case of bifundamental fields living on intersecting 7-branes. To simplify the discussion we 
consider the above simple $U(3)$ example with no magnetization, although the results are valid for more realistic (e.g. SU(5), see section \ref{applic}) 
group theory structures. We slightly generalize the setting by considering the three D7-branes in the original stack to be tilted an arbitrary angle, so that the gauge 
group is fully broken to $U(1)^3$. As before, 4d bifundamental scalars arise from  $U(3)$ off-diagonal fluctuations of the adjoint fields
\begin{equation}
\Phi=\left(\begin{array}{ccc}\Phi_{\mathbf{1}^0}&\Phi_{a^+}&\Phi_{c^-}\\ \Phi_{a^-}&\Phi_{\mathbf{1}^0}'&\Phi_{b^+}\\\Phi_{c^+}&\Phi_{b^-}&\Phi_{\mathbf{1}^0}''\end{array}\right)\ , \qquad A=\left(\begin{array}{ccc}A_{\mathbf{1}^0}&A_{a^+}&A_{c^-}\\ A_{a^-}&A_{\mathbf{1}^0}'&A_{b^+}\\ A_{c^+}&A_{b^-}&A_{\mathbf{1}^0}''\end{array}\right)
\end{equation}
In absence of magnetic fluxes the three sectors $a$, $b$ and $c$ are vector-like and contain
massless chiral matter fields $a^\pm$, $b^\pm$, $c^\pm$ that are described by wavefunctions of the form (\ref{wave1}). For concreteness we 
take the curves $\Sigma_a$, $\Sigma_b$ and $\Sigma_c$ to be given by
\begin{equation}
\Sigma_a \ =\ \{z_1=0\} \ , \qquad  \Sigma_b \ =\ \{z_2=0\}\ , \qquad \Sigma_c \ =\ \{z_1=z_2\}
\end{equation}
as in the $U(3)$  model presented in section 2.3 of \cite{Aparicio:2011jx}.

One important effect of turning on a background for the transverse scalar $\Phi$ is that the eigenstates (\ref{wave1}) that solve the equations of motion in the internal space are generically a combination of $A^1$, $A^2$ and $\Phi$. Since the rotation induced in the space of internal wavefunctions commutes with dimensional reduction, we can think of the following three-step procedure to obtain the 4d Lagrangian of bifundamental fields. We first dimensionally reduce the 8d Lagrangian (\ref{8df}) to obtain a 4d Lagrangian for bulk fields, as we have already done in section \ref{sec2}. Next, we trace over the gauge indices in order to express this Lagrangian in terms of bifundamental fields. Last, we rotate the 4d fields to a new basis that diagonalizes eqs.~(\ref{ddeq}) and decouple massive modes that are at the string scale. Note that the rotation is different for each of the sectors of the theory, $a$, $b$ and $c$, involved in a Yukawa coupling.

For instance, for the matter fields localized in curve $\Sigma_a=\{z_1=0\}$ the rotation in the space of wavefunctions is given by
\begin{equation}
 \left(\begin{array}{c}\varphi_{a^+_1}\\ \varphi_{a^+_2}\end{array}\right)=\frac{1}{\sqrt{2}}\left(\begin{array}{cc}-1&1\\ 1&1\end{array}\right)\left(\begin{array}{c}A^1_{a^+}\\ \Phi_{a^+}\end{array}\right)\quad ;\quad \varphi_{a_3^+}=A_{a^+}^2
\label{rotation}
\end{equation}
Neglecting the effect of closed string fluxes on the internal wavefunctions (based on the
assumed  large hierarchy of scales $M_{ss}\ll M_{s}$), the fields $\varphi_{a_i}$ 
correspond to mass eigenstates with $m^2_{\varphi_{a^+_1}}=0$, $m^2_{\varphi_{a^+_2}}=2|m_a|^2$ and $m^2_{\varphi_{a^+_3}}=|m_a|^2$, as we saw in the previous subsection. For the sector $a^-$ the rotation is equivalent but the role of the fields $\varphi_{a_1}$ and $\varphi_{a_2}$ is interchanged. Moreover, by supersymmetry the same rotation also acts on the auxiliary fields, namely
\begin{equation}
\left(\begin{array}{c}F_{A_{a^+}}\\ F_{\Phi_{a^+}}\end{array}\right)
  =\frac{1}{\sqrt{2}}\left(\begin{array}{cc}-1&1\\ 1&1\end{array}\right) 
\left(\begin{array}{c} F_{\varphi_{a^+_1}}\\  F_{\varphi_{a^+_2}}\end{array}\right) \ .\label{frot}
\end{equation}

Since the fields $\varphi_{a^+_2}$, $\varphi_{a^-_1}$ and $\varphi_{a^\pm_3}$ are very heavy (with masses of order the string scale), 
correct decoupling in the effective theory dictates that in the effective 4d Lagrangian we should set
\begin{equation}
F_{\varphi_{a^+_2}}\ =\ F_{\varphi_{a^-_1}}\ =\ F_{\varphi_{a^\pm_3}}\ =\ 0
\end{equation}
along with 
\begin{equation}
\varphi_{a^+_2}\ =\ \varphi_{a^-_1}\ =\ \varphi_{a^\pm_3}\ =\ 0
\end{equation}
Thus, we can make use of the following replacements  in the effective action (\ref{noscaleadjoint}),
\begin{equation}
F_{A_{a^+}}\ = \ -\frac{F_{\varphi_{a^+_1}}}{\sqrt{2}}\ , \quad F_{\Phi_{a^+}}\ =\ \frac{F_{\varphi_{a^+_1}}}{\sqrt{2}}\ , \quad \Phi_{a^+}\ = \ \frac{\varphi_{a^+_1}}{\sqrt{2}}
\end{equation}
and the analogous ones for $a_2^-$ and for the sectors $b$ and $c$. This leads to a scalar potential of the form
\begin{equation}
V 
\  =\   \sum_{\alpha =a,b,c} \left(
\frac {1}{2} \vert  M^*\varphi_{\alpha^+_1}^* + F_{\varphi_{\alpha^+_1}}\vert ^2\ +\ \frac {1}{2} \vert  F_{\varphi_{\alpha^+_1}}\vert^2
\ +\  (\alpha^+_1\leftrightarrow \alpha^-_2)\right)\ 
\label{potbif}
\end{equation}
where the first term in this expression originates from $F_\Phi$ whereas the second term comes from $F_A$. 
To see explicitly how the soft terms for matter fields arise from this expression we expand the squared sum that appears in the above potential,
\beq
V\ =\ \sum_{\alpha =a,b,c}  \left(  \frac12\vert M\vert ^2\vert \varphi_{\alpha^+_1}\vert ^2\  +\ \vert F_{\varphi_{\alpha^+_1}}\vert ^2\ +\ \frac12 MF_{\varphi_{\alpha^+_1}}\varphi_{\alpha^+_1}\ + \ \textrm{h.c.}\ +\ (\alpha^+_1\leftrightarrow \alpha^-_2)\right)\ .\label{potbi}
\eeq
The first term corresponds to a soft mass for the scalar fields $\varphi_{\alpha_1^+}$, which is a factor $1/2$ smaller than the one that we had for adjoint fields.
Moreover, in absence of magnetization the superpotential contains $\mu$-terms proportional to $\varphi_{\alpha_1^+}\varphi_{\alpha_2^-}$ and hence we can express the auxiliary field of $\varphi_{a^+_1}$ as 
\begin{equation}
F_{\varphi_{a^+_1}}\ =\ \frac{1}{\sqrt{2}}\left( F_{\Phi_{a^+}}\ +\ F_{A_{a^+}}\right)=\ 
\frac{\mu \Phi_{a^-}}{\sqrt{2}}\ + \ \ldots \ = \ \frac{\mu }{2} \varphi_{a^-_2}\ + \ \ldots \ = \ \mu_{\rm bif}^a\varphi_{a^-_2}\ +\ \ldots
\end{equation}
where the dots represent higher-order superpotential terms such as Yukawa couplings. Therefore, $\mu_{\rm bif}^a=\mu/2$ 
and similarly for the other two matter curves, if they host vector-like states.
The second term in eq.~(\ref{potbi}) hence gives rise to supersymmetric masses for the scalar fields, given by $|\mu_{\rm bif}^\alpha|^2$, and to the usual supersymmetric trilinear coupling, that can be written as a product of $\mu_{\rm bif}^\alpha$ and the effective Yukawa coupling. Finally, the last term in eq.~(\ref{potbi}) gives rise to $B$-terms and SUSY-breaking trilinear couplings $A$. Comparing with the case of adjoint fields, they are also suppressed by a factor $1/2$.

Summing over the three curves $a$, $b$ and $c$ we therefore obtain soft masses of the form $|M|^2/2$ for each of the 4d scalars. In addition, we get $\mu$-terms, $B$-terms and a supersymmetric trilinear coupling for each non-chiral curve. Recall that for the soft SUSY-breaking trilinear coupling we get the same result three times (one for each curve), leading to an extra multiplicative factor $3$. 

Summarizing, we have obtained the following set of  soft terms for bifundamental fields in a system of intersecting non-magnetized D7-branes with ISD 3-form fluxes,
\beqa
m_{{\rm bif},\alpha}^2\ & =& \  \frac {|M|^2}{2} \ +\ \frac {|\mu|^2}{4}\ =\
\frac{g_s}{4}\left(\vert G_{\bar{1}\bar{2}\bar{3}}
\vert^2+\frac{1}{8}\vert S_{\bar{3}\bar{3}}\vert^2\right)\ 
\nonumber
\\
(B_{{\rm bif}}\, \mu_{\rm bif})^\alpha &=& \frac {1}{2} B_{\Phi}\, \mu_{\rm bif}^\alpha \ =\   M\, \mu_{\rm bif}^\alpha\ =\  - \ \frac{g_s}{8}(G_{{\bar 1}{\bar 2}{\bar 3}})^*(S_{{\bar 3}{\bar {3}} })^*
\nonumber
\\
A_{\rm bif}^{ijk}\ &=&\ \frac {3A_\Phi}{2} \ =\   - \ \frac {3}{2}Mh^{ijk}\ =\  - {{3g_s^{1/2}}\over {2\sqrt{2}}}\,
(G_{\bar{1}\bar{2}\bar{3}})^* h^{ijk}.
\label{softbif1}
\eeqa
where $\alpha = a,b,c$ and we have now factored out in this expression explicitly the $\mu_{\rm bif}$ factor from the definition of the $B$-parameter. Gaugino masses remain unaltered since they are not localized by the non-trivial background of $\Phi$. Hence, for the 
 fermonic masses we have
\beqa
M\ &=& \  {{g_s^{1/2}}\over {\sqrt{2}}}
(G_{\bar{1}\bar{2}\bar{3}})^* \nonumber\\
\mu _{\rm bif}^\alpha \ & =& \  \frac {\mu}{2} \ =\ -{{g_s^{1/2}}\over {4\sqrt{2}}} (S_{{\bar 3}{\bar 3}})^*
 \label{softbif2}
\eeqa
We can also guess the contribution to soft scalar masses coming from the IASD fluxes $S_{ii}$, $i=1,2$. 
Indeed, looking at the results  for the bulk D7-brane fields in eq.~(\ref{softgen7}), we expect an additional dependence on
IASD fluxes through the replacement 
\beq
|S_{{\bar 3}{\bar 3}}|^2 \ \rightarrow \ |S_{{\bar 3}{\bar 3}}|^2+|S_{11}|^2+|S_{22}|^2
\eeq
in the mass squared and
\beq 
G_{{\bar 1}{\bar 2}{\bar 3}}S_{{\bar 3}{\bar 3}}\ \rightarrow \ G_{{\bar 1}{\bar 2}{\bar 3}}S_{{\bar 3}{\bar 3}}+\frac {1}{2} (S_{11}S_{22})^* \ .
\eeq
for the B-term. This is suggested by symmetry arguments similar to those  used in section 3.1 of \cite{Camara:2004jj}.
 However, no contribution to trilinear couplings or fermion masses is expected 
from $S_{ii}$ IASD fluxes, since there are no holomorphic gauge components $A_{1,2}$ present in the
chiral matter fields.

We now turn to discuss how magnetization modifies this pattern of soft terms for matter fields.

\subsection{Effect of magnetic fluxes on soft terms for fields at matter curves}
\label{magnetico}

Magnetization leads to 4d chiral spectra, as reviewed in section \ref{matterfields}, with total chirality determined by the integral of $\langle F_2 \rangle$ over the various matter curves of the theory. For concreteness, let us assume that the curves $\Sigma_a$ and $\Sigma_b$ in our $U(3)\to U(1)^3$ toy model above are now charged under the flux, such that only the modes $a^+$ and $b^+$ survive in the 4d spectrum. We take the matter curve $\Sigma_c$ however to be neutral under the flux, and so the spectrum arising from this curve is unaffected, containing the vector-like pair given by $c^+$ and $c^-$. In this toy model we may think of the non-chiral sector localized in $\Sigma_c$ as the Higgs sector, whereas the chiral sectors localized in the curves $\Sigma_a$ and $\Sigma_b$ can be thought as MSSM chiral sectors. A more realistic example is given in section \ref{applic}, where we apply the results of this section to study the hypercharge dependence of soft terms in a local F-theory SU(5) GUT model.

As we have already mentioned, magnetic fluxes affect the 4d soft SUSY-breaking Lagrangian in two ways. On one side, 
the presence of a non-trivial background for $F_2$ leads to new renormalizable couplings in 4d which can be traced back to higher-dimensional couplings in the 8d theory where some of the fields-strengths present in the coupling are replaced by the background flux. These corrections were computed in subsection \ref{magnet} for the case of bulk D7-brane fields. They are quadratic in the magnetic flux density and from the point of view of the 4d effective supergravity correspond to renormalizable thresholds to the K\"ahler potential and/or the gauge kinetic function of the 4d effective theory. The other effect of magnetic fluxes, relevant for matter fields, is to modify the profile of the internal wavefunctions, as it has been described in subsection \ref{matterfields}, and therefore also the rotation in the space of internal wavefunctions. For instance, in our $U(3)$ example above the internal wavefunctions for 4d charged fields in presence of magnetic fluxes were given in eq.~(\ref{wave3}). The rotation eq.~(\ref{rotation}) in the space of fields is thus modified such that
\begin{equation}
\begin{pmatrix}\varphi_{a^+_1}\\ \varphi_{a^+_2}\end{pmatrix}=\begin{pmatrix}\frac{F_-^a-\lambda_a}{\sqrt{2\lambda_a(\lambda_a-F_-^a)}}&\frac{m_a}{\sqrt{2\lambda_a(\lambda_a-F_-^a)}}\\ \frac{F_-^a+\lambda_a}{\sqrt{2\lambda_a(\lambda_a+F_-^a)}}&\frac{m_a}{\sqrt{2\lambda_a(\lambda_a+F_-^a)}}\end{pmatrix}\begin{pmatrix}A^1_{a^+}\\ \Phi_{a^+}\end{pmatrix}\quad ;\quad \varphi_{a_3^+}=A_{a^+}^2
\label{rotation_flux}
\end{equation}
Mass eigenstates therefore still originate from a mixture between Wilson lines and transverse scalars, but this mixture now depends on the magnetic flux in the curve $\Sigma_a$. Only in the case without magnetic flux, $\Phi$ and $A$ contribute equally to the mass eigenstates. Moreover, this correction begins at linear order on the magnetic flux, and therefore for bifundamental fields the quadratic corrections described in section \ref{magnet} are sub-leading. We thus ignore those and just consider the leading effect coming from the modification of the $\Phi-A$ mixing induced by magnetic fluxes.

In order to compute the soft terms of matter fields localized in the curves $\Sigma_a$, $\Sigma_b$ and $\Sigma_c$, we follow the same procedure described in the previous subsection. The rotation matrix for the fields localized in the matter curve $\Sigma_a$ is now given by eq.~(\ref{rotation_flux}). A similar rotation also applies to the fields localized in curve $\Sigma_b$, after interchanging $A^1\leftrightarrow A^2$. The rotation for auxiliary fields is modified in the same way than for the scalar fields.  Therefore, for the fields localized in the matter curves $\Sigma_{a,b}$ we can make the replacements
\begin{equation}
F_{A_{\alpha^+}}\ =\ -\sqrt{\frac{\lambda_\alpha-F_-^\alpha}{2\lambda_\alpha}}F_{\varphi_{\alpha_1^+}}\ , \qquad F_{\Phi_{\alpha^+}}\ =\ \sqrt{\frac{\lambda_\alpha-F_-^\alpha}{2\lambda_\alpha}}\frac{\lambda_\alpha+F_-^\alpha}{m_\alpha}F_{\varphi_{\alpha_1^+}}\ ,  \qquad \alpha=a,b
\end{equation}
where we have already set the auxiliary fields of massive modes to zero. In these expressions $\lambda_\alpha =\sqrt{(F_-^\alpha)^2+m_\alpha^2}$. Note that only the fields coming from the sector $a^+$ and $b^+$ are normalizable and therefore those coming from the sectors $a^-$ and $b^-$ are not present in the low energy spectrum of the theory. This implies that $\mu$- and $B$-terms are absent in the matter curves $\Sigma_a$ and $\Sigma_b$. Curve $\Sigma_c$ on the other hand is not affected by magnetic fluxes, and therefore the same expression (\ref{frot}) for the rotation of auxiliary fields in absence of magnetic fluxes still applies for $\Sigma_c$. 

Making all these substitutions in the potential (\ref{noscaleadjoint}) we get
\begin{multline}
V \  =\   \sum_{\alpha =a,b} \left(
\frac{\lambda_\alpha-F_-^\alpha}{2\lambda_\alpha}\frac{(\lambda_\alpha+F_-^\alpha)^2}{m_\alpha^2} \vert  M^*\varphi_{\alpha^+_1}^* + F_{\varphi_{\alpha^+_1}}\vert ^2\ +\ \frac{\lambda_\alpha-F_-^\alpha}{2\lambda_\alpha} \vert  F_{\varphi_{\alpha^+_1}}\vert^2
\ \right)\ +\\
+\ \frac {1}{2} \vert  M^*\varphi_{c^+_1}^* + F_{\varphi_{c^+_1}}\vert ^2\ +\ \frac {1}{2} \vert  F_{\varphi_{c^+_1}}\vert^2
\ +\  (c^+_1\leftrightarrow c^-_2)\ 
\label{potbif_flux}
\end{multline}
and expanding perturbatively in powers of the ratio $F_-^\alpha/m_\alpha$ between the magnetization and the intersection parameter, the contribution to the scalar potential that comes from the sectors $\alpha=a,b$ becomes
\begin{multline}
V_\alpha \  =\   
\frac12 \left[1-\left|\frac{F_-^\alpha}{m_\alpha}\right|+\mathcal{O}\left(\left|\frac{F_-^\alpha}{m_\alpha}\right|^3\right)\right] \vert  M\vert^2|\varphi_{\alpha^+_1}|^2  +\ \vert F_{\varphi_{\alpha^+_1}}\vert ^2\ +\\
+\ \frac12 \left[1-\left|\frac{F_-^\alpha}{m_\alpha}\right|+\mathcal{O}\left(\left|\frac{F_-^\alpha}{m_\alpha}\right|^3\right)\right] M F_{\varphi_{\alpha^+_1}}\varphi_{\alpha^+_1}
\ 
\label{potbif_flux2}
\end{multline}
The $\mu$-term $\mu_{\rm bif}^c$ is not modified to linear order on the fluxes, since the curve $\Sigma_c$ is neutral under the magnetic flux. Hence, to leading order we still have $\mu_{\rm bif}=\mu/2$, as in the case with no magnetization discussed in the previous subsection.

Summarizing, from eqs.~(\ref{potbif_flux}) and (\ref{potbif_flux2}) we have therefore derived the following set of flux-induced soft terms for intersecting magnetized 7-branes in the above $U(3)$ toy model, for fields localized in each of the matter curves $\Sigma_\alpha$,
\begin{align}
m_{{\rm bif},\alpha}^2\ & \ = \  \frac {|M|^2}{2}\left(1-\left|\frac{F_-^{\alpha}}{m_\alpha}\right|\right)\ = \ \frac{g_s}{4}|G_{\bar 1\bar 2\bar 3}|^2\left(1-\left|\frac{F_-^\alpha}{m_\alpha}\right|\right)\   \qquad \alpha = a,b
\nonumber
\\
m_{{\rm bif},c}^2\ & \ =\  \frac {|M|^2}{2} \ +\ \frac {|\mu|^2}{4} \ = \ \frac{g_s}{4}\left(|G_{\bar 1\bar 2\bar 3}|^2+\frac18|S_{\bar 3\bar 3}|^2\right)\nonumber \\
(B_{\rm bif}\, \mu_{\rm bif})^c &\ =\ \frac {1}{2} B_{\Phi}\, \mu_{\rm bif}^c \ =\ M\mu_{\rm bif}^c\ = \ -\frac{g_s}{8}(G_{\bar 1\bar 2\bar 3})^*(S_{\bar 3\bar 3})^* 
\nonumber
\\
A_{\rm bif}^{ijk}\ &\ =\ -\frac {M}{2}  \left(3-\left|\frac{F_-^a}{m_a}\right|-\left|\frac{F_-^b}{m_b}\right|\right)h^{ijk}\ = \ -\frac{g_s^{1/2}h^{ijk}}{2\sqrt{2}}(G_{\bar 1\bar 2\bar 3})^*\left(3-\left|\frac{F_-^a}{m_a}\right|-\left|\frac{F_-^b}{m_b}\right|\right)
\label{softbif_flux}
\end{align}
Note in particular that $\mu$- and $B$-terms do not receive corrections linear in the magnetic fluxes, as we have already mentioned.
This is because in this particular case there are no magnetic fluxes along the curve c, which is the one hosting the Higgs fields.
On the other hand there would appear  corrections quadratic  in the magnetic fields, analogous to those appearing for adjoint
fields in the previous chapter.

Let us mention for completeness that 
there is also a third possibility for brane distributions with consistent Yukawa couplings, even  in compactifications with a rigid $S$ divisor.
Indeed, we can have a coupling of the form (I-I-A) involving two fields coming from the intersection of D7-branes and one field coming from the reduction of the gauge field $A$ living in the 7-brane  worldvolume. If this is the case, it is natural to assume that the Higgs field arises
 from the worldvolume of the D7-branes while the MSSM chiral matter arise from intersections (labelled by $a,\, b$). Hence the soft mass and the $B$-term for the Higgs are forbidden by gauge invariance, whereas the soft masses for the chiral fields take the form described above.
We can summarize this structure in the following scalar potential,
\beq
V=\sum_{\alpha=a,b} V_\alpha + |F_A|^2
\eeq
where $V_\alpha$ is described in eq.~(\ref{potbif_flux2}) and $F_A$ is the auxiliary field for the $A$ field. Recall that it does not include a $\mu-$term by gauge invariance.
Thus the trilinear coupling will have only two contributions coming from $V_{a,b}$. To sum up, this brane distribution leads to the following flux-induced soft SUSY-breaking terms,
\begin{align}
m^2_H&\ =\ 0\ , \qquad B_H\ =\ 0\label{IIA}\\
m^2_{{\rm bif}, \, a,b}&\ =\ \frac{|M|^2}{2}\left(1-\left|\frac{F_-^{a,b}}{m_a}\right|\right)\nonumber\\
A^{Hij}&\ =\ -\frac {M}{2}  \left(2-\left|\frac{F_-^a}{m_a}\right|-\left|\frac{F_-^b}{m_b}\right|\right)\, h^{Hij}\  .
\nonumber
\end{align}

\subsection{Comparison with effective $\mathcal{N}=1$ supergravity}
\label{compsugra}

As emphasized in refs.~\cite{Camara:2003ku,Camara:2004jj}, the pattern of flux-induced soft terms that arise in the worldvolume of
D3/D7-branes for ISD 3-form fluxes can be also understood in terms of effective $\mathcal{N}=1$ supergravity. For the case of adjoint fields with no magnetization, discussed in section \ref{sec21}, soft terms agree with those obtained from a simple no-scale 
K\"ahler potential for a single K\"ahler modulus $T$ and a gauge kinetic function of the form 
\beq
K\ =\ -3\, \textrm{log}(T+T^*) \ ,\qquad   f_a\ =\ T \ ,
\eeq
as well as a K\"ahler metric for  matter fields
\beq
K_{ij}\ =\  \frac {\delta_{ij}} { (T+T^*)^{\xi_i}}\ \label{kahlmet}
\eeq
with  $\xi_i$ the so-called {\it modular weight} of the scalar field $\phi_i$. This structure is more than a toy model. Indeed, one obtains such a simple structure 
in isotropic toroidal orientifolds in which $T$ is the overall K\"ahler modulus with $T=T_1=T_2=T_3$, and a stack of D7-branes wraps a 4-torus $T^2\times T^2$ within the $T^6$. The modular weight of 4d adjoint fields that descend from $\Phi$ and $A$ is given respectively by $\xi = 0$ and $\xi = 1$. We ignore the dependence of these expressions on the complex axion-dilaton, the complex structure moduli
and the other K\"ahler moduli present in the theory, since those are not relevant for the computation of soft terms below.

Assuming that the F-term auxiliary field $F_T$ of the modulus $T$ is non-vanishing ({\it modulus dominance}), the standard 
$\mathcal{N}=1$ supergravity formulae  (see e.g. \cite{Brignole:1997dp}) yield
\begin{align}
m_{\phi_i}^2&\ =\ |M|^2(1-\xi_i) \\
A_{ijk} & \ =\  -  Mh_{ijk}\sum_{\alpha=i,j,k}\ (1-\xi_\alpha)\nonumber\\ 
B^i & \ =\ M\sum_{\alpha=\phi_u^i,\phi_d^i}(1-\xi_\alpha) \nonumber
\end{align}
where $\phi_{u,d}^i$ represent possible vector-like states allowing for a supersymmetric $\mu$-term. In particular, for adjoint fields 
that descend from $\Phi$ and $A$ we get
\begin{equation}
m_{\Phi_i}^2 \ =\ {|M|^2}\ , \qquad A_{ijk} \ =\  -  Mh_{ijk}\ , \qquad B_\Phi \ =\ 2M
\end{equation}
and $B_A=m_A^2= 0$. This is consistent with the more general result shown in eq.~(\ref{softgen7}) particularized to case in which only 
ISD fluxes  $G_{\bar 1\bar 2\bar 3}$  and $S_{\bar 3\bar 3}$ are present. 
In the case on which magnetic fluxes are also present, the K\"ahler metric (\ref{kahlmet}) is suitably corrected by the magnetization as
\cite{Aparicio:2008wh,Aparicio:2012ju}
\beq
K_{ij}\ =\  \frac {\delta_{ij}} { t^{\xi_i}} \left(1\ +\ c_\xi t^{\xi-1}\right) \ \label{kahler_flux}
\eeq
with $t=T+T^*$ and 
 $c_\xi$ some flux-dependent constant whose value will depend on the modular weight and the flux quanta. The corrections to the soft terms that arise from this K\"ahler metric are in agreement with those found in eqs.~(\ref{corr1}) and (\ref{corr2}) particularized to the case of ISD 3-form fluxes and anti self-dual magnetic flux $F_-$ once we identify the flux correction of (\ref{kahler_flux}) with our microscopic description of the flux density, $\rho\equiv\frac{c_\Phi}{t}=g_s^{-1}\sigma^2F_-^2$. 
 One also finds that for $A$ fields, which have $\xi=1$, one has
  $c_A=0$ and the fields that descend from $A$ remain massless even after the addition of magnetic fluxes.

Similarly, the soft terms for matter fields in intersecting D7-branes given in eqs.~(\ref{softbif1}) and (\ref{softbif2}) can be also reproduced by the above  $\mathcal{N}=1$ supergravity formulae. Indeed, the modular weight of chiral fields localized at intersecting D7-branes is given by $\xi=1/2$. In absence of magnetic fluxes, standard supergravity formulae then leads to
\beq
m_{{\rm bif},i}^2\ =\    \frac {|M|^2}{2}\ ,\qquad  A_{ijk} \ =\   -  \frac {3M}{2}h_{ijk}\ , \qquad
B^i_{\rm bif} \ =\ M \ 
\eeq
in agreement with eqs.~(\ref{softbif1}). The corrections from magnetic fluxes arising from (\ref{kahler_flux}) to the different soft terms are parametrized for the case of fields with modular weight $\xi=1/2$ by $\rho_{\text{bif}}\equiv \frac{c_{\text{bif}}}{t^{1/2}}$. Note that in the large $t$ limit (corresponding to the flux diluted regime) these corrections are dominant since $\rho_{\text{bif}}>\rho$. This is consistent with the linear (instead of quadratic) dependence on the fluxes found in (\ref{softbif_flux}).

Finally we can also derive the structure of soft terms in a (I-I-A)-type configuration using the K\"ahler metric above. In this case we have two matter fields coming from D7-brane intersections with modular weight $\xi=1/2$ and one adjoint field that descends from $A$ with modular weight $\xi=1$. The standard $\mathcal{N}=1$ supergravity formulae yield
\beq
m_f^2\ =\    \frac {|M|^2}{2}\ ,\qquad  A_{ijk} \ =\   -  Mh_{ijk}\ , \qquad
B_H\ = \ m^2_H\ =\ 0 \ 
\eeq
in agreement with eq.(\ref{IIA}) as expected. The flux correction for the matter fields will be also parametrized by $\rho_f=\frac{c_f}{t^{1/2}}$ consistent with the linear dependence found in (\ref{IIA}).

The above structure of soft terms does not only arises in toroidal settings but also in {\it swiss-cheese} compactifications 
\cite{LARGEvolume}
in which 
a stack of 7-branes containing the SM fields wraps a small cycle of size $t_s=\textrm{Re}(T_s)$ inside a large-volume CY manifold with overall volume modulus $T_b$. This is also the type of configurations that one expects in local  F-theory  GUT models, 
where $T_s$ would correspond to the local K\"ahler modulus associated to the local divisor $S$. 

In the simplest type IIB swiss-cheese examples the K\"ahler potential for the moduli $T_s$ and $T_b$ is given by
\cite{cremades}
\beq
K\ =\ -2\log\,(t_b^{3/2} \ -\ t_s^{3/2}) \, ,
\eeq
with $t_b=\textrm{Re}(T_b)\gg t_s$, whereas to leading order the gauge kinetic function is given by $f=T_s$. The K\"ahler metric for
the matter fields reads
\beq
 K_\alpha \ =\ \frac {t_s^{(1-\xi_\alpha )}
}{t_b}\,  , 
\eeq
with $\xi_\alpha$ the corresponding modular weights. Expanding the action in powers of $t_s/t_b$ and assuming $F_{t_b,t_s}\not= 0$
we obtain the same patterns of soft-terms as in the above toroidal case, where now $M=F_{t_s}/t_s$ \cite{Aparicio:2008wh}.

Note however  that
 the microscopic derivation of soft-terms in section \ref{sec2} and in this section  go beyond  these $\mathcal{N}=1$ supergravity results in various respects. In particular they do not assume any form for the $\mathcal{N}=1$  K\"ahler potential but
give explicit expressions for the soft-terms in terms of the underlying general closed string background. In this regard,
they are expected to be valid in more complicated non-toroidal settings and may also include the effect of
IASD sources.  Obtaining the closed string background around the D7-branes in a general 
compactification is usually a too complicated task, but once the closed string background is known, the techniques developed in the above
sections allow to obtain the soft-breaking patterns for the fields in the worldvolume of D7-branes. This approach might be particularly useful 
for fields localized at D7-brane intersections, since their K\"ahler metrics are only fully known in the case of toroidal compactifications, whereas for the case of local systems like the swiss-cheese kind of setting discussed above the structure of the K\"ahler metrics for matter fields can at present only be guessed
in terms of scaling arguments \cite{cremades,Aparicio:2008wh}.

\section{Effect of distant  branes on the local soft terms}
\label{sec4}

\begin{figure}[!ht]
\begin{center}
\includegraphics[scale=1.0]{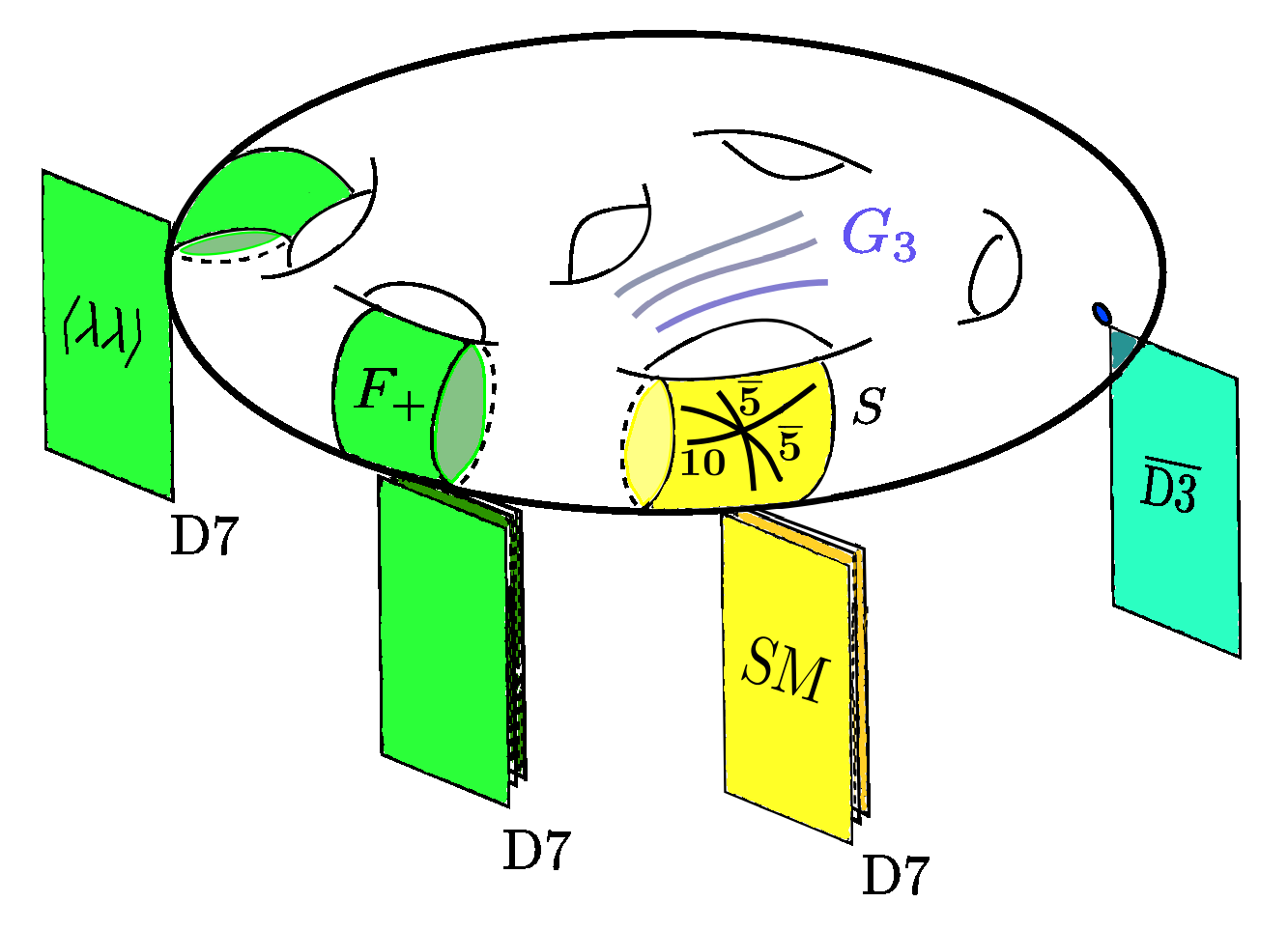}
\caption{\small  Summary of the type of sources that are present in a standard phenomenological IIB orientifold compactification. The SM is located in a stack of intersecting D7-branes with a higher dimensional SU(5) GUT structure. Apart from topologically non-trivial closed string 3-form fluxes $G_3$ , there are distant localized sources that may also contribute to SUSY-breaking and/or moduli stablization. These include gaugino condensation in the worldvolume of D7-branes, self-dual magnetic fluxes also in the worldvolume of D7-branes and/or anti-D3-branes. The effect of distant sources
in the effective theory on the worldvolume of the SM D7-branes can be studied in terms of their backreaction in the local patch.\label{fig1}}
\end{center}
\end{figure}

When building phenomenological type IIB orientifold compactifications the degrees of freedom of the SM typically are located 
in the worldvolume of D7-branes and/or  D3-branes subject to closed string and open string fluxes. The type of settings that are typically considered is shown in figure \ref{fig1}. Apart from the branes of the SM sector, there
may also be additional localised sources at other regions of the compact space. For instance, there could be distant D7-branes giving rise to
gaugino condensation and stabilizing some of the K\"ahler moduli of the compactification. There might also be anti-D3-branes, as in the KKLT setting \cite{Kachru:2003aw}, required to uplift
the vacuum from AdS to dS. Alternatively, this role might also be played by distant D7-branes with self-dual magnetic fluxes in their
worldvolume \cite{Burgess:2003ic}. The effect of distant localised sources on the SM branes may be discussed in terms of their backreaction near the SM branes, as discussed e.g.~in \cite{Marchesano:2009rz,Baumann:2010sx} for the particular case of gaugino condensation on D7-branes. In this section we discuss the effect of distant localised sources on the pattern of soft breaking terms by computing the backreaction of localised sources on the local geometry. 

%%%%%%%%%%%
%%%%%%%%%%%

For concreteness we focus on the case of distant anti-D3-branes. In general these backreact the metric and the RR 5-form field-strength through the equations of motion. We have seen in previous sections that in absence of magnetization soft-terms for fields on D7-branes (both bulk and on intersections) do not depend on the metric nor on the RR 5-form and thus the presence of distant   D3- or anti-D3-branes does not modify D7-brane soft terms within this approximation. This is expected, since unmagnetised D7-branes have no net D3-brane charge. However, once anti self-dual or self-dual magnetic fluxes are switched on in the worldvolume of D7-branes, some D3- or anti-D3-brane charge is respectively induced in their worldvolume. This implies that distant   anti-D3-branes (or D3-branes, respectively) are now expected to give rise to corrections for the soft-terms in the worldvolume of magnetized D7-branes. Indeed, we saw in section \ref{magnet} that magnetization leads 
to corrections to D7-brane soft terms that depend on the background for the metric and the RR 5-form. Although these corrections are quadratic in the magnetic fluxes, they can lead to relevant physical effects if 3-form fluxes are suppressed or in the context of fine-tuned scalar potentials,
in which minute effects become important.

We begin this section by reviewing the computation of soft scalar masses induced on the worldvolume of D3-branes by distant anti-D3-branes
in flat space \cite{Camara:2003ku}.  We then move to the same computation for magnetised D7-branes in flat space. Finally, we consider compactification effects in these computations.

\subsection{Scalar masses for D3-branes in the presence of distant anti-D3-branes}
\label{sec41}

We first consider the case of a probe D3-brane located in (non-compact) locally flat space and 
a distant stack of $N$ anti-D3-branes, and compute the induced soft scalar masses in the worldvolume of the 
D3-brane. This computation was addressed  in Ref.~\cite{Camara:2003ku}, but we revisit it here with the 
aim of extending it to other settings in the next subsections.

In general, anti-D3-branes backreact the metric and the RR 4-form potential through the following type IIB supergravity equations of motion
\begin{align}
-\tilde\nabla^2 Z &=\frac{g_s}{12}G_{mnp}\bar G^{\widetilde{mnp}}+(2\pi\sigma)^ 2\tilde{\rho}_3^{\rm loc}+Z^{-1}\left[\partial_m Z\tilde\partial^m Z-(Z\chi)^4\partial_m\chi^{-1}\tilde\partial^m\chi^{-1}\right]\label{eq1}\\
-\tilde\nabla^2 \chi^{-1}&=\frac{ig_s(Z\chi)^2}{12}G_{mnp}*_6\bar G^{\widetilde{mnp}}+(2\pi\sigma)^2\tilde Q_3^{\rm loc}+2\left[Z^{-1}\partial_m\chi^{-1}\tilde\partial^mZ-\chi\partial_m\chi^{-1}\tilde\partial^m
\chi^{-1}\right]\nonumber
\end{align}
where tilded quantities are taken with respect to the unwarped metric and $\tilde{\rho}_3^{\rm loc}(z)$ and $\tilde{Q}_3^{\rm loc}(z)$ are the energy density and D3/$\overline{\rm D3}$-brane charge density associated to localized sources. These equations are easily solved for backgrounds that only involve same-sign D3-brane charges (recall that we are taking the D3-branes as probes). For the particular case of a stack of N anti-D3-branes and vanishing 3-form fluxes
\begin{equation}
\tilde{Q}_3^{\rm loc}(z)=-\tilde\rho_3^{\rm loc}=-N \frac{\delta(\vec z_0-\vec z)}{\sqrt{\tilde g}} \ , \qquad Z=-\chi^{-1}
\label{solution}
\end{equation}
where $\vec z_0$ denotes the position of the stack of anti-D3-branes in the internal space  
Eqs.~(\ref{eq1}) are then proportional to each other and reduce to a standard Poisson equation in the internal space
\begin{equation}
-\tilde\nabla^2 Z =(2\pi\sigma)^2\tilde{\rho}_3^{\rm loc}\label{pois}
\end{equation}
When the internal space is non-compact flat space this leads to the standard 
supergravity solution for anti-D3-branes in asymptotically flat space, namely
\begin{equation}
Z=1-\frac{g_s N \sigma^2}{\pi\, |\vec{z}-\vec{z}_0|^4}\label{zantid3}
\end{equation}

Soft terms in the worldvolume of the probe D3-branes are fully determined in terms of the local backreaction around the D3-branes. Concretely, for the soft scalar masses \cite{Camara:2003ku}
\begin{align}
m_{i\bar{j}}^2=2K_{i\bar{j}}-\chi_{i\bar{j}}+g_s(\textrm{Im}\,\tau)_{i\bar{j}}\label{mBB}\\
B_{ij}=2K_{ij}-\chi_{ij}+g_s(\textrm{Im}\,\tau)_{ij} \ .
\nonumber
\end{align}
where $K$, $\chi$ and $\tau$ were defined in eq.(\ref{expan}).
For concreteness we take the probe D3-branes to be located at the origin of coordinates. Expanding eq.~(\ref{zantid3}) around the origin leads in real coordinates to
\begin{equation}
Z\ =\ 1-\frac{g_s N \sigma^2}{\pi\, r_0^6}\left[r_0^2+4x_0^mx^m+2\left(\frac{6x_0^mx_0^n}{r_0^2}-\delta^{mn}\right)x^mx^n+\ldots \right].\ \label{zzd}
\end{equation}
where $r_0^2=\sum_n(x_0^n)^2$. The linear term shows the expected instability due to the attraction between branes and anti-branes. We assume in what follows that
such a term is absent, leading to a static configuration. That may originate in a variety of ways  like e.g.~an orbifold projection,
a particularly symmetric configuration or the D3-branes being fractional and stuck at a singularity.  

Comparing to eqs.~(\ref{expan}) and making use of eqs.~(\ref{solution}) and (\ref{mBB}) we then obtain the following  scalar masses and $B$-term in the worldvolume of the probe D3-brane
\begin{equation}
m_{i\bar{j}}^2=\text{const.}\left(\frac{6}{r_0^2}z_0^i\bar{z_0}^j-\delta^{ij}\right)\ , \qquad
B_{ij}=\text{const.}\,\frac{6}{r_0^2}z_0^iz_0^j
\label{BB}
\end{equation}
where the proportionality constant is $ 8g_s N \sigma^2/(\pi r_0^6-g_s N \sigma^2 r_0^2)$. We would have obtained the same result if we have instead considered the reverse situation, namely soft terms induced on a anti-D3-brane by the presence of a distant D3-brane.

These mass terms by themselves may easily trigger instabilities for the scalars on the D3-brane, since they may
 be tachyonic. For instance, if along the $i$-th complex plane $|z_0^i|\ll |z_0^j|$ with $i\neq j$, the second piece in the first equation (\ref{BB}) dominates and the D3-brane scalar field $\Phi^i$ becomes tachyonic. 
 In the isotropic case, where $z_0^1=z_0^2=z_0^3$ and thus $\frac{z_0^i}{r_0}=\frac{1}{\sqrt{6}}$, one gets
\begin{equation}
m_{i\bar{j}}^2\ =\ \text{const.}\,(1-\delta^{ij})\ , \qquad B_{ij}\ =\ \text{const.} \label{mB2}
\end{equation}
Hence, in that case diagonal  masses vanish and off-diagonal ones are equal to the B-term. 
Still, there are tachyonic mass eigenstates, since the mass matrix is traceless.
Note that, as emphasized in \cite{Camara:2003ku} this source of SUSY breaking {\it by itself } would lead 
in addition to
no gaugino masses  nor $\mu$-terms and  would therefore not be  phenomenologycally viable for
MSSM soft terms without the addition of further ingredients.

\subsection{Scalar masses for magnetized D7-branes  in the presence of distant anti-D3-branes}

We can perform the same analysis as above for the case of magnetised D7-branes in the presence of 
distant anti-D3-branes in asymptotically flat space. To simplify the presentation let us consider only 
a non-vanishing anti self-dual magnetic flux $F_-$ in the worldvolume of some probe D7-branes. From 
eq.~(\ref{corr1}) we get for the scalar bilinears 
\begin{equation}
m_{3\bar 3}^2 =  \sigma^2\left( 2g_s^{-1}K_{3\bar 3}-\chi_{3\bar 3} \right)F_-^2\ , \qquad
B_{33} = \sigma^2\left(2g_s^{-1}K_{33}-\chi_{33} \right)F_-^2 \ .\label{ffh}
\end{equation}
Note that  $K_{{ 3}{\bar 3}}$, $K_{33}$, $\chi_{3{\bar 3}}$  and $\chi_{33}$ obtained in the previous subsection depend on the 
 coordinates of the D7-branes along the internal space, $z^1$ and $z^2$, and dimensional reduction to 4d is therefore non-trivial. However, if the 
 wavefunctions of the 4d fields are strongly localized in the internal space, as occurs for instance for fields localized at Yukawa coupling
 enhancement points in F-theory GUTs, we can approximate wavefunctions by a delta function. Here we take for instance the case of a vector-like 
 pair of scalars localized at the origin of coordinates. Then, making use of eqs.~(\ref{solution}) and (\ref{zzd}) in (\ref{ffh}) we find for $|z_0^3|\gg |{ z}_0^1|,|{ z}_0^2|$
\beq
m_{3\bar 3}^2\ =\ \text{const.}\, g_s^{-1}\sigma^2 F_-^2\ \ ,\qquad
B_{33}\ =\ \frac34 \text{const.}\, g_s^{-1}\sigma^2 F_-^2 \ 
\eeq
where we have included an extra factor 1/2 with respect to eq.~(\ref{ffh}) to account for the fact that we are now considering 
 a vector-like pair of bifundamental scalars, according to what we found in section \ref{sec32}.

\subsection{Compactification effects}

\begin{figure}[!ht]
\begin{center}
\includegraphics[scale=.7]{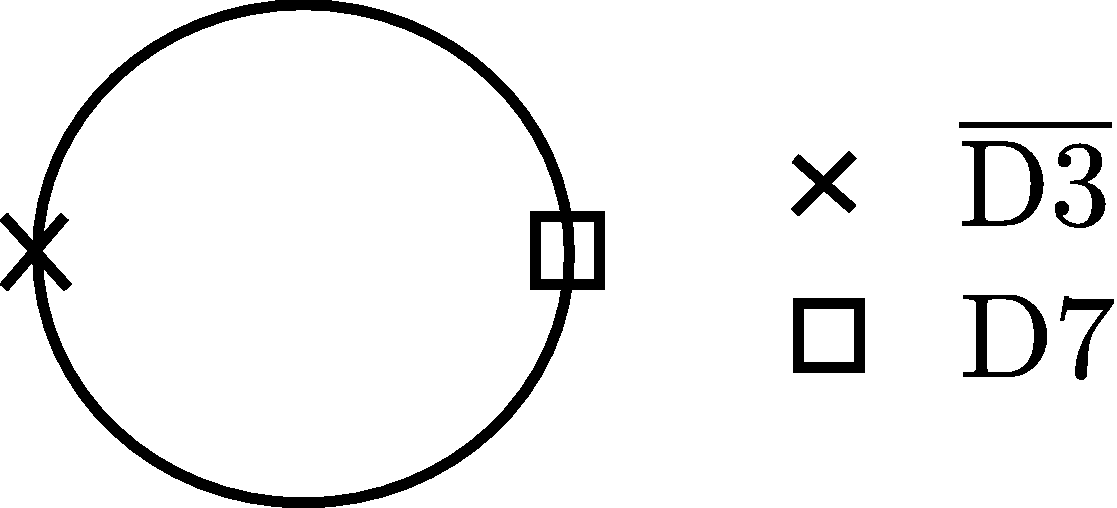}
\caption{\small  A stack of $\overline{\rm D3}$-branes on the opposite side of the magnetized D7-branes.\label{fig:d3d72}}
\end{center}
\end{figure}

The situations discussed in this section  
so far are unrealistic in that they are non-compact. However, they served us to illustrate how the expressions
that we found in sections \ref{sec2} and \ref{sec3} for the soft breaking terms can capture the contributions from distant localised sources that break supersymmetry. We would like now to consider a slightly more complete toy model on which the internal space is taken to be compact, in order to illustrate compactification effects on soft terms. Thus, we consider again the case of a stack of magnetized probe D7-branes and $N$ distant anti-D3-branes, but we now solve eq.~(\ref{pois}) in a 2-torus transverse to the D7-branes (and we smear the D3-brane charge along the remaining internal directions). Concretely, we take the magnetized D7-brane to be at the origin of coordinates and the anti-D3-branes exactly at the opposite point in the transverse $T^2$, as depicted in figure \ref{fig:d3d72}. In that case, linear terms automatically vanish due to the balance between the attraction forces on the two sides of the D7-branes. 
%%%%%%%%%%%
%%%%%%%%%%%

Following \cite{Baumann:2006th} we can express the solution to eq.~(\ref{pois}) in terms of the Green's function $G(x - y)$ on the transverse space to the D7-branes as
\begin{equation}
Z(z)=(2\pi\sigma)^2 N G(z_0^3-z^3)\label{zze}
\end{equation}
The Green function for a 2-torus with unit volume is
\begin{equation}
G(z)=\frac{1}{2\pi} \log \left|\frac{\vartheta_1(z;U)}{\vartheta'_1(0;U)}\right|^2-\frac{(\textrm{Im }z)^2}{\textrm{Im }U}
\label{green}
\end{equation}
where $U$ is the complex structure of the torus and $\vartheta_i$ are the usual Jacobi theta functions. Expanding  around $z_k=1/2$ reads
\begin{multline}
G(z)=-\frac{1}{\pi}\left(\log \pi + \log |\vartheta_4(0;\, 2U)|^2 \right)-\frac{\left|z-\frac12\right|^2}{2\, \textrm{Im } U}-\frac{\pi}{12}(\hat E_2(U)+\vartheta_3^4(0;\, U)+\vartheta_4^4(0;\, U))\left(z-\frac12\right)^2\\
-\frac{\pi}{12}(\hat E_2(U)+\vartheta_3^4(0;\, U)+\vartheta_4^4(0;\, U))^*\left(\bar z-\frac12\right)^2+\ldots
\end{multline}
where $\hat E_2$ is the modified second Eisenstein series defined as
\begin{equation}
\hat E_2(U) = E_2(U)-\frac{3}{\pi \textrm{Im }U}
\end{equation}
and
\begin{equation}
E_2(U)=1-24\sum_{n=1}^\infty\frac{nq^n}{1-q^n}
\end{equation}
and $q=e^{2\pi iU}$. From eqs.~(\ref{zze}) and (\ref{corr1}) then we get in this case the following set of scalar bilinears in the worldvolume of magnetized D7-branes
\begin{align}
m_{3\bar 3}^2 &= -\frac{g_s^{-1}\sigma^4 \pi^2 F_-^2 N}{4\, \textrm{Vol}(T^6)\, \textrm{Im}\,U}\ , \\
B_{33}&= \frac{g_s^{-1}\sigma^4 \pi^2 F_-^2 N}{4\,\textrm{Vol}(T^6)\,\textrm{Im}\, U}\left[1-\frac{3\,\textrm{Im}\, U}{\pi}(E_2(U)+\vartheta_3^4(0;\, U)+\vartheta_4^4(0;\, U))\right]\nonumber
\end{align}
where we have introduced the volume of the internal space back in these expressions. Note that soft-terms in particular now depend on the complex structure of the transverse 2-torus. It is also interesting to recall the interpretation of the different terms in these expressions from an effective field theory point of view. Indeed, scalar masses and the first contribution in the $B$-term are tree-level contributions similar to those computed in
the previous subsections. However, in the present compact case the $B$-term receives in addition loop threshold corrections that are exponential in the complex structure of the 2-torus. Those come from  integrating out heavy modes that propagate in the transverse $T^2$ and stretch between the D7-branes and the anti-D3-branes. 

This example, as it stands, is a toy model with no direct phenomenological interest. In particular, scalar masses are tachyonic, showing the instability of D7-branes under small fluctuations. The tachyonic instability in this setting was expected, since once the anti-D3-branes move a bit from their original position, the attractive forces on the two sides of the anti-D3-branes are no-longer balanced and they quickly decay towards the magnetized D7-branes. In this regard, it might be interesting to extend this example by including closed string fluxes and see whether it is possible to make it stable.

\section{Applications}
\label{applic}

In this section we study several applications of the previous results. 
In particular we  study the effect of fluxes on soft terms for fields on local F-theory SU(5) models
with enhanced $SO(12)$ and $E_6$ symmetries. We are particularly interested in the dependence on the hypercharge flux, 
required for the breaking from SU(5) to the SM gauge group. We also study the generation of flavor violating soft terms
in these settings and in models of D3-branes at singularities. We finally discuss the different sources of
soft masses for a Higgs system and explore the geometrical origin in string theory of a possible fine-tuning of its mass.

\subsection{Hypercharge dependence of soft terms in F-theory  SU(5) unification}

In previous sections we have considered corrections of open string magnetic fluxes to the soft terms of
7-brane fields, including also fields localized at intersections. To simplify the discussion, we considered 
a toy model with an underlying $U(3)$ gauge symmetry. The generalization to gauge symmetries of phenomenological interest 
is however straightforward. Indeed, we now apply the results of the above sections  
to SU(5) unification in the context of type IIB/F-theory compactifications. More precisely, in this subsection we concentrate in 
the case of local F-theory  SU(5) GUTs with gauge symmetry enlarged to $SO(12)$ at complex co-dimension 3 singularities.
This is the gauge symmetry enhancement that is relevant for the presence of local Yukawa couplings of the form
$\mathbf{\bar 5}\times \mathbf{10}\times \mathbf{\bar{5}_H}$ that lead to masses for charged leptons and D-type quarks. 
In particular, we identify the possible  (e.g. hypercharge) magnetic flux dependence of the scalar soft masses, as it might be of 
phenomenological relevance.

We consider the same local $SO(12)$ F-theory structure as introduced
 in  Ref.~\cite{yukawasSO(12)} (see also \cite{Camara:2011nj}). To avoid expressions with two many indices, throughout this section we use the alternative notation $x\equiv z_1$ and $y\equiv z_2$ to denote the two local complex coordinates in the 
4-cycle $S$. The vev for the transverse 7-brane position field is given by
\beq
\langle \Phi_{xy} \rangle \ =\ m^2(xQ_x\ +\ yQ_y)
\eeq
where $m$ is related to the intersection slope of the 7-branes, as we have already discussed, and it is generically of order the string scale. $Q_x$ and $Q_y$ are $SO(12)$ Cartan generators breaking the symmetry respectively down to $SU(6)\times U(1)$ 
and $SO(10)\times U(1)$. As in the $U(3)$ toy model of previous sections, we have matter curves $\Sigma_a,\,\Sigma_b$ and $\Sigma_c$ at $x=0$, $y=0$ and $x=y$ respectively.
Matter curves $\Sigma_a$ and $\Sigma_b$ host respectively $5$-plets and $10$-plets associated to quarks and leptons, while  $\Sigma_c$ hosts
$5_H$-plets that include the Higgs multiplets.  In order to get chiral matter and family replication  we must add magnetic fluxes to this setting.
We follow Ref.~\cite{yukawasSO(12)} and consider the above local system of matter curves subject to  approximately constant magnetic fields, 
that break the gauge symmetry down to that of the SM and give rise to chirality. The magnetic flux background comes in three pieces,
\begin{equation}
\langle F_2\rangle\, = \, \langle F_{(1)} \rangle \, + \, \langle F_{(2)} \rangle\, + \, \langle F_Y \rangle\\  
\end{equation}
with
\beqa
\langle F_{(1)} \rangle\,& = & \,i \left( M_x\, dx \wedge d\bar{x} + M_y\, dy \wedge d\bar{y}\right) \, Q_F \\
\langle F_{(2)} \rangle\, & =& \, i  \left( dx \wedge d\bar{y} +  dy \wedge d\bar{x}\right) \left(N_a Q_x + N_b Q_y  \right)\nonumber\\
\langle F_Y \rangle\, & =& \, i \left[ \left( dx \wedge d\bar{y} +  dy \wedge d\bar{x}\right) N_Y + \left(dy \wedge d\bar{y} - dx \wedge d\bar{x} \right)  \tilde{N}_Y \right]Q_Y \ .\nonumber
\eeqa
and $Q_F=-Q_x-Q_y$. The first piece leads to chirality (and matter replication) for fields that are localized in the matter curves $\Sigma_{a,b}$. 
The second piece leads to chirality for the Higgs fields, localized in the matter curve $\Sigma_c$. This is interesting in order to obtain
doublet-triplet splitting and a suppressed $\mu$-term. Finally, the third piece corresponds to a magnetic flux along the hypercharge direction, that 
breaks SU(5) down to the SM gauge group and  for the particular choice $N_Y=3(N_a-N_b)$ is consistent with doublet-triplet splitting. We refer to 
\cite{yukawasSO(12)} for further details on this configuration. 

Thus, putting all pieces together the complete flux may be written as
\begin{equation}
\langle F_2\rangle=i(dy\wedge d\bar{y}-dx\wedge d\bar{x})Q_P+i(dx\wedge d\bar{y}+dy\wedge d\bar{x})Q_S+i(dy\wedge d\bar{y}+dx\wedge d\bar{x})M_{xy} Q_F
\label{F}
\end{equation}
where
\beq
Q_P={\tilde M}Q_F+\tilde N_Y Q_Y\ \ ,\ \ 
Q_S=N_a Q_x+N_b Q_y +N_Y Q_Y
\label{QpQs}
\eeq
and
\begin{equation}
{\tilde M}=\frac{1}{2}(M_y-M_x)\ , \qquad M_{xy}=\frac{1}{2}(M_y+M_x)
\label{M}
\end{equation}
The local D-term SUSY condition would imply $M_{xy}=0$.

The local  internal wavefunctions of the fields must satisfy the system of differential equations that we have described in section \ref{sec3} and were solved in \cite{yukawasSO(12)} for this particular case. The zero modes for each sector are given  (in the {\it holomorphic gauge}) by  
\begin{equation}
\Psi_\rho=\begin{pmatrix} -\frac{i \lambda_x}{m^2}\\ \frac{i \lambda_y}{m^2}\\1 \end{pmatrix} \chi_\rho^i E_\rho\ , \quad  \quad \chi^i_\rho=e^{-q_\Phi (\lambda_x\bar{x}-\lambda_y\bar{y})   }f_i(\lambda_x y+\lambda_y x) \ \qquad \rho =a^\pm,b^\pm,c^\pm
\label{wavefunctions}
\end{equation}
where $E_\rho$ are the corresponding 
$SO(12)$ step  generators for fields in $\Sigma_{a,b,c}$. The values of the parameters $q_\Phi$ and $\lambda_{x,y}$ are given in table \ref{tablalambdas} for each of the fields in the curves $\Sigma_{a,b,c}$. 
The physical, normalizable,  wavefunctions in the {\it real gauge} can be obtained from
\begin{equation}
\chi_\rho^{\text{real}}\, =\, e^{i\Omega} \chi_\rho^{\text{hol}}
\label{realwave}
\end{equation}
where
\beq
\begin{array}{rl}
\Omega \  =\, \frac{i}{2}\left[ \left( |y|^2-|x|^2 \right) Q_P + \left ( x\bar y+ y\bar x \right )Q_S + M_{xy} \left( |y|^2+|x|^2 \right) Q_F \right] \ .
\end{array}
\label{omeguilla}
\eeq
\begin{table}[htb] 
\begin{center}
	\begin{tabular}{|c|c|c|c|c|}
\hline
		$\rho$ & $q_\Phi$ & $\lambda_x$ & $\lambda_y$ & SU(5) rep. \\
\hline
		$a_p^+$ & $-x$ & $\lambda_+$ & $-q_s\frac{\lambda_+}{\lambda_+-q_p}$ & $\mathbf{\bar 5}$ \\
		$a_p^-$ & $x$ & $\lambda_-$ & $q_s\frac{\lambda_-}{\lambda_-+q_p}$ & $\mathbf{5}$ \\
		$b_q^+$ & $y$ & $-q_s\frac{\lambda_+}{\lambda_++q_p}$ & $\lambda_+$ & $\mathbf{10}$ \\
		$b_q^-$ & $-y$ & $q_s\frac{\lambda_-}{\lambda_--q_p}$ & $\lambda_-$ & $\mathbf{\overline{10}}$ \\
		$c_r^+$ & $x-y$ & $\frac{q_s\lambda_+-m^4}{\lambda_++q_p-q_s}$ & $-\lambda_+-\frac{q_s\lambda_+-m^4}{\lambda_++q_p-q_s}$ & $\mathbf{\bar 5}$ \\
		$c_r^-$ & $-(x-y)$ & $-\frac{q_s\lambda_-+m^4}{\lambda_--q_p+q_s}$ & $-\lambda_++\frac{q_s\lambda_-+m^4}{\lambda_--q_p+q_s}$ & $\mathbf{5}$ \\
\hline
	\end{tabular}
	\end{center}
	\caption{Wavefunction parameters.}
	\label{tablalambdas}
	\end{table}
The constants $\lambda_\pm$ that appear in table \ref{tablalambdas} are defined as the lowest eigenvalue for the sectors $a^\pm_p,b^\pm_q$ and $c^\pm_r$ and satisfy the following cubic equations
\cite{yukawasSO(12)}
\begin{align}
(\lambda_i^a)^3-(m^4+(q_p^a)^2+(q_s^a)^2)\lambda_i^a+m^4q_p^a&\ =\ 0\\
\label{lambda_a}
%\end{equation}
(\lambda_i^b)^3-(m^4+(q_p^b)^2+(q_s^b)^2)\lambda_i^b-m^4q_p^b&\ =\ 0\nonumber\\
(\lambda_i^c)^3-(2m^4+(q_p^c)^2+(q_s^c)^2)\lambda_i^c+2m^4q_s^c&\ =\ 0\nonumber
\end{align}
where for simplicity we have assumed the D-term condition
$M_{xy}=0$. To first order in the fluxes the constants $\lambda_\pm$ are given by 
\begin{equation}
\lambda_\pm ^a= \mp m^2-\frac{1}{2} q_p^a +\dots  \ ;\ 
\lambda_\pm^b=\mp m^2+\frac{1}{2} q_p^b +\dots \ \ ;\ 
\lambda_\pm^c=\mp \sqrt{2} m^2-\frac{1}{2} q_s^c +\dots
\label{lam}
\end{equation}
In order to compute the physical soft masses we need to normalize these local wave functions.
 It is useful to factorize the normalization of the vector in eq.~(\ref{wavefunctions}) from the normalization of the scalar function $\chi_\rho^i$ so that
\begin{equation}
\langle \Psi_\rho^i|\Psi_\rho^j\rangle=m^2_*\int_S \textrm{Tr}(\Psi_\rho^i\Psi_\rho^j)\, d\textrm{vol}_S=2m_*^2||\vec{v}_\rho||^2\int_S\chi_\rho^i(\chi_\rho^j)^*\, d\textrm{vol}_S=\delta_{ij}
\end{equation}
where
\begin{equation}
\vec{v}_\rho=\left(\begin{array}{c} -\frac{i \lambda_x}{m^2}\\ \frac{i \lambda_y}{m^2}\\1 \end{array}\right)_\rho
\label{v}
\end{equation}
and hence $||\vec{v}_\rho||^2=1+\frac{\lambda_x^2}{m^4}+\frac{\lambda_y^2}{m^4}$. By using the definition of $\lambda_x$ and $\lambda_y$ in table \ref{tablalambdas} and eq.~(\ref{lam}), we get
\begin{align}
||\vec{v}_{a^\pm}||^{-2}&\ \simeq\ \frac{1}{2}\left(1\mp\frac{q_p^{a^\pm}}{2m^2}+\dots\right)\\
||\vec{v}_{b^\pm}||^{-2}&\ \simeq\ \frac{1}{2}\left(1\pm\frac{q_p^{b^\pm}}{2m^2}+\dots\right)\nonumber\\
||\vec{v}_{c^\pm}||^{-2}&\ \simeq\ \frac{1}{2}\left(1\mp\frac{q_s^{c^\pm}}{2\sqrt{2}m^2}+\dots\right)\nonumber
\end{align}
Having the normalized internal wavefunctions, we can compute the soft masses for these fields by making use of the results of previous sections. For simplicity we only consider soft scalar masses in the presence
of an ISD (0,3)-form closed string background. 
We expand the non-Abelian DBI+CS action of the 7-brane in powers of the transverse adjoint $SO(12)$ scalar $\Phi$ in the presence of a non-trivial $G_{(0,3)}$ flux, as we did in section \ref{sec2}, obtaining
\begin{equation}
\mathcal{L}_{\rm 8d}\ =\ \textrm{Tr}\left(D_a\Phi D^a\bar \Phi-\frac{1}{4}F_{ab}F^{ab}-\frac{g_s}{2}|G|^2|\Phi|^2+\dots\right)
\label{lagrangiano}
\end{equation}
The local flux density induces a 8d mass term for the transverse scalar $\Phi$. Upon dimensional reduction in the presence of non-trivial backgrounds $\langle \Phi \rangle$ and $\langle F_2\rangle$ this leads to 4d soft masses for the fields localized at the matter curves. The scalar $\Phi$ transforms in the adjoint representation of $SO(12)$ and can thus be decomposed as 
\begin{equation}
\textrm{Tr}(|\Phi|^2)\ =\ |\Phi_{a_p^+}|^2+|\Phi_{a_p^-}|^2+|\Phi_{b_q^+}|^2+|\Phi_{b_q^-}|^2+|\Phi_{c_r^+}|^2+|\Phi_{c_r^-}|^2 \ +\ \dots
\end{equation}
where $\Phi_\rho$ corresponds to the third component of eq.~(\ref{wavefunctions}), namely the internal wavefunction of the transverse scalar that solves the equation of motion in each sector.
The induced 4d soft masses for the matter fields living in the sector $a^+$ therefore read
\begin{equation}
(m_{ij}^{a_p^+})^2\ =\ \frac{g_s}{2\,\textrm{Vol}(S)}\int_S d\textrm{vol}_S\, |G|^2 |\Phi_{a_p^+}|^2\ =\ \frac{g_s}{2\textrm{Vol}(S)||\vec{v}_{a_p}||^2}\int_S d\textrm{vol}_S\, |G|^2 \chi^i_{a_p^+}(\chi^j_{a_p^+})^*
\end{equation}
Using the definition of $\lambda_x$ and $\lambda_y$ in table \ref{tablalambdas} we obtain
\begin{equation}
(m^{ij}_{a_p^+})^2\ =\ \frac{g_s}{4\textrm{Vol}(S)}\int_S d\textrm{vol}_S\, |G|^2 \left(1-\frac{q_p^{a^+}}{2m^2}\right)\chi^i_{a_p^+}(\chi^j_{a_p^+})^* \ .
\label{masaconflux}
\end{equation}
Note that in the presence of magnetic fluxes, scalar kinetic terms get also flux corrections. However, those start at quadratic order
in the magnetic flux, so that they only give rise to subleading corrections to this expression for the soft masses. When the flux $G$ is constant over the 4-fold we recover the results of eq.~(\ref{softbif_flux}), extended to the SU(5) GUT case here considered.
Analogously for the sector $a^-$ we get,
\begin{equation}
(m^{ij}_{a_p^-})^2\ =\ \frac{g_s}{4\textrm{Vol}(S)}\int_S d\textrm{vol}_S\, |G|^2 \left(1+\frac{q_p^{a^-}}{2m^2}\right)\chi^i_{a_p^-}(\chi^j_{a_p^-})^*
\end{equation}
Taking into account that the zero mode in the sector $a_p^+$ ($a_p^-$) is normalizable only if $q_p^{a^+}>0$ ($q_p^{a^-}<0$), we can rewrite these expressions as
\begin{equation}
(m^{ij}_{a_p})^2\ =\ \frac{g_s}{4\textrm{Vol}(S)}\int_S d\textrm{vol}_S\, |G|^2 \left(1-\frac{|q_p^{a_p}|}{2m^2}\right)\chi^i_{a_p}(\chi^j_{a_p})^*
\label{ma}
\end{equation}
assuming  that only one of the two modes is actually present in the massless spectrum.
The result for the sector $b_q^\pm$ reads
\begin{equation}
(m^{ij}_{b_q})^2\ =\ \frac{g_s}{4\textrm{Vol}(S)}\int_S d\textrm{vol}_S\, |G|^2 \left(1-\frac{|q_p^{b_q}|}{2m^2}\right)\chi^i_{b_q}(\chi^j_{b_q})^*
\label{mb}
\end{equation}
where we have used that the zero mode in the sector $b_q^+$ ($b_q^-$) is normalizable only if $q_p^{b^+}<0$ ($q_p^{b^-}>0$).
Finally for the sector $c^\pm_r$ we obtain
\begin{equation}
(m^{ij}_{c_r})^2\ =\ \frac{g_s}{4\textrm{Vol}(S)}\int_S d\textrm{vol}_S\, |G|^2 \left(1-\frac{|q_s^{c_r}|}{2\sqrt{2}m^2}\right)\chi^i_{c_r}(\chi^j_{c_r})^* \ .
\label{mc}
\end{equation}
Thus we observe that only fields with different absolute value of hypercharge have different soft masses at the unification scale.
In particular, both Higgs fields $H_u$ and $H_d$ have equal soft masses as long as they 
feel the same amount of hypercharge flux. 

Making use of the SO(12) chiral spectrum summarized in table \ref{spectrum}, we can express the above results in a more compact form
\begin{equation}
(m^{ij})^2\ =\ \frac{g_s}{4\textrm{Vol}(S)}\int_S d\textrm{vol}_S\, |G|^2 \left(1-\frac{1}  {2m^2} \left|  \eta \tilde M - q_Y   {\tilde N}_Y\right| 
\right)\chi^i_{c_r}(\chi^j_{c_r})^*
\end{equation}
where $\eta =+1,-1,0$ respectively for matter fields in the $\mathbf{\overline 5}$, $\mathbf{10}$ and $\mathbf{{\overline 5}_H}$ multiplets, and $q_Y$ is the usual SM hypercharge
(i.e.$Y(E_R)=1$).
Moreover, for the case of the Higgs doublets the
replacement ${\tilde N}_Y\rightarrow \frac{5}{3\sqrt{2}}N_Y$ should be also made in this expression.

If the fluxes are approximately constant over the 4-cycle $S$ we can perform the integral over the normalized wave functions, getting
\begin{equation}
(m^{ij})^2 \ =\  \frac {M^2\delta_{ij}}{2}\   \left(1-\frac{1}  {2m^2} \left|  \eta \tilde M - q_Y   {\tilde N}_Y\right| \right) \ 
\label{escalarfinal}
\end{equation}
where we have expressed $G$ in terms of the gaugino mass $M$, see eq.~(\ref{softgen7}).

\begin{table}[htb] 
\begin{center}
	\begin{tabular}{|c|c|c|c|c|}
\hline
Sector & Chiral mult. & $SU(3)\times SU(2)$ & $q_Y$ & $q_p$ \\
\hline
$a_1^+$ & $D_R$ & $3\mathbf{(\bar{3},1)}$ & $\frac13$ & $\tilde M-\frac13 \tilde N_Y$  \\
$a_2^+$ & $L$ & $3\mathbf{(1,2)}$ & $-\frac12$ & $\tilde M+\frac12 \tilde N_Y$ \\
$b_1^+$ & $U_R$ & $3\mathbf{(\bar{3},1)}$ & $-\frac23$ & $-\tilde M+\frac23 \tilde N_Y$  \\
$b_2^+$ & $Q_L$ & $3\mathbf{(3,2)}$ & $\frac16$ & $-\tilde M-\frac16 \tilde N_Y$  \\
$b_3^+$ & $E_R$ & $3\mathbf{(1,1)}$ & $1$ & $-\tilde M- \tilde N_Y$  \\
\hline
Sector & Chiral mult. & $SU(3)\times SU(2)$ & $q_Y$ & $q_s$ \\
\hline
$c_1^+$ & $D_d$ & $\mathbf{(\bar{3},1)}$ & $\frac13$ &  0 \\
$c_2^+$ & $H_d$ & $\mathbf{(1,2)}$ & $-\frac12$ & $\frac56 N_Y$\\
\hline
	\end{tabular}
	\end{center}
	\caption{SO(12) chiral spectrum.}
	\label{spectrum}
\end{table}

The possible phenomenological relevance of the magnetic flux contributions depend on the size of the fluxes. A naive estimate shows that
these corrections are potentially important. Indeed, flux quantization imply
$\int_{\Sigma_2} \langle F_2\rangle \simeq 2\pi$, so that we expect $\tilde M\simeq N_Y\simeq {\tilde N}_Y\simeq (2\pi)/\textrm{Vol}_S^{1/2}$.
On the other hand we know that  $\alpha_G\simeq 1/(M_{s}^4\textrm{Vol}_S)\simeq 1/24$, so that flux contributions are expected to be of order
$\sim 0.2\, M_{s}^2$.  

We can extract some additional information on the structure and size of the fluxes from other phenomenological
considerations. Indeed, 
magnetic  fluxes have also been shown to play an important role in the computation of Yukawa
couplings in local F-theory models. 
In \cite{yukawasSO(12)} it was found an expression relating ratios of second and third generation 
quark/lepton masses to local fluxes in an F-theory  $SO(12)$ setting,
\beq
\frac{m_\mu/m_\tau}{m_s/m_b}=\left ( \frac{({\tilde M}+\frac{1}{2}\tilde N_Y)({\tilde M}+\tilde N_Y)}{({\tilde M}-\frac{1}{3}\tilde N_Y)({\tilde M}+\frac{1}{6}\tilde N_Y)} \right )^{1/2}
\label{ratioratio}
\eeq
This expression is independent  of the hierarchical (non-perturbative) origin of Yukawa couplings and is based on the fact that holomorphic Yukawas must respect the SU(5) gauge symmetry, even after flux-breaking to the SM gauge group. 
The difference in Yukawas of charged leptons $\tau,\mu$ and  $b,s$ quarks originates exclusively from the different (hypercharge-dependent)
fluxes present at the matter curves, which appear through wavefunction normalisation.
Eq.~(\ref{ratioratio}) applies at the unification scale. Including the RG running and uncertainties one finds agreement with low-energy data 
for $\frac{m_\mu/m_\tau}{m_s/m_b}=3.3\pm 1$ at the GUT scale, therefore implying ${\tilde N}_Y/{\tilde M}=1.2-2.4$ \cite{yukawasSO(12)}.

In order to see the implications of this relation on the structure of soft terms, let us demand without loss of generality that the local zero modes arise from the sectors $a^+$, $b^+$ and $c^+$. In terms of the local flux densities, that requires
\beq
-\tilde M-q_Y\tilde N_Y\ <\ 0\ <\ \tilde M-q_Y \tilde N_Y\quad \text{and} \quad N_Y\ >\ 0
\label{fluxes_norm}
\eeq
for every possible value of the hypercharge $q_Y$. Eq.~(\ref{escalarfinal}) then implies a hierarchy of soft scalar masses for each generation
\beq
m_E^2\ <\ m_L^2 \ <\ m_Q^2 \ < \ m_D^2 \ < \ m_U^2  \ ,\label{softhier}
\eeq
at the unification scale. This non-degenerate structure is different from those induced by the RG running or D-terms in
the MSSM, and may have interesting phenomenological consequences.  
Moreover, the average scalar squared mass for fields in the 5-plet and 10-plet of
each generation, $m_0^2$,  
 is independent of the hypercharge flux, 
\beq
m_0^2\ =\ \frac {1}{5}\left(3m_D^2+2m_L^2\right)\ =\ \frac {1}{10}\left(6m_Q^2+3m_U^2+m_E^2\right)\ =\ \frac {M^2}{2}\left(1-\frac {\tilde M}{2} \right) \ 
\eeq
where fluxes are written in units of $m^2$.
Thus, we can  write soft masses for the 5-plet, the 10-plet and the Higgs $H_d$  respectively as
\beqa
m_{\mathbf{\bar 5}}^2 & = & m_0^2 \ + \  \frac {q_Y}{4}{\tilde N}_Y M^2 \\
m_{\mathbf{10}}^2 & = & m_0^2 \ - \  \frac {q_Y}{4}{\tilde N}_Y M^2 \nonumber\\
m_{H_d}^2 & = &  m_0^2 \ + \ \frac {M^2}{2} \left( \frac {\tilde M}{2}\ -\ \frac {5}{6\sqrt{2}} |q_Y N_Y|\right)\nonumber
\eeqa
These equations neatly show the linear dependence of
the soft masses on the hypercharge fluxes.  

In \cite{yukawasSO(12)} it was also  shown that certain choices of the magnetic fluxes 
lead to $h_b/h_\tau$ Yukawa
ratios that are consistent with the experimentally observed values, for example,  ${\tilde M}\simeq 0.3$,\, ${\tilde N}_Y\simeq 0.4$ and $N_Y\simeq 0.6$
in units of $m^2$. 
With the above expressions, such values lead to the following pattern of soft masses at the unification scale
\beq
m^2(Q,U,D,L,E,H_d)\ =\ \frac {M^2}{2}(0.82,\, 0.98,\, 0.92,\, 0.75,\, 0.65,\, 0.82) .
\eeq
Thus, we observe that squark squared masses and slepton and Higgs squared masses become respectively $10-20$\% and $25-35$\% smaller  than the hypercharge-uncorrected value. Note however that the precise results  depend on the particular values for the fluxes,  and there are other flux choices also leading to Yukawa 
couplings consistent with experimental constraints. It would we interesting to do a full scan over flux parameters giving consistent  Yukawa results to
see their impact on the obtained soft masses.

It is interesting to note how  in this scheme the fermion mass spectrum 
gives information on the structure of sfermion masses, whereas in the standard context of supersymmetric field theory these would be independent quantities. We have not studied in detail the phenomenology of a MSSM model subject to a hierarchy of soft scalar masses of the form (\ref{softhier}), but we note that a particularly interesting feature is that in such a scheme the stau has the smallest soft mass (after taking into account the
running of the gauge and Yukawa couplings) and may easily be the next-to-lightest SUSY particle (NLSP). This in particular might be relevant for having the appropriate amount of neutralino dark matter through stau-neutralino coannihilation. It would be interesting to perform a RGE analysis and  study the generation of EW radiative
symmetry breaking in a model with this structure, including this new  hypercharge degree of freedom. This would correspond to an extension of the 
work in \cite{Aparicio:2008wh,Aparicio:2012iw,Aparicio:2012vk}.

We now turn to describe the effect of magnetic fluxes on the trilinear couplings, in the context of this local $SO(12)$ F-theory setting.
As we discussed in section \ref{magnetico}, the leading effect of the magnetization results from the modification of the $\Phi-A$ mixing. Since by supersymmetry this modification is the same for the scalar and auxiliary fields, we can factorize the correction induced by the fluxes in the scalar potential (see eq.~(\ref{potbif_flux})). Consequently, both scalar masses and trilinear couplings receive the same correction, that we have already derived in eqs.~(\ref{ma})-(\ref{mc}). After summing over the three matter curves that are involved in the coupling, the soft trilinear coupling takes the form
\beq
A=-\frac{M}{2}\left(3-\frac{|q_p^{a_p}|}{2m^2}-\frac{|q_p^{b_q}|}{2m^2}-\frac{|q_s^{c_r}|}{2\sqrt{2}m^2}\right)
\label{tril_SO(12)}
\eeq
where the parameters $q_p$ and $q_s$ are given in table \ref{spectrum} for the relevant sectors of the theory (see also table 2 in \cite{yukawasSO(12)} for the complete spectrum, including the non-normalizable modes). 
Requiring the flux densities to satisfy eq.~(\ref{fluxes_norm}) is equivalent to imposing $q_p^{a^+}>0$, $q_p^{b^+}<0$ and $q_s^{c^+}>0$. Thus, making use of table \ref{spectrum} in eq.~(\ref{tril_SO(12)}), we can recast the down- and lepton-type trilinear couplings as
\begin{align}
A_d&\ =\ -\frac{M}{2}\left(3-\tilde M+\frac{\tilde N_Y}{12}-\frac{5 N_Y}{12\sqrt{2}}\right)\\
A_l&\ =\ -\frac{M}{2}\left(3-\tilde M-\frac{3\tilde N_Y}{4}-\frac{5 N_Y}{12\sqrt{2}}\right)\nonumber
\end{align}
where fluxes have been expressed in units of $m^2$.

A more complicate issue is that of the induced $B$-terms. Indeed, in these local F-theory $SO(12)$ and $E_6$ settings the Higgs fields
$H_u$ and $H_d$ are chiral and live on different matter curves. A $\mu$-term would have to be generated by some e.g. non-perturbative effect.
The final physical $\mu$-term is related to the integral of the two wavefunctions and is only non-vanishing 
if the $H_u$ and $H_d$ matter curves overlap.  It would be interesting to study a local configuration in which,  in addition,  the two Higgs
matter curves intersect at a point with $SU(7)$ enhancement, leading to an effective $\mu$-term from the coupling to a singlet,
as suggested e.g. in \cite{Beasley:2008kw}.

\subsection{Soft terms at $E_6$ enhancement points in F-theory SU(5) unification}
\label{sec52}

In the above subsection we have considered F-theory SU(5) unification with an underlying $SO(12)$ gauge symmetry enhancement at the point where the internal wavefunctions localize. Such configuration is incomplete in that up-type $\mathbf{10}\times \mathbf{10}\times \mathbf{5_H}$ Yukawa couplings are not generated, as those require an $E_6$ gauge symmetry enhancement \cite{Beasley:2008dc}. In order to reproduce the desired rank-one structure of Yukawas, one must take into 
account non-trivial 7-brane monodromies, which may be conveniently described in terms of T-brane configurations \cite{Cecotti:2010bp}. From the point of view of the effective 8d theory, this amounts to considering non-Abelian profiles for the transverse scalar \cite{Hayashi:2009ge}. This approach was in particular used in \cite{Font:2013ida} to perform the explicit computation of up-type Yukawa couplings in local F-theory SU(5) GUTs. 

In this subsection we address the computation of soft masses for fields localized at a $\mathbf{10}$ matter curve near a point of $E_6$ gauge symmetry enhancement. The novel feature with respect to the $SO(12)$ case discussed above is that the profile of the transverse scalar $\langle \Phi\rangle$ does not necessarily commute with other elements of the background and, in particular, $[\langle \Phi\rangle,\langle \bar \Phi\rangle ]\neq 0$. Thus, in order to satisfy the D-term condition,
\begin{equation}
\omega\wedge F\, +\, \frac12 [\Phi,\bar \Phi]\ =\ 0
\end{equation} 
with $\omega$ the K\"ahler form, we must turn on a non-primitive background flux $\langle F_{\rm NP}\rangle$. This non-trivial background can be parametrized in terms of a real function $f$ such that
\begin{equation}
[\langle \Phi\rangle,\langle \bar \Phi\rangle]\ =\ \hat m^2(e^{2f}-\hat{m}^2|x|^2e^{-2f})P\ , \qquad \langle F_{\rm NP}\rangle=-i\partial\bar\partial f P
\label{background}
\end{equation}
where $P$ is some combination of the Cartan generators of $E_6$ and $x$ a local coordinate of the 4-cycle S. At short distances the function $f$ can be expanded as
\begin{equation}
f(r)\ =\ \text{log}\, c\, +\, c^2\hat{m}^2|x|^2\, +\, \hat{m}^4|x|^4\left( \frac{c^4}{2}-\frac{1}{4c^2}\right)\, +\, \dots
\end{equation}
Hence, we can parametrize the solution in terms of a real dimensionless constant $c$ that encodes the details of the global embedding of the 7-brane local model.

Near the Yukawa point we can approximate $f(r)\, =\, \text{log}\, c+c^2\hat{m}^2|x|^2+\dots$ such that the flux $F_{\rm NP}$ is constant and we can compute analytically the wavefunctions around that point.
The two $\mathbf{10}$ matter curves, although coming from the same smooth curve $\Sigma_{\mathbf{10}}$, seem locally different. They have a different local zero mode associated to each curve, given by
\begin{equation}
\psi_{\mathbf{10}^+}^j=\frac{1}{||\vec v_{\mathbf{10}}||}\begin{pmatrix}\frac{i\lambda_{\mathbf{10}}}{\hat{m}^2}\\ -\frac{i\lambda_{\mathbf{10}}\xi_{\mathbf{10}}}{\hat{m}^2}\\0\end{pmatrix}e^{f/2}\chi_{\mathbf{10}}^j\ ,\qquad
\psi_{\mathbf{10}^-}^j=\frac{1}{||\vec v_{\mathbf{10}}||}\begin{pmatrix}0\\0\\1\end{pmatrix}e^{-f/2}\chi_{\mathbf{10}}^j
\label{functions_10}
\end{equation}
where $||\vec v_{\mathbf{10}}||$ is the normalization factor of the wavefunction across the entire $\Sigma_{\mathbf{10}}$ matter curve and $\lambda_{\mathbf{10}}$ is the negative root of the equation
\begin{equation}
\hat{m}^4(\lambda_{\mathbf{10}}-q_p)\, +\, \lambda_{\mathbf{10}}c^2\left(c^2\hat{m}^2(q_p-\lambda_{\mathbf{10}})-\lambda_{\mathbf{10}}^2+q_p^2+q_s^2\right)\ =\ 0
\label{lambda_10}
\end{equation}
and $\xi_{\mathbf{10}}=-q_s/(\lambda_{\mathbf{10}}-q_p)$. The scalar wavefunction $\chi_{\mathbf{10}}^j$ takes the same form than in the $SO(12)$ model above. Indeed the only difference with respect to the above local model resides in the value of $\lambda_{\mathbf{10}}$ due to the presence of the parameter $c$ in eq.~(\ref{lambda_10}). Solving that equation for small magnetic fluxes $q_p,q_s$ we find that to first order in the fluxes $\lambda_{\mathbf{10}}$ is given by
\begin{multline}
\lambda_{\mathbf{10}}\ =\ -\hat{m}^2g_1(c)\, -\, g_2(c)\frac{q_p}{2}\, +\, \dots\ =\ -\frac{\hat{m}^2}{2c}\left(c^3+\sqrt{4+c^6}\right)\, -\, \frac12\left(1+\frac{c^3}{\sqrt{4+c^6}}\right)q_p\, +\, \ldots\\
=\ -\hat{m}^2\left(\frac1c +\frac{c^2}{2}+\dots\right)\, -\, \left(1+\frac{c^3}{2}+\dots\right)\frac{q_p}{2}\, +\, \dots
\label{lambda_sol}
\end{multline}
where in the last line we have expanded for small $c$. Thus, to linear order $\lambda_{\mathbf{10}}\, \simeq\, -\frac{\hat{m}^2}{c}-\frac{q_p}{2}$ and the wavefunctions (\ref{functions_10}) are well approximated by
\begin{equation}
\psi_{\mathbf{10}^+}^j\ \simeq\ \frac{1}{||\vec v_{\mathbf{10}}||}\begin{pmatrix}-i\left(1+\frac{q_pc}{2\hat{m}^2}\right)\\ i\frac{q_s c}{\hat{m}^2}\\0\end{pmatrix}\frac{1}{\sqrt{c}}\, \chi_{\mathbf{10}}^j\ , \qquad
\psi_{\mathbf{10}^-}^j\ \simeq\ \frac{1}{||\vec v_{\mathbf{10}}||}\begin{pmatrix}0\\0\\1\end{pmatrix}\frac{1}{\sqrt{c}}\, \chi_{\mathbf{10}}^j
\label{functions2}
\end{equation}
To first order in the fluxes, the normalization factor reads
\begin{equation}
||\vec{v}_{10}||^{-2}=\frac12\left(1-\frac{q_p c}{2\hat{m}^2}+\dots\right)
\end{equation}

Soft masses for fields living in the $\mathbf{10}$ matter curve are (for constant fluxes) given by
\begin{equation}
m^2_{\mathbf{10}}\ =\ |M|^2\int_S d\textrm{vol}_S\, |\Phi_{\mathbf{10}}|^2 \ = \ |M|^2\int_S d\textrm{vol}_S\, \left(|\Phi_{\mathbf{10}^+}|^2+|\Phi_{\mathbf{10}^-}|^2\right)
\end{equation} 
where $M$ is the gaugino mass and $\Phi_{\mathbf{10}^\pm}$ is the lower entry of the vectors (\ref{functions2}), including the normalization factor.  Therefore, we obtain the following result for the soft masses
\begin{equation}
m^2_{\mathbf{10}}\ =\ |M|^2\frac{1}{1+c^2 g_1^2}\left(1-\frac{c^2g_1g_2}{1+c^2g_1^2}\frac{q_p}{\hat{m}^2}+\ldots\right) \ \simeq \ \frac{|M|^2}{2}\left(1-\frac{q_p c}{2\hat{m}^2}\right).
\end{equation}
where we have kept only the leading contribution of the primitive fluxes and taken the limit for small $c$ in the last step. Note that the magnetic flux correction depends now on the parameter $c$, that parametrizes the non-primitive flux. Moreover, note that the limit $c\rightarrow 0$ does not correspond to the result that we obtained in the previous section for the curve $\Sigma_{b}$ in the $SO(12)$ case. This is in fact something expected. Indeed, looking at the commutator in eq.~(\ref{background})
\begin{equation}
[\langle \Phi\rangle ,\langle \bar \Phi\rangle]\ =\ \hat{m}^2\left(c^2-\hat{m}^2|x|^2\frac{1}{c^2}\right)\, +\, \dots
\end{equation}
we observe that there is not a continuous way to make $[\langle \Phi\rangle ,\langle \bar \Phi\rangle ]\rightarrow 0$ by turning off $c$, as it diverges for $c\rightarrow 0$. Hence, this T-brane configuration gives rise to a new qualitative behaviour that is encoded in the non-trivial dependence of the soft masses  on $c$. From a phenomenological point of view though this parameter can be seen just as a redefinition of the flux density that modifies the soft masses. In particular, the hierarchy between the masses for the fields living in the $\mathbf{\bar 5}$ curve or the $\mathbf{10}$ curve depends on the value of $c$. Interestingly, extending the solution for $f(r)$ to all the real axis and requiring absence of poles leads to $c\sim 0.73$. If this is the case, there is only a small suppression on the flux correction and the scalars living in the $\mathbf{10}$ matter curve are only slightly heavier than those in the $\mathbf{\bar 5}$ curve.

Let us conclude with the soft mass corresponding to $H_u$. In this setup the Higgs sector is chiral and both Higgses $H_d$ and $H_u$ live in different matter curves, $\mathbf{\bar 5}$ and $\mathbf{5}$ respectively. In the previous section we studied the soft mass for $H_d$ near a point of SO(12) gauge symmetry enhancement. However, in order to allow for an up-type Yukawa coupling we have seen that we need to go to a point of $E_6$  gauge symmetry enhancement. Fortunately, unlike the $\mathbf{10}$ curve, the $\mathbf{5}$ curve does not feel the presence of the non-primitive flux $\langle F_{NP}\rangle$ so the wavefunctions are the same than those obtained in the previous section for the $\mathbf{\bar 5}$ curve but with the opposite chirality. We can borrow then the result for the soft mass obtaining
\begin{equation}
m^2_{H_u}\ =\ \frac{|M|^2}{2} \left(1-\frac{|q_s|}{2\sqrt{2}m^2}\right)=\ \frac{|M|^2}{2} \left(1-\frac{5|N_Y|}{12\sqrt{2}m^2}\right)\ .
\label{soft_up}
\end{equation}
We can see that the soft mass does not depend on the hypercharge sign, so in this setup the soft Higgs masses are universal whenever they feel the same amount of hypercharge flux density $N_Y$. This is a good approximation since both curves $\mathbf{5}$ and $\mathbf{\bar 5}$ can not be very far away from each other in order to reproduce the known flavor structure and CKM matrix of the SM. It would be interesting though to apply these results to a more realistic F-theory compactification with $E_7$ or $E_8$ enhancement in which we could consider both Yukawa points and all the matter curves simultaneously.

\subsection{Flavor non-universalities}
\label{nonuniv}

The soft terms found in the context of type IIB/F-theory SU(5) GUTs in the previous two subsections are not universal. However, for constant 3-form flux $G_3$ over the 4-cycle $S$, they are flavor-independent. 
On the other hand, as remarked in \cite{Camara:2013fta}, if $G_3$ is not constant, departures from generation independence may arise. 
Indeed, consider  eq.~(\ref{masaconflux}) and let us discuss the case of sfermion masses. To simplify the 
discussion we set $q_s=0$, since this is only required to be non-vanishing for having doublet-triplet splitting of the Higgss multiplet, but it plays no
role in the sfermion sector. For the wavefunction of $\mathbf{\bar 5}$ matter fields then we have $\lambda_y=0$ and
$\lambda_x=\lambda_+\simeq -m^2-\frac {1}{2}q_p^a$, and zero modes read
\begin{equation}
\Psi_{a_i^+}\ =\ \begin{pmatrix} -\frac{i \lambda_x}{m^2}\\  0 \\1 \end{pmatrix}\, \chi^{\rm real}_{a_i^+}
\ =	 \
\begin{pmatrix}  i +\frac {iq_p^a}{2m^2} \\  0 \\1 \end{pmatrix}\, \chi^{\rm real}_{a_i^+}
\end{equation}
where $i=1,2,3$ labels the three SM generations and
\beq
\chi^{\rm real}_{a^+_i}\ =\ \gamma_{a^+}^i\ m^{4-i}y^{3-i}\ e^{-m^2|x|^2} e^{-\frac {q_p}{2}|y|^2} \ ,
\eeq
The normalisation factors $\gamma_{a^+}^i$ are given by
\beq
||\gamma_{a^+}^i||^2 \ =\ 
\frac {1}{\pi^2 (3-i) ! } \frac {m^4}{m^4+\lambda_+^2}      \left( \frac {q_p}{m^2}\right)^{4-i}  \ .
\eeq
where we have extended the domain of integration to $\mathbb{C}^2$. This is indeed a good approximation in the limit on which the volume of the 4-cycle $S$ is large. As we saw in the previous subsection, scalar soft masses for this sector are given by eq.~(\ref{ma}). Allowing for a non-constant flux $G_{(0,3)}$ we can make a local expansion 
\beq
|G|^2\, =\, |G_0|^2 \left(1\, +\,  G_{y}^*\,y+G_{y}\,\bar y  \, +\,   G_{y\bar y}  \,  |y|^2 \, +\, \ldots \right)
\eeq
where $G_0$ and $G_{y}$  are complex constants and $G_{y\bar y}$ is real.
We are only displaying terms of the expansion that contribute to the flavor dependence of
the two lightest families. In particular, we omit the expansion in $x$ since it has no consequences for the  flavor dependence of soft-masses in the $a^\pm$ sector. 
Extending the domain of integration of eq.~(\ref{ma}) to $\mathbb{C}^2$, we therefore get 
\begin{multline}
m_{ij}^2 = \frac{g_s \gamma_i\gamma_j}{4}\int_{\mathbb{C}^2}  d^2x\, d^2y  \, \left[ |\hat G_0|^2 \left(1\, +\,  G_{y}^*y+G_{y}\bar y  \, +\,   G_{y\bar y} |y|^2 \, +\, \ldots \right)\right.\\
\left.y^{3-i}\bar{y}^{3-j}\, e^{-2m^2|x|^2-|q_p| |y|^2}\right]
\label{m2}
\end{multline}
where we have defined
\begin{equation}
|\hat G_0|^2\ \equiv\ |G_0|^2\left(1-\left|\frac{q_p}{2m^2}\right|\right)  \ .
\end{equation}

Sizeable flavor non-diagonal transitions  $\delta_{ij}^{RR}$ or  $\delta_{ij}^{LL}$ that do not mix the left and right sectors generically arise from soft mass terms. 
In particular, the leading contributions to FCNC transitions come from the off-diagonal mass terms. 
For  $\Delta F = 1$ soft masses we have
\begin{multline}
m_{ij}^2\ =\ \frac{g_s \gamma_i\gamma_j}{4}\int_0^\infty 2\pi x\, dx\int_0^\infty 2\pi y\, dy \, |\hat G_0|^2 \left(G_y^*y  + G_y\bar y\right) e^{-2m^2|x|^2-|q_p||y|^2}y^{3-i}\bar{y}^{3-j}\\
=\frac{g_sk}{4}\frac{|\hat G_0|^2G_y}{\sqrt{|q_p|}}\ , \qquad \textrm{where} \qquad k\equiv\left\{\begin{tabular}{c}$\sqrt{2}\quad$ for $\quad i=1,\ j=2$\\
$1\quad$ for $\quad i=2,\ j=3$
\end{tabular}\right.
\label{mij}
\end{multline}
The off-diagonal $\Delta F = 2$ mass term $m_{13}^2$  is proportional to higher derivatives of the 3-form flux and is therefore  subleading with respect to $m_{12}^2$ and $m_{23}^2$.
The relevant quantity in the generation of FCNC effects in the Kaon system is 
\begin{equation}
\delta_{12}^{\rm d}\ =\ \frac{ m_{12}^2}{ m_{ \tilde q}^2}\ =\ \frac{{\sqrt{2}}G_y} {\sqrt{|q_p|}}= \frac{\sqrt{2}G_y} { \sqrt{|\tilde M-\frac {1}{3}{\tilde N}_Y |}}
\label{difmas}
\end{equation}
whereas for the left-handed leptons we have
\begin{equation}
\delta_{12}^{\rm L}\ =\ \frac{ m_{12}^2}{ m_{\tilde L}^2}\ =\ \frac{\sqrt{2}G_y} {\sqrt{|q_p|}}= \frac{\sqrt{2}G_y} { \sqrt{|\tilde M+\frac {1}{2}{\tilde N}_Y |}} \ .
\label{difmaslep}
\end{equation}
Hence,  flavor violation induced by non-constant 3-form fluxes in this context is slightly larger for sleptons than for squarks. 

Making use of the standard type IIB formulae for the mass scales and the unified fine-structure constant we can estimate the natural order of magnitude of these effects. In \cite{Camara:2013fta} we showed that this type of corrections are strongly constrained from FCNC transitions and CP violation, requiring 
squark and slepton masses in the multi-TeV range for natural values of the background parameters. 
Indeed, from 
\begin{equation}
M_{\rm Pl} = \frac{\sqrt{2\textrm{Vol}(B_3)}}{4\pi^3 g_s\alpha'^2} \ ,\qquad 
M_{\rm G}=  \left(\frac{2\alpha_{\rm G}}{\alpha'^2 g_s}\right)^{\frac14} \ , \qquad \alpha_{\rm G}^{-1} = \frac{\textrm{Vol}(S)}{8\pi^4 g_s \alpha'^2}
\label{scaleIIB}
\end{equation}
and noticing that $G_y$ and $q_p$ scale as 
\beq 
G_y\ \sim\  \frac  {c_{y,G}} {\textrm{Vol}(B_3)^{1/6}}\  \ ,\qquad    q_{p,s}\ \sim\ \frac {n}{\textrm{Vol}(S)^{1/2}}  \  \label{scale1}
\eeq 
where $c_{y,G}$ and $n$ are adimensional parameters of order $\mathcal{O}(1)$, we obtain 
\begin{equation}
\delta_{12}\ \sim\ %\frac{m_{12}^2}{\tilde m_{\tilde q}^2}\ \sim \ 
\frac {c_{y,G}}{\sqrt{|n|}}  \left(\frac {M_{\rm G}}{M_{\rm Pl}\,\alpha_{\rm G}}\right)^{1/3} \ .
\label{deltafinalG}
\end{equation}
This is parametrically suppressed by the ratio between unification and Planck scales. However, with the standard unification scale $M_{\rm G}\simeq 10^{16}$ GeV
the suppression is very mild. 
The size of these flavor-violating terms also depends on the variation of the closed string fluxes over $S$ through  $c_{y,G}$ and  is inversely 
proportional to the square root of the open string flux, which is what determines the width of the wavefunctions.  As expected, 
the more localised the wavefunction is, the smaller the amount of flavor violation. 
However, within the current approximation it is not possible to suppress the size of flavor-violating effects by making the open string flux  $q_p$ large, since the perturbative flux expansion that we
are assuming in our computations would break down. Also in that limit the soft scalar masses in eq.~(\ref{ma}) may become tachyonic. On the other hand, 
these flavor-violating effects may be suppressed if the closed string fluxes $G$ vary slowly over $S$, namely if $c_{y,G}$ is small.

Flavor non-diagonal transitions  $\delta_{ij}^{LR}$ mixing left and right also generically appear from soft trilinear scalar
couplings with non-constant closed string fluxes.
By reducing the DBI+CS action in the presence of closed string fluxes and backgrounds for $\Phi$ and $F_2$ we obtain
\begin{equation}
A_{ijk}\ =\ -3g_{\rm YM}\, g_s^{1/2}\int G^*\ \textrm{det}(\vec{v}_\alpha, \vec{v}_\beta, \vec{v}_\gamma)\ f_{\alpha \beta \gamma}\ \chi^i_{a_p^+}\chi^j_{b_q^+}\chi^k_{c_r^+}\, d\textrm{vol}_S
\label{Asoft}
\end{equation}
where 
$f_{\alpha\beta\gamma}=-i\text{Tr}([E_\alpha,E_\beta],E_\gamma)$. When $G_{\bar 1\bar 2\bar 3}$ varies over the 4-cycle $S$, flavor-dependent trilinear couplings appear. Once the Higgs boson takes a vev at the EW scale, these give rise to flavor-violating soft masses of the form $\delta m^2_{LR}$. Although they are suppressed by the Higgs vev, they still might be relevant since the experimental constraints for $\delta m^2_{LR}$ are rather strong. Indeed, the relevant terms in the local expansion of the $G_{(0,3)}$ flux around the triple intersection point are in this case
\begin{equation}
G^*\ =\ G_0^*\sum_{n,m}G_{nm}\bar x^n\bar y^m\, +\, \ldots
\label{Gexpansion}
\end{equation}
When performing the integral (\ref{Asoft}) the rest of the terms in the expansion vanish. To leading order in the magnetic fluxes, the above expression becomes
\begin{equation}
A_{ij}\ \simeq\ \textrm{const.}\, G_0^* \sum_{nm}G_{nm}\int\bar x^n\bar y^m\ \chi^i_{a_p^+}\chi^j_{b_q^+}\chi_{c_r^+}\, d\textrm{vol}_S
\label{Asoft2}
\end{equation}
where all the flavor independent factors have been absorbed in the constant in front of this expression.
Note that we have set $k=1$ since the matter curve $\Sigma_c$ only has one single generation corresponding to the Higgs, while the matter curves $\Sigma_{a,b}$ must accommodate three generations corresponding to the three chiral families of the SM ($i,j=1,2,3$). The computation of this integral is cumbersome but we can easily estimate the order of magnitude of the flavor non-universalities that appear. Since $G_{nm}$ scales as
\begin{equation}
G_{nm} \ \sim\ \frac{c_{nm}}{\textrm{Vol}(B_3)^{\frac{n+m}{6}}}
\end{equation}
with $c_{nm}$ an adimensional parameter, making use of eqs.~(\ref{scaleIIB}) and (\ref{scale1}) we find for $c_{nm}\simeq 1$,
\beq
A_{ij} \ \simeq \ G_0^*\ \left(\frac {M_{\rm G}}{\alpha_{\rm G}\, M_{\rm Pl}\, n^{3/2}}\right)^{2-\frac {i+j}{3}}
\eeq
The induced flavor-violating soft masses are given by
\begin{equation}
(\delta m_{LR}^2)_{ij}\simeq\frac{A_{ij}\langle v\rangle}{m^2_{\rm soft}} 
\end{equation}
where $\langle v\rangle$ is the EW vacuum expectation value of the Higgs and $m^2_{\rm soft}\sim |G_0|^2$. Thus, from the above expressions we obtain that flavor-violating soft masses mixing the first two generations scale as
\beq
(\delta m_{LR}^2)_{12}\ \sim\ \frac{\langle v\rangle}{\sqrt{m^2_{\rm soft}}} \frac{c_{12}}{n^{3/2}}\left(\frac{M_{\rm G}}{\alpha_{\rm G} M_{\rm Pl}}\right)\ \sim\  \frac{6}{\sqrt{m^2_{\rm soft}(\textrm{GeV})}}
\eeq
where in the last equation we have used $M_{\rm G}\simeq 10^{16}$ GeV, $\alpha_{\rm G}=1/24$, $M_{\rm Pl}\simeq 10^{19}$ GeV, $\langle v\rangle=246$ GeV and $n,c\sim \mathcal{O}(1)$. If the SUSY breaking scale is of order $\sqrt{m^2_{\rm soft}}\sim 1$~TeV, then these flavor non-universalities are of order $10^{-2}-10^{-3}$, whereas experimental bounds from 
$\mu\rightarrow e\gamma$ require $(\delta m_{LR}^2)_{e\mu}< 10^{-5}-10^{-6}$ for slepton masses of order $1$~TeV,
see \cite{flavour}.
This suggests again sefermion masses  should be in the multi-TeV range.
It would be interesting to perform a more detailed phenomenological analysis along the lines of \cite{flavour} of flavor violation induced also
by trilinear couplings, both for squarks and sleptons.

\subsection{D3-branes at singularities and flavor non-universalities}
\label{sec53}

There are essentially two options for embedding the SM in IIB/F-theory compactifications. 
We have described one in the previous sections, with the SM gauge group living in the worldvolume of magnetized intersecting 7-branes 
and matter fields localized at 7-brane intersections.  The other possibility is to have the SM fields living in the worldvolume of D3-branes that are on
top of singularities of the compact manifold. The localization on the singularity leads to chiral fermions. In view of the danger of flavor violation
induced by varying fluxes in the case of magnetized 7-branes, it is worth exploring whether D3-branes at singularities are safer in what concerns flavor violation.

The construction  of  MSSM-like models from D3-branes at singularities has been abundantly pursued in the literature, see e.g. \cite{BOOK}
for an introduction and references.
SUSY-breaking soft terms induced by closed string 3-form fluxes have also been worked out \cite{Camara:2003ku,Grana:2002nq,Grana:2003ek,McGuirk:2012sb}. As occurs with D7-branes, 
soft terms for D3-brane fields  only depend on the closed string background in a local transverse patch around the D3-branes. However, contrary to what happens with D7-branes, D3-branes do not span any direction in the compact manifold and non-constant 3-form fluxes in principle do not
induce any flavor violation on the D3-brane scalars.  Here, however we argue 
 that the backreaction of the fluxes on the metric and the 5-form flux does give rise in general to non-universalities also
in the case of D3-brane fields.

The spectrum of matter fields in phenomenological models with D-branes at singularities 
is composed of bi-fundamentals with respect to the
gauge group $U(3)\times U(2)\times U(1)^n$  or some extension of it. It may be represented by a quiver diagram in
which simple groups are represented by nodes and bifundamental fields by links.  The simplest class of such models are
those obtained from local $\mathbf{Z}_N$ orbifold singularities, with $\mathbf{Z}_3$ the simplest example.  
In fact $\mathbf{Z}_3$ singularities are the unique type of $\mathbf{Z}_N$ singularities that lead to 3 generations of quarks and leptons in a supersymmetric context. We consider this case 
as a prototype example. 

It is possible to construct local models with 3 generations and gauge group $U(3)\times U(2)\times U(1)$ based on a stack of 6 D3-branes on a $\mathbf{Z}_3$ singularity.  Out of the 3 U(1)'s, only hypercharge remains massless after
the Green-Schwarz mechanism gives mass to the two orthogonal (anomalous) U(1)'s. The model has three generations 
because there are three local complex coordinates transverse to the D3-branes. 

In Ref. \cite{Camara:2004jj} it was found that 3-form fluxes do not directly lead to soft masses for D3-brane scalars, but only
through the backreaction  of the fluxes on the local metric and 5-form field, as we have reviewed in section \ref{sec41}.  The soft masses that are induced in the presence of backgrounds have the form eq.~(\ref{mBB}), where 
subindices $i,j$ label the local complex coordinates but, within the current context, also label the three SM generations. Type IIB supergravity equations of motion lead to a constraint on the trace of $m_{ij}^2$ \cite{Camara:2003ku}. In particular, if only ISD 3-form fluxes are present, the trace of the scalar mass matrix must vanish 
\beqa
m_1^2 \, +\,  m_2^2\,  +\, m_3^2\, =\, 0
\label{genmass}
\eeqa
This equation implies that there is at least one tachyonic sfermion; or alternatively all masses vanish, as it would occur for instance  
if there is some local permutation symmetry among the three coordinates.  This is the tacit assumption that it is made 
when stating that D3-branes get no masses in the presence of only ISD fluxes. From the point of view of effective supergravity, eq.~(\ref{genmass}) 
is a consequence of K\"ahler modulus domination
and the no-scale structure of the K\"ahler potential.
Soft terms are then generated by subleading corrections to the no-scale structure. In particular, the rhs of eq.~(\ref{genmass}) no longer vanishes in the presence of IASD fluxes \cite{Camara:2003ku}. 
However, we see  from eqs.~(\ref{mBB}) that from a microscopic point of view there is no reason for those corrections to be flavor universal. Even if a particularly symmetric local configuration guaranteeing that $m_i^2=0$ to leading order  is assumed, we also expect additional
sources of non-universalities coming from e.g.~distant localized sources. Those may be 
anti-D3-branes (and/or magnetized D7-branes with a non-trivial D-term) required for uplifting the vacuum from AdS to a (slight) dS vacuum, as in the KKLT approach or generalisations of it. The presence of these extra sources  affect through backreaction the masses of
the SM fields at the singularity, as we already saw in section \ref{sec4}. For instance, the soft scalar masses for D3-brane fields
in the presence of distant anti-D3-branes, computed in that section, are not necessarily flavor-diagonal,
\beq
m_{i\bar{j}}^2\ =\ \text{const.}\left(\frac{6}{r_0^2}z_0^i\bar{z_0}^j-\delta^{ij}\right)\label{m} \ .
\eeq
Thus, the flavor structure depends on the particular geometric distribution of the 
distant sources. Even in the isotropic case, there may be off-diagonal mass terms, see eq.~(\ref{mB2}), that are left invariant by the 
$\mathbf{Z}_3$ symmetry and that lead to flavor violation.

In summary, although D3-branes  at singularities are not directly sensitive to variations of the closed string 3-form fluxes,
 the fact
that family replication in that context is related to the existence of three local transverse complex dimensions easily leads to 
non-universal soft masses for the D3-brane scalars. These non-universal effects can come from the local  backreaction of the 3-form fluxes 
but also from the generic presence of other SUSY-breaking localized sources in specific compactifications. 
Analogous effects are expected for models of D3-branes at  del Pezzo singularities as explored e.g. in ~\cite{pezzo}
and references therein.

\subsection{Fine-tuning the Higgs mass}
\label{sec54}

Low energy SUSY is still the most prominent candidate to explain the stability of the Higgs mass
against quantum corrections. Nevertheless, since no trace of SUSY particles has
been observed at LHC(8\,TeV) so far, it is becoming more and more plausible that some
fine-tuning,  of yet unknown origin,  is at the root of the hierarchy of fundamental scales.
A different type of  fine-tuning, based on anthropic arguments,  was  previously  put forward by Weinberg as a potential explanation
of the smallness of the cosmological constant (c.c.). In that case the existence of a huge landscape of
string theory vacua, parametrized by a large number of discrete choices for fluxes in type IIB string theory,
makes plausible the existence of  vacua with small (and slightly positive) c.c. \cite{Bousso:2000xa}. In the simple KKLT setting  \cite{Kachru:2003aw}
such fine-tuning is possible because of two ingredients: i) there is a large number of 3-form flux choices, making possible to fine-tune
a constant superpotential in the effective action and ii) there is an uplift mechanism provided by anti-D3 branes trapped
on flux throats with tunable wrapping factor. The latter might be replaced by D7-branes with self-dual 
magnetic fluxes, such that they carry an effective anti-D3-brane charge.  This scheme has been generalised in different directions,
as in the LARGE volume scenario \cite{LARGEvolume}.   Although at the moment there is not a complete example fulfilling  all the
phenomenological requirements, it is reasonable to think that type IIB string theory
vacua with fluxes and D-brane sources is sufficiently rich to allow for a landscape  solution to the 
c.c.~problem.  

It is then natural to ask whether type IIB string theory also allows for a simultaneous fine-tuning of the 
Higgs mass.  
If so, has anything to do with the c.c.~fine-tuning in e.g.~the
KKLT scheme?  What would be in this case the microscopic description of the tuning? 
There is of course an obvious difference with the fine-tuning of the c.c., namely the smaller amount of tuning that is required. Indeed, the fine-tuning of the electroweak scale is much less severe,
with  $(M_{\rm EW}/M_{X})^2$ of order 
$(10^{-2})^2 - (10^{-14})^2$,  where $M_X$ is either the string scale $M_s$ or the 
SUSY-breaking scale $M_{ss}$.  This is to be compared to the   $ (10^{-30})^4$  tuning
required for the (almost)  cancellation of the c.c.

It is very difficult at present to give a complete answer to the above questions.  
More modestly, 
in this subsection
we would like to display the different leading microscopic contributions to the Higgs mass matrix that we
can envisage in the bottom-up context of the present paper. 
For definiteness we assume that  a  MSSM-like structure exists at some high scale, anywhere below the 
string scale \cite{Hall:2009nd,hebecker1,imrv,iv,hebecker2,iv2}.
Having SUSY at such high scales does not help with the hierarchy problem but facilitates the
stability of the scalar potential through the absence of potential tachyons. Assuming such a structure,  the Higgs mass matrix at the unification/string scale $M_s$ is
\begin{equation}
m_{\rm Higgs}^2\ =\   \begin{pmatrix}
    m_{H_u}^2   &
     m_3^2 \\
 m_3^2&
 m_{H_d}^2
\end{pmatrix}
\label{matrizmasas}
\end{equation} 
In order to have a light Higgs scalar $h_{\rm SM}$  at scales $M_{\rm EW}\ll M_s$  we need to fine-tune
$m_3^4=m^2_{H_u}m^2_{H_d}$ at the scale $M_{ss}$ at which SUSY is broken, with
soft terms of order $M_{ss}$.  The idea is that at the string scale $m_{H_u}^2m_{H_d}^2>m_3^4$,  and 
MSSM loop corrections lead eventally to  $m_3^4=m^2_{H_u}m^2_{H_d}$ at the SUSY-breaking scale 
$M_{ss}$, so that a massless Higgs doublet survives  \cite{hebecker1,imrv,iv,hebecker2}.
The question that we would like to address is whether flux-induced soft masses allow for
this structure.

We consider the case in which the SM fermions and scalars  are localised on
matter curves (or intersecting 7-branes) on a 4-fold $S$ wrapped by a stack of 7-branes, as in local SU(5) F-theory
models.  For simplicity we take the case of a non-chiral Higgs matter curve in which a $\mu$-term is
generated by an ISD 3-form flux $S_{(0,2)}$ and the dominant source of SUSY-breaking also comes from ISD fluxes.
According to our results in previous sections, at the string scale the Higgss mass matrix has a
qualitative structure of the form
\begin{multline}
m_{\rm Higgs}^2   \ =\  \frac {g_s}{8} \left(\begin{array}{cc}
    2 |G_{(0,3)}|^2+ \frac {1}{4} |S_{(0,2)}|^2   &
     -G_{(0,3)}S_{(0,2)} \\ 
-G_{(0,3)}^*S_{(0,2)} ^*&
 2 |G_{(0,3)}|^2+ \frac {1}{4} |S_{(0,2)}|^2
\end{array}
\right) \, +\, \mathcal{O}(\langle F_2\rangle^2) \\ 
\ +\ \mathcal{O}(S_{(2,0)},G_{(3,0)})\ +\ \ldots
\label{finetunematrix}
\end{multline}
The first term shows the structure of soft terms that we found in section \ref{sec32} for 
the Higgs field on a matter curve with ISD fluxes.  
The reader can check that this matrix is positive definite and
hence has only positive eigenvalues. Thus, at the unification scale there is no zero eigenvalue, and 
thus no light Higgs, for arbitrarily large 3-form fluxes.  On the other hand, as we have already argued,
the RG running from $M_s$ down to the SUSY-breaking scale $M_{ss}$ should lead to det$(M_{\rm Higgs})^2=0$ so that a light Higgs scalar becomes possible
\footnote{In fact in \cite{iv} it was shown that this choice of soft terms when applied to the 
MSSM leads to a massless Higgs upon running 
down to an intermediate SUSY breaking scale $M_{ss}\simeq 10^{10}$ GeV, for certain ranges 
of the $\mu$-parameter. In \cite{iv2} it was also noted that such intermediate scale could be consistent with the large cosmological
tensor fluctuations found at BICEP2 \cite{bicep2}.}.   

There are various types of corrections in eq.~(\ref{finetunematrix}). The factor $\mathcal{O}(\langle F_2\rangle^2)$ denotes corrections quadratic in the magnetic
fluxes that appear in non-chiral Higgs matter curves, such as those computed in section \ref{magnet}.
 Those corrections may have
different origins, as we have already discussed. For instance, they may encode contributions induced by distant anti-branes, computed in section \ref{sec4}. 
We can illustrate those by summing over
the contributions of $n$ distant stacks of  $N_i$ anti-D3-branes located at distances $r_{0i}$ from the SM
7-branes,
\begin{equation}
\delta m_{\rm Higgs}^2 \  =\ \frac {2\sigma ^4}{\pi } \sum_i^n 
\frac {N_i} {Z_{0i}} \frac {F^2_- }{r_{0i}^6} 
 \left(\begin{array}{cc}
4 & 3 \\
3 & 4 
\end{array}
\right)\label{higgsad3}
\end{equation}
with $Z_{0i}=1-g_s N \sigma^2 \pi^{-1}r_{0i}^{-4}$. These corrections are higher order in the magnetic flux since, as we have already mentioned,
only in the presence of magnetic flux $F_-$ in the worldvolume of the 7-branes the backreaction  of  anti-D3-branes is felt  by
 7-branes.  Analogous contributions could be induced by distant 7-branes with self-dual magnetic fluxes $F_+$ 
in their worldvolume.  

There may be also contributions from IASD closed string 3-form fluxes, denoted by $\mathcal{O}(S_{(2,0)},G_{(3,0)})$
in eq.~(\ref{finetunematrix}).
In fact, specific scenarios of moduli fixing include additional 7-branes with gaugino condensation 
or instanton effects that generate superpotentials which are crucial in fixing the K\"ahler moduli of the compactification. 
It was shown in \cite{Baumann:2006th} that such non-perturbative effects generate
both ISD  and  IASD 3-form fluxes as part of their backreaction.  

The size of the various contributions to eq.~(\ref{finetunematrix}) is very model-dependent.
For instance, in certain class of LARGE volume compactifications the
main source of SUSY-breaking is modulus domination
\cite{LARGEvolume}, being locally given by the contribution of ISD 3-form fluxes above.
In others, including the original KKLT scenario, the contribution of distant anti-D3-branes and
IASD fluxes turns out to be non-negligible. 
We can make a naive estimate of the relative size of ISD 3-form 
flux contribution  with respect to that of distant anti-D3-branes.
Considering uniform fluxes $G_{(0,3)}\simeq {\alpha}' /R^3$, we expect flux-induced soft terms of order
\begin{equation}
m_{\rm Higgs}^2\ %=\ \frac{g_s}{2}|G_3|^2\ =\ 
\simeq\ \frac{g_s\alpha'^2}{2R^6}   
\ \simeq \  \frac{M_s^4}{g_sM_{\rm Pl}^2}
\end{equation}
where we have used the type IIB equation
\begin{equation}
M_{\rm Pl}^2\ =\ \frac{8\textrm{Vol}(B_3)}{(2\pi)^6g_s^2\alpha'^4}
\label{Mp}
\end{equation}
with $\textrm{Vol}(B_3)\approx (2\pi R)^6$ the volume of the compact space. 
On the other hand, 
assuming  that the distance between the branes $r_{0i}$  is of the order of the size of the CY,  we can replace $r_{0i}\sim R$ and from eq.~(\ref{higgsad3}) we obtain that the contribution of anti-D3-branes to the Higgs mass matrix scales as
\begin{equation}
\delta m^2_{\rm Higgs}%\sim B\ \sim\ \frac{(4\pi g_s/M_s^4)}{r_0^6}\times \left(N\sigma^2F_«^2\right)=
\ \sim \ \frac{M_s^4}{g_sM_p^2} \  
\times \left( nN\sigma^2F_-^2\right) \ .
\label{scaleB}
\end{equation}
The contribution of distant anti-D3-branes to soft masses is thus comparable to that of 3-form fluxes, except for the fact 
that the first are suppressed by
the magnetic flux factor $\sigma^2F_-^2$. The latter is assumed to be small if the open string fluxes are diluted, so that 
the 3-form flux contribution is expected to dominate in many situations. These scalings also show that a string scale $M_s\sim 10^{15}$~GeV naturally leads to a SUSY-breaking scale of the order $M_{ss}\sim 10^{11}$~GeV, assuming that fluxes are uniform and that adimensional parameters are set to one. 
Nevertheless 3-form fluxes might be diluted at the position of the SM 7-branes or alternatively the local 3-form flux  could be fine-tuned, and therefore   low-energy SUSY with $M_{ss}\sim 1$~TeV can be achievable even with a large string scale. 

The above discussion  shows the abundance  of possible contributions to
the fine-tuning of the Higgs mass. Even in cases where ISD 3-form fluxes dominate SUSY-breaking, the contributions from
open string magnetic fluxes, distant anti-branes or IASD fluxes can probably not be neglected in what concerns 
the Higgss fine-tuning.  All of them are important, along with loop corrections,
as long as the SUSY breaking scale is much above $1-10$~TeV. For instance, if $M_{ss}\simeq 10^{11}$~GeV,
a fine-tuning of 16 orders of magnitude is required in which all these effects can potentially become important. 
In particular, the same anti-D3-branes which play a role in (almost) cancelling the c.c. in
KKLT and related scenarios,  generically influence the fine-tuning of the Higgss mass. 
In this regard, one important point to remark is that the Higgss mass is really directly sensitive to the {\it local } values of
closed and open string  flux densities, rather than to the integrated fluxes. Of course, in a putative
compactification with all moduli fixed, the full geometry (including also the local values of fluxes near the SM branes) depend on the 
global features of the compactification such as the integer flux quanta, and therefore the Higgs mass, like the c.c., will eventually depend on the flux integers.

\section{Discussion}
\label{sec6}

Type IIB orientifolds with intersecting D7-branes and their F-theory extensions constitute a most promising avenue
for the embedding of the observed SM physics within string theory.  In particular, local F-theory SU(5) GUTs 
allow for an embeding of gauge coupling unification within string theory consistent with  the required
structure of Yukawa couplings. In addition closed string fluxes, combined with non-perturbative effects, 
can potentially fix all the moduli of the theory while also breaking supersymmetry.

An important phenomenological question is what are the SUSY-breaking soft terms in such class of
string compactifications.  Trying to answer this question we follow in the present paper a 
{\it bottom-up approach} and concentrate only on the SUSY-breaking effects which are relevant 
for the set of intersecting 7-branes and matter curves in which the SM fields are localised.
These SUSY-breaking effects appear through closed string fluxes, which may be topological or
induced by other distant localized sources in the compactification. The backreaction of these  distant sources
also affect the closed string backgrounds  felt by the SM sector.  No matter how complicated the
structure of the compactification is, the idea is that we can parametrize our ignorance in terms of general 
{\it local} values for the ISD and IASD $G_3$ fluxes as well as $F_5$, dilaton  and metric backgrounds.

This kind of computations for the case of bulk (adjoint) matter fields,  with no magnetic fluxes, 
were performed in Ref.~\cite{Camara:2004jj}. In the present paper we generalize these computations to the case of
chiral matter fields, which are of more direct phenomenological interest.
To compute the effect of all these backgrounds on the soft terms of chiral matter fields
 we use a mixed approach,
making use of information from the DBI+CS action combined with that obtained from the local equations of motion which
describe the wave functions of zero modes on the intersecting matter curves. 
We also study the effect of open string magnetic fluxes on the obtained soft terms.

We present some applications of these results. We consider first  the  local 
setting of F-theory SU(5) matter curves  studied in Ref.~\cite{yukawasSO(12)}. This includes a choice  of
local magnetic fluxes consistent with SU(5) chirality and hypercharge fluxes breaking
the symmetry SU(5) down to the SM gauge group, while allowing for doublet-triplet splitting.  We compute the
SUSY-breaking soft terms induced by ISD $G_3$ fluxes, including also the contribution from
magnetic fluxes.  We find that magnetic fluxes may give rise to substantial non-universal corrections 
which are hypercharge dependent.  This we also this computation for the case of the $E_6$ local configuration
of \cite{Font:2013ida}.
This hypercharge dependence of the soft terms is the soft SUSY-breaking
analogue of  the hypercharge flux corrections found for the gauge kinetic functions in Ref.~\cite{Blumenhagen:2008aw}. 

Another interesting application is the computation of flavor non-universalities. Indeed, for
non-uniform closed string backgrounds the obtained soft terms are flavor non-universal.
We estimate these flavor corrections which appear not only on the scalar masses but
also for  the trilinear scalar couplings. They are generically large, suggesting that sfermions
should have masses at least in  the multi-TeV range to avoid experimental FCNC constraints.
We also argue that the presence of these 
non-universalities do not only appear in the context of intersecting 7-brane models, but also 
in the alternative models in which SM fields live on D3-branes located at singularities.
In the latter case it is not the non-uniformity of fluxes but the generic non-isotropy of the compactifications 
which are the cause of non-universalities.

We  finally briefly discuss the different contributions to the mass matrix of a SUSY Higgs pair of scalars. This we do it for the purpose of giving a geometrical microscopic description 
of the fine-tuning required to get a light Higgs in the context of high scale SUSY-breaking. One finds that the tuning depends on a delicate interplay between the 
closed string flux densities and the presence and location of additional 
brane sources in the compactification.

It would be interesting  to apply in other contexts the soft terms for chiral matter fields that we have obtained.
In particular, the fact that the soft terms explicitly depend on the hyperharge of each sfermion
could leave an imprint on the low-energy spectrum of the MSSM. It would be interesting to perform
a detailed study of the spectra, radiative EW symmetry breaking  and  LHC constraints for a MSSM model with hypercharge dependent
 non-universalities as described in the present paper. As we said, one  stau-lepton would be typically the NLSP 
 and could play an important role in getting viable neutralino dark matter from satu coannihilations. 
 It would also  be interesting to extend the analysis of \cite{Camara:2013fta}, on FCNC limits from non-universal 
 scalar masses, to the case here considered with explicit hypercharge-dependent masses as well as 
 trilinear scalar masses contributing to $\delta{\tilde m}_{LR}$.  Finally, it would be important 
 to apply the results in this paper to a fully semirealistic F-theory compactification in which 
 the full structure of intersecting matter curves and magnetic fluxes would be available.

 \vspace{1.5cm}

\bigskip

\centerline{\bf \large Acknowledgments}

\bigskip

\noindent We thank   F.~Marchesano, M. Montero, D. Regalado   and A.~Uranga for useful discussions.  
This work has been supported by the ERC Advanced Grant SPLE under contract ERC-2012-ADG-20120216-320421; 
by  the grants FPA 2009-09017, FPA 2009-07908, FPA 2010-20807-C02, AGAUR 2009-SGR-168 and ERC Starting Grant ``HoloLHC-306605''. We also thank the 
spanish MINECO {\it Centro de excelencia Severo Ochoa Program} under grant SEV-2012-0249. I.V. is supported through the FPU grant AP-2012-2690.

 \vspace{1.5cm}

\appendix

\end{document}